\documentclass[%
reprint,
amsmath,amssymb,
aps,
prx,
twocolumn,
]{revtex4-2}

\usepackage{macros-design}
\usepackage{macros-phys}
\usepackage[normalem]{ulem}

\newcommand{\mytitle}{Quantum resource trade-offs of mixed sparse single and double factorization for qubitization-based phase estimation
}

\begin{document}

\title{
\mytitle
}

\author{Troy Losey}
\author{Jacob Z. Williams}
\author{Avanish Mishra}
\author{Jian-Xin Zhu}
\author{Yu Zhang}
\email{zhy@lanl.gov}
\affiliation{%
Theoretical Division, Los Alamos National Laboratory, Los Alamos, NM, 87545
}%

\date{\today}

\begin{abstract}
We study a mixed block encoding of quantum-chemistry Hamiltonians for quantum phase estimation (QPE) on fault-tolerant quantum computers. Following a Cholesky decomposition of the electron repulsion integral tensor, we assign each Cholesky tensor to either a double-factorization (DF) branch or a sparse-single-factorization (SPSF) branch. This freedom permits a trade-off between reducing the block-encoding normalization factor $\lambda$ with DF and reducing the coefficient data loaded through quantum read-only memory (QROM) with SPSF. Resource estimates across molecular benchmarks show, however, that the mixed circuit does not generally reduce the Toffoli count relative to the less expensive pure circuit. The main obstacles are the overhead of loading substantial data through separate branch-specific QROMs and the limited benefit of mixing when the pure DF and SPSF circuits have dissimilar costs. Modest savings occur only in restricted parameter regimes, most clearly when logical-qubit requirements are included in the cost metric. These results identify the circuit-level conditions that a useful mixed block encoding must satisfy.
\end{abstract}

\maketitle

\section{Introduction}

Quantum devices such as quantum computers and analog quantum simulators offer promising platforms for simulating quantum systems, as anticipated in foundational proposals for quantum simulation~\cite{feynman:1982qc,Lloyd1073} and developed in subsequent reviews~\cite{RevModPhys.92.015003,chemrev.8b00803,garnet2020cr,Alexeev:2025aa}. Of particular interest are fault-tolerant quantum computers (FTQCs), which exploit entanglement and superposition to address calculations that are intractable for classical computers. One of the most widely studied applications of quantum computers is electronic structure calculations in quantum chemistry~\cite{Aspuru-Guzik1704,PhysRevX.6.031007}. A leading approach is the quantum phase estimation (QPE) algorithm, which estimates the eigenvalues of a Hamiltonian with asymptotically better scaling than the best known exact classical methods~\cite{kitaev1995qpe,PRL835162,Aspuru-Guzik1704}. A long-standing challenge in quantum chemistry is to calculate relative energies with errors below 1.6 millihartrees (commonly called chemical accuracy). This level of precision is essential for predicting chemical reaction rates, reaction barrier heights, and excitation energies. However, it remains prohibitively expensive for large molecular systems using the best currently available classical methods~\cite{szabo1996modern,Friesner6648,C2CP42695A,10.1039/d4sc04755a,10.1021/acs.chemrev.0c00006}. Because chemically relevant QPE requires deep coherent circuits, it is generally regarded as a fault-tolerant rather than a noisy intermediate-scale quantum application~\cite{preskil2018,PRXQuantum.2.017001}. Its advantageous scaling makes QPE a prime candidate for simulating large molecular systems, but the best known implementations remain too resource intensive for early FTQCs~\cite{Lee2021THC,yngp-5fpm,doi:10.1073/pnas.1619152114}.

The leading QPE algorithms utilize qubitization~\cite{Low2019hamiltonian} and have seen significant improvements via more efficient block encoding schemes. The choice of Hamiltonian decomposition affects both the amount of data that needs to be loaded into the quantum circuit and the coefficient 1-norm $\lambda$ of the Hamiltonian in a linear combination of unitaries (LCU) form. Orbital transformations and symmetry-based preprocessing can further reduce this normalization~\cite{Koridon:2021aa,BLISSTHC2025}. Sparse encodings truncate small entries of the electron repulsion integral (ERI) tensor and load the remaining coefficients~\cite{Berry2019QubitizationChemistry}. Single-factorization block encoding decomposes the ERI tensor with a Cholesky decomposition that keeps a limited number of Cholesky tensors~\cite{berry2019qubitization}. This reduces the amount of data loaded into the circuit, but generally increases $\lambda$. Double-factorization block encoding also begins with a Cholesky decomposition. However, the Cholesky tensors are diagonalized to reduce $\lambda$~\cite{PhysRevResearch.3.033055}. Applying the diagonalizing unitaries in the quantum circuit can be expensive, but the associated reduction in $\lambda$ can offset this cost. Sparse single factorization (SPSF), in contrast, retains selected elements of the Cholesky tensors directly and can reduce the amount of coefficient data that must be loaded, although it generally produces a larger $\lambda$ than double factorization (DF). These competing normalization and data-loading costs motivate block encodings that combine the strengths of SPSF and DF.

We consider a mixed block encoding scheme that leverages the flexibility of assigning each Cholesky tensor obtained from a Cholesky decomposition of the ERI tensor to either a DF or an SPSF branch to study resource trade-offs for simulating molecules with QPE. A similar mixed scheme for decomposing the ERI tensor has been shown to reduce computational scaling for auxiliary-field quantum Monte Carlo or exchange energy calculation at alrge~\cite{zhang_2026}, but has yet to be applied to block encoding schemes. We show that this additional flexibility does not generally reduce the Toffoli count because the two branches require quantum read-only memories (QROMs) with substantially different output formats. Furthermore, when the individual DF and SPSF circuits do not have similar costs, the mixed block encoding circuit provides no advantage over the less expensive pure circuit. Our benchmarks find only modest advantages in restricted parameter regimes, most often when logical-qubit savings are included in the cost metric. This paper therefore identifies the circuit-level barriers to mixed block encoding and the conditions under which a mixed SPSF--DF circuit can outperform its pure counterparts.

This paper is organized as follows. In Sec.~\ref{sec:hybrid_qubitization}, we explain the mixed decomposition of the Hamiltonian and motivate potential improvements for block encoding costs. In Sec.~\ref{sec:benchmark}, we explicitly decompose the ERI tensor for different circuits, give an overview of the mixed circuit schematic, compare the mixed-circuit cost with the pure-circuit baselines, and explain barriers to realizing block encoding cost reductions with mixed circuits. In Sec.~\ref{sec:conclusion}, we summarize the results of this work.

\section{Mixed sparse single- and double-factorization scheme for qubitization}
\label{sec:hybrid_qubitization}

\subsection{Motivation}
\label{subsec:motivation_hybrid}

In qubitization-based QPE, the leading Toffoli cost is determined jointly by the Hamiltonian subnormalization $\lambda$ and the cost $C_{\walk}$ of one application of the qubitized walk operator~\cite{Low2019hamiltonian,Gilyen2019QSVT,Berry2019QubitizationChemistry,Lee2021THC}. At fixed target energy precision $\epsilon$, a useful cost model is
\begin{equation}
    T_{\tot}
    \;\propto\;
    \frac{\lambda}{\epsilon}\, C_{\walk}.
    \label{eq:ttot_general_intro}
\end{equation}
Consequently, reducing either $\lambda$ or the amount of coefficient data loaded by the PREPARE and SELECT oracles can lower the total cost, but improving one factor at the expense of the other need not be beneficial.

After a common Cholesky decomposition, individual Cholesky tensors can differ substantially in their LCU coefficient 1-norms and in the number of coefficients retained after truncation. This heterogeneity suggests assigning each tensor either to an SPSF branch, which can reduce the amount of data loaded by QROM, or to a DF branch, which diagonalizes the tensor and generally reduces its contribution to $\lambda$. The resulting mixed circuit is advantageous only if these reductions outweigh the cost of coherently implementing both branches.

\subsection{Electronic Hamiltonian and Cholesky decomposition}
\label{subsec:hamiltonian}

We begin from the electronic Hamiltonian in second quantization,
\begin{equation}
    \Ham = \hat{T} + \hat{V},
    \label{eq:h_split_basic}
\end{equation}
with
\begin{align}
    \hat{T} &= \sum_{\sigma\in\{\uparrow,\downarrow\}} \sum_{p,q=1}^{N/2} T_{pq} \cre{a}_{p,\sigma} \ann{a}_{q,\sigma},
    \label{eq:h1def}
    \\
    \hat{V} &= \frac{1}{2}\sum_{\alpha,\beta\in\{\uparrow,\downarrow\}} \sum_{p,q,r,s=1}^{N/2} V_{pqrs}  \cre{a}_{p,\alpha} \ann{a}_{r,\beta} \cre{a}_{q,\alpha}  \ann{a}_{s,\beta}.
    \label{eq:h2def}
\end{align}
Here $N$ denotes the number of spin-orbitals, $V_{pqrs}$ are the ERI, and $T_{pq}$ are given by
\begin{equation}
    T_{pq} = h_{pq}-\frac{1}{2}\sum_{r=1}^{N/2}V_{prrs},
\end{equation}
where $h_{pq}$ are the one-electron matrix elements.

For the positive-semidefinite electron--electron Coulomb tensor, a Cholesky decomposition (CD) gives
\begin{equation}
    V_{pqrs} \approx \sum_{\gamma=1}^{N_\gamma} L^\gamma_{pr} L^\gamma_{qs},
    \label{eq:chol_decomp}
\end{equation}
where $N_\gamma=\mathcal{O}(N)$. We apply this decomposition to the electron--electron Coulomb tensor $V_{\mathrm{ee}}$. The decomposition is standard in both classical and quantum algorithms for electronic structure~\cite{Aquilante:2007uj, Beebe:1977ve, Koch:2003vj, Roeggen:2008wh, Motta2019THCAFQMC}. Substituting Eq.~\ref{eq:chol_decomp} into Eq.~\ref{eq:h2def} gives
\begin{equation}
    \hat{V} \approx \hat{W} = \frac{1}{2} \sum_{\gamma=1}^{N_\gamma}
    \Bigg(\sum_{\sigma\in\{\uparrow,\downarrow\}} \sum_{p,q=1}^{N/2}L^\gamma_{pq}\cre{a}_{p,\sigma}\ann{a}_{q,\sigma}\Bigg)^2.
    \label{eq:h2_chol}
\end{equation}
Equation~\ref{eq:h2_chol} is the common starting point for SPSF and DF and permits a branch assignment at the level of an individual Cholesky tensor.

\subsection{Mixed SPSF--DF representation}
\label{subsec:mixed_partition}

We partition the Cholesky tensors into two disjoint subsets,
\begin{equation}
  \{L^\gamma\}_{\gamma=1}^{N_\gamma}
  =\mathcal G_{\mathrm{SPSF}}\sqcup\mathcal G_{\DF},
  \label{eq:partition}
\end{equation}
where $\mathcal{G}_{\text{SPSF}}$ contains the Cholesky tensors retained in a block-sparse representation and $\mathcal{G}_{\DF}$ contains the Cholesky tensors further compressed in a low-rank (DF) form. 
For $\gamma\in\mathcal G_{\DF}$, the real symmetric Cholesky matrix is diagonalized,
\begin{equation}
    L^\gamma_{pq}\approx
    \sum_{i=1}^{\Xi^\gamma_{\DF}}
    U^\gamma_{pi}f^\gamma_i U^\gamma_{qi},
    \label{eq:df_local}
\end{equation}
where $f_i^\gamma$ and $U^\gamma$ are the retained eigenvalues and eigenvector matrices, respectively, and $\Xi^\gamma_{\DF}$ is the number of eigenvalues retained after truncation. 
The mixed two-body operator is then
\begin{align}
    \hat W_{\mathrm{mix}}
    =&\frac{1}{2}\sum_{\gamma\in\mathcal G_{\mathrm{SPSF}}}
    \left(\sum_{\sigma,p,q}L^\gamma_{pq}
    \cre{a}_{p,\sigma}\ann{a}_{q,\sigma}\right)^2
    \nonumber\\
    &+\frac{1}{2}\sum_{\gamma\in\mathcal G_{\DF}}U_\gamma
    \left(\sum_{\sigma,i}f^\gamma_i
    \Op{a}^{\dagger}_{i,\sigma}\Op{a}_{i,\sigma}\right)^2U^\dagger_\gamma.
    \label{eq:mixed_operator}
\end{align}
Thus, the DF branch uses basis rotations and retained eigenvalues.

\subsection{Competing normalization and QROM costs}
\label{subsec:mixed_tradeoff}

After the Jordan--Wigner transformation and oblivious amplitude amplification (OAA) used by the circuits in Sec.~\ref{sec:benchmark}, the contribution of tensor $L^\gamma$ to the two-body LCU coefficient 1-norm is
\begin{align}
    \lambda^{\mathrm{SPSF}}_\gamma
    &=\frac{1}{4}
      \left(\sum_{p,q}|L^\gamma_{pq}|\right)^2,
    \\
    \lambda^{\DF}_\gamma
    &=\frac{1}{4}
      \left(\sum_i|f^\gamma_i|\right)^2.
    \label{eq:branch_lambda}
\end{align}
Without branch-dependent truncation, $\sum_i|f_i^\gamma|=\|L^\gamma\|_*$ is the nuclear norm and is no larger than the entrywise 1-norm $\sum_{p,q}|L^\gamma_{pq}|$. DF therefore does not increase the tensorwise normalization in this comparison. In the numerical data, DF likewise normally provides the smaller $\lambda_\gamma$, especially for the Cholesky tensors with the largest norms. The possible benefit of the SPSF branch instead comes from reducing the coefficient data loaded by QROM.

For the circuit analyzed below, a convenient tensorwise measure of loaded data is
\begin{align}
    D^\gamma_{\mathrm{SPSF}}
    &=\Xi^\gamma_{\mathrm{SPSF}}b_{\mathrm{SPSF}},
    \\
    D^\gamma_{\DF}
    &=\Xi^\gamma_{\DF}(b_{\DF}+b_{\mathrm{rot}}),
    \label{eq:branch_data}
\end{align}
where $\Xi^\gamma_{\mathrm{SPSF}}$ counts retained matrix elements, $\Xi^\gamma_{\DF}$ counts retained eigenvalues, and the $b$'s are the corresponding QROM output widths. In particular, $b_{\mathrm{rot}}=N\beth/2$ accounts for the orbital-rotation angles and can make the DF data load large even when relatively few eigenvalues are retained.

The mixed circuit does not load the sum of these data through a single QROM. It uses separate SPSF and DF QROMs with significantly different output widths, so their optimized Toffoli costs add. This loss of QROM efficiency is the central overhead of the mixed construction. Ancilla reuse can reduce the logical-qubit count because the two branch QROMs are not active simultaneously, but that saving does not generally compensate for the additional Toffoli cost when Toffolis are the primary optimization target.

\subsection{Global branch optimization and benchmark-informed expectation}
\label{subsec:hybrid_routing}

A branch cannot be selected reliably from the rank, sparsity, $\lambda_\gamma$, or $D_\gamma$ of one Cholesky tensor alone. Moving one tensor changes the aggregate $\lambda$, the dimensions of both branch QROMs, and hence the marginal cost of moving every other tensor. We therefore minimize the complete resource estimate by sweeping over pairs of branch assignments. The optimization is repeated from several initial assignments, including tensorwise heuristics and the two pure limits, as described in the Supplemental Material.

Equation~\ref{eq:ttot_general_intro} makes the necessary condition for a useful mixed circuit transparent. The pure SPSF and DF circuits must first have comparable costs. Otherwise, assigning any appreciable number of tensors to the more expensive branch provides too little reduction in $\lambda$ or QROM data to overcome the cost of the additional branch. Even near this crossover, the mixed circuit is penalized because two comparably sized QROMs cost more than one QROM containing the same total amount of data. Accordingly, the mixed SPSF--DF construction should be viewed as a conditional optimization rather than as a generally lower-cost factorization.

This expectation is borne out by Sec.~\ref{sec:benchmark}: the mixed circuit usually does not reduce the Toffoli count relative to the cheaper pure circuit. Modest improvements occur only over narrow ranges of SPSF truncation thresholds, and are more common for the combined Toffoli--qubit metric than for Toffoli count alone. These results shift the main conclusion from a proposed asymptotic advantage to a characterization of the circuit-design conditions required for a mixed representation to be beneficial.

\section{Benchmarks and quantum resource trade-offs}
\label{sec:benchmark}

We benchmark the mixed SPSF--DF circuit against its two pure limits using the total Toffoli count $T$ and the composite metric $TQ$, where $Q$ is the number of logical qubits. The central question is not whether individual Cholesky tensors exhibit different sparsity and rank, but whether exploiting that heterogeneity reduces the complete QPE resource estimate after accounting for $\lambda$, all QROMs, basis rotations, branch controls, and reusable ancillas.

Our benchmark set contains chemically challenging localized active-space Hamiltonians for synthetic \ce{[Fe2S2]} and \ce{[Fe4S4]} clusters \cite{li_chan_2017, li_2026a}. The iron-molybdenum cofactor (FeMoco) \cite{li_etal_2019a, li_2026} and P-cluster \cite{li_etal_2019b, li_2024} of nitrogenase; and the heme center of cytochrome P450 \cite{goings_etal_2022, goings_etal_2022a}. These active spaces have been deposited as publicly available datasets (see the preceding references) in the FCIDUMP format \cite{knowles_handy_1989}.  In this work, we choose the active spaces corresponding to the oxidized (spin-4) state of the nitrogenase P-cluster  and the largest active space (58 orbitals) of the P450 heme unit. The active spaces are stored in a localized and orthonormal representation obtained by a three-step procedure given in the Supplementary Information of \cite{li_etal_2019b} and summarized here. First, the partially occupied active space and fully occupied core are obtained by partitioning the natural orbitals from a spin-unrestricted mean-field computation. Pipek--Mezey localization \cite{pipek_mezey_1989} is applied separately to the core and active spaces. Finally, some localized natural orbitals from the core are added manually to the active space based on chemical intuition (for instance, those resembling doubly occupied \ce{d} orbitals on, or orbitals bonding to, the metal centers), and the Pipek--Mezey procedure is repeated on the enlarged set.

To investigate the system size dependence of costs, we also apply our scheme to linear $n$-acenes ($n = 1,\dots,6$), from benzene to hexacene. We construct and approximately relax the structures with \textsc{Avogadro} \cite{hanwell_etal_2012} and obtain the Hartree--Fock Hamiltonian of the acenes---in the relatively localized basis of orthogonalized atomic orbitals---using the correlation-consistent double-zeta Dunning basis set \cite{dunning_1989} orthogonalized by L\"owdin (symmetric) procedure \cite{lowdin_1950}, with the \textsc{PySCF} software package \cite{sun_etal_2018, sun_etal_2020, pu_sun_2025a, sun_etal_2026}, version 2.11.0. We also leverage \textsc{PySCF} to process the iron-sulfur clusters' FCIDUMP files.

The Cholesky decomposition of the electron repulsion integrals, sparse representation of the Cholesky tensors, double factorization, and quantum resource estimation are performed by in-house code \textsc{QubitChem}.
Unless stated otherwise, the Cholesky decomposition uses a threshold of $10^{-4}$. The branch-dependent truncation budgets are specified below. Rank and sparsity diagnostics for every system are provided in the Supplemental Material.

The localized metal-cluster active spaces produce Cholesky tensors that are comparatively dense, whereas the spatially extended acenes exhibit substantially greater sparsity. These differences provide a useful test of the mixed-scheme hypothesis: if tensorwise structural heterogeneity were sufficient, the acenes would be natural candidates for robust SPSF--DF savings. As shown below, however, sparsity and rank alone do not predict the circuit optimum because the two branches contribute differently to $\lambda$ and require QROMs with different output widths.

\subsection{LCU coefficient 1-norm}

The LCU coefficient 1-norm $\lambda$ fixes the normalization of the encoded Hamiltonian and, at fixed energy precision, the number of qubitized-walk applications. Because SPSF and DF produce different LCU representations, a reduction in coefficient data cannot be evaluated independently of the corresponding change in $\lambda$. We next give the SPSF, DF, and mixed expressions consistently used for all resource estimates.

Following the derivations in the Appendices of Ref.~\cite{Lee2021THC}, we perform a Jordan-Wigner transformation~\cite{JW_transformation} to rewrite the creation and annihilation operators as unitary Pauli strings according to
\begin{equation}
    \cre{a}_{p,\sigma}\ann{a}_{q,\sigma} + \cre{a}_{q,\sigma}\ann{a}_{p,\sigma} = \frac{X_{p,\sigma}\vec{Z}X_{q,\sigma} + Y_{p,\sigma}\vec{Z}Y_{q,\sigma}}{2}
\end{equation}
and
\begin{equation}
    \cre{a}_{p,\sigma}\ann{a}_{p,\sigma} = \frac{\mathbb{I} - Z_{p,\sigma}}{2}
\end{equation}
for Pauli operators $X$, $Y$, and $Z$, and $A_p\vec{Z}A_q = A_pZ_{p+1}\cdots Z_{q-1}A_q$. We can then rewrite the approximated ERI tensor $\hat{W}$ from Eq.~\ref{eq:h2_chol} in LCU form as
\begin{align}
     \hat{W} &= \frac{1}{8}\sum_{\gamma=1}^{N_\gamma}\Bigg(\sum_{\sigma\in\{\uparrow,\downarrow\}} \sum_{p,q=1}^{N/2}L^\gamma_{pq}\big(\cre{a}_{p,\sigma}\ann{a}_{q,\sigma} + \cre{a}_{q,\sigma}\ann{a}_{p,\sigma}\big)\Bigg)^2 \nonumber \\
     &= \frac{1}{8}\sum_{\gamma=1}^{N_\gamma}\Bigg(\sum_{\sigma\in\{\uparrow,\downarrow\}} \sum_{p,q=1}^{N/2}L^\gamma_{pq}Q'_{pq\sigma}\Bigg)^2,
\end{align}
where
\begin{equation}
    Q'_{pq\sigma} = 
    \begin{cases}
        X_{p,\sigma}\vec{Z}X_{q,\sigma}, & p<q, \\
        Y_{p,\sigma}\vec{Z}Y_{q,\sigma} & p>q, \\
        \mathbb{I} - Z_{p,\sigma} & p=q.
    \end{cases}
\end{equation}
The same procedure allows us to write the one-body operator from Eq.~\ref{eq:h1def} in LCU form as
\begin{equation}
    \hat{T} = \frac{1}{2}\sum_{\sigma\in\{\uparrow,\downarrow\}} \sum_{p,q=1}^{N/2}T_{pq}Q'_{pq\sigma}.
\end{equation}

We also implement the improvement from Ref.~\cite{Lee2021THC} by removing the identity matrix from $Q'_{pq\sigma}$ when $p=q$. Using
\begin{equation}
    Q_{pq\sigma} = 
    \begin{cases}
        X_{p,\sigma}\vec{Z}X_{q,\sigma}, & p<q, \\
        Y_{p,\sigma}\vec{Z}Y_{q,\sigma} & p>q, \\
        -Z_{p,\sigma} & p=q,
    \end{cases}
\end{equation}
the approximated two-body operator $W$ becomes
\begin{align}
    \hat{W} &= \frac{1}{8}\sum_{\gamma=1}^{N_\gamma}\Bigg(\sum_{\sigma\in\{\uparrow,\downarrow\}} \sum_{p,q=1}^{N/2}L^\gamma_{pq}Q_{pq\sigma} + 2\sum_{r=1}^{N/2}L_{rr}^\gamma\mathbb{I}\Bigg)^2 \\
    \label{eq:expandedW}
    &= \frac{1}{8}\sum_{\gamma=1}^{N_\gamma}\Bigg(\sum_{\sigma\in\{\uparrow,\downarrow\}} \sum_{p,q=1}^{N/2}L^\gamma_{pq}Q_{pq\sigma}\Bigg)^2 \\
    &+ \frac{1}{2}\sum_{\sigma\in\{\uparrow,\downarrow\}} \sum_{p,q,r=1}^{N/2}V_{pqrr}Q_{pq\sigma} + \frac{1}{2}\sum_{\gamma=1}^{N_\gamma}\Bigg(\sum_{r=1}^{N/2}L_{rr}^\gamma\mathbb{I}\Bigg)^2, \nonumber
\end{align}
where we use $\sum_{\gamma=1}^{N_\gamma}L_{pq}^\gamma L_{rr}^\gamma \approx V_{pqrr}$ for the second term in Eq.~\ref{eq:expandedW} when $N_\gamma$ is sufficiently large, so that this term can be added to the one-body operator. The effective one-body operator used by the SPSF block encoding is then
\begin{equation}
    \hat{T}'_{\text{SPSF}} = \frac{1}{2}\sum_{\sigma\in\{\uparrow,\downarrow\}} \sum_{p,q=1}^{N/2}T'_{pq}Q_{pq\sigma},
\end{equation}
where
\begin{equation}
    T'_{pq} = T_{pq} + \sum_{r=1}^{N/2}V_{pqrr}.
\end{equation}
The two-body operator used by the SPSF circuit is what remains of Eq.~\ref{eq:expandedW} after removing the one-body contribution and the term with $\mathbb{I}$ that only contributes a constant energy shift:
\begin{equation}
    \hat{V}'_{\text{SPSF}} = \frac{1}{8}\sum_{\gamma=1}^{N_\gamma}\Bigg(\sum_{\sigma\in\{\uparrow,\downarrow\}} \sum_{p,q=1}^{N/2}L^\gamma_{pq}Q_{pq\sigma}\Bigg)^2.
\end{equation}

The LCU coefficient 1-norm for the SPSF circuit can then be calculated as
\begin{equation}
    \lambda^{\text{SPSF}} = \lambda^{\text{T,SPSF}}+\lambda^{\text{V,SPSF}},
\end{equation}
where
\begin{equation}
    \lambda^{\text{T,SPSF}}=\sum_{p,q=1}^{N/2}\big|T'_{pq}\big|
\end{equation}
and
\begin{equation}
    \lambda^{\text{V,SPSF}} = \frac{1}{4} \sum_{\gamma=1}^{N_\gamma} \Bigg( \sum_{p,q=1}^{N/2} \big|L^\gamma_{pq}\big| \Bigg)^2.
\end{equation}
The one-body term is not truncated, but the sparse truncation sets some elements of $L^\gamma$ to zero, so they won't contribute to $\lambda^{\text{V,SPSF}}$ or be loaded into the circuit. Note that the sums over spin contribute a factor of $2$ to $\lambda^{\text{T,SPSF}}$ and a factor of $4$ to $\lambda^{\text{V,SPSF}}$. However, OAA allows us to divide $\lambda^{\text{V,SPSF}}$ by a factor of $2$.

For the DF block encoding, the decomposition of the Hamiltonian is largely the same as what was outlined for the SPSF circuit above, except that the one-body operator and Cholesky tensors are diagonalized to further reduce the LCU coefficient 1-norm. The resulting two-body term for the DF circuit is
\begin{equation}
    \hat{V}'_{\text{DF}} = \frac{1}{8}\sum_{\gamma=1}^{N_\gamma}U_\gamma\Bigg(\sum_{\sigma\in\{\uparrow,\downarrow\}} \sum_{p=1}^{\Xi^\gamma}f^\gamma_{p}Z_{p,\sigma}\Bigg)^2U_\gamma^\dagger,
\end{equation}
where $f_p^\gamma$ are the eigenvalues of $L^\gamma$ and $U_\gamma$ are eigenvector matrices of the diagonalization. $\Xi^\gamma$ is the number of eigenvalues that are kept in the $\gamma$th Cholesky tensor after truncation, and since we now only have diagonal elements, $Q_{pq\sigma}$ is always $-Z_{p,\sigma}$. Similarly,
\begin{equation}
    \label{eq:1BodyDF}
    T'_{\text{DF}} = -\frac{1}{2}U_0\Bigg(\sum_{\sigma\in\{\uparrow,\downarrow\}} \sum_{p=1}^{N/2}f^{(0)}_{p}Z_{p,\sigma}\Bigg)U^\dagger_0,
\end{equation}
Here, the index $\gamma=0$ for $f_p^\gamma$ and $U_\gamma$ just refers to the diagonalization of the matrix with elements $T'_{pq}$.

The LCU coefficient 1-norm for the DF block encoding can then be calculated as

\begin{equation}
    \lambda^{\text{DF}} = \lambda^{\text{T,DF}}+\lambda^{\text{V,DF}},
\end{equation}
where
\begin{equation}
    \label{eq:lambdaTDF}
    \lambda^{\text{T,DF}}=\sum_{p=1}^{N/2}\big|f_p^{(0)}\big|
\end{equation}
and
\begin{equation}
    \lambda^{\text{V,DF}} = \frac{1}{4} \sum_{\gamma=1}^{N_\gamma} \Bigg( \sum_{p=1}^{\Xi^\gamma} \big|f^\gamma_{p}\big| \Bigg)^2.
\end{equation}

For the mixed circuit, we always place the one-body component on the DF branch and split the Cholesky tensors between the two branches. Since the one-body term isn't truncated, there is little to gain by placing it on the SPSF branch. Instead, placing it on the DF branch minimizes $\lambda^{\text{T}}$. Therefore, the one-body term and $\lambda^{\text{T}}$ for the mixed circuit are given by Eq.~\ref{eq:1BodyDF} and Eq.~\ref{eq:lambdaTDF}, respectively.

The two-body term for the mixed circuit is
\begin{align}
    \hat{V}'_{\text{mix}} &= \frac{1}{8}\sum_{\gamma\in \mathcal{G}_{\text{SPSF}}}\Bigg(\sum_{\sigma\in\{\uparrow,\downarrow\}} \sum_{p,q=1}^{N/2}L^\gamma_{pq}Q_{pq\sigma}\Bigg)^2 \nonumber \\
    &+ \frac{1}{8}\sum_{\gamma\in \mathcal{G}_{\text{DF}}}U_\gamma\Bigg(\sum_{\sigma\in\{\uparrow,\downarrow\}} \sum_{p=1}^{\Xi^\gamma}f^\gamma_{p}Z_{p,\sigma}\Bigg)^2U_\gamma^\dagger,
\end{align}
and the contribution to $\lambda$ from the two-body term is
\begin{align}
    \lambda^{\text{V,mix}} &= \frac{1}{4} \sum_{\gamma\in \mathcal{G}_{\text{SPSF}}} \Bigg( \sum_{p,q=1}^{N/2} \big|L^\gamma_{pq}\big| \Bigg)^2 \nonumber \\
    &+ \frac{1}{4} \sum_{\gamma\in \mathcal{G}_{\text{DF}}} \Bigg( \sum_{p=1}^{\Xi^\gamma} \big|f^\gamma_{p}\big| \Bigg)^2.
\end{align}
The total LCU coefficient 1-norm for the mixed block encoding circuit is then
\begin{equation}
    \lambda^{\text{mix}} = \lambda^{\text{T,DF}}+\lambda^{\text{V,mix}}.
\end{equation}

\subsection{Truncation scheme}

Elementwise truncation of the original ERI tensor is straightforward, but truncation after Cholesky decomposition is nonlocal in the reconstructed Hamiltonian. Removing one entry $L_d^\gamma$ eliminates every LCU term containing that entry, with a total coefficient weight equal to $|L_d^\gamma|(2\sum_{p=1}^{\Xi^\gamma}|L_p^\gamma|-|L_d^\gamma|)$. A useful truncation rule must therefore account for the current contents of the entire Cholesky tensor rather than only the magnitude of the candidate entry.

We therefore minimize the retained coefficient data at fixed discarded LCU weight, defined as
\begin{equation}
    \lambda_{\text{disc}} = \lambda_{\text{full}} - \lambda_{\text{trunc}},
\end{equation}
where $\lambda_{\text{full}}$ and $\lambda_{\text{trunc}}$ are the LCU coefficient 1-norms of the Hamiltonian before and after truncation, respectively.

The truncation scheme that we use is much like the one proposed by \citet{PhysRevResearch.3.033055}, where the values $L_n^\gamma$ in the $\gamma$th Cholesky tensor are truncated if
\begin{equation}
    \Bigg(\sum_{p=1}^{N/2}|L_p^\gamma|\Bigg)|L_n^\gamma| < \text{threshold}.
\label{eq:vBTruncation}
\end{equation}
After performing a Cholesky decomposition on $V$, this rule accounts for the cost of discarding each Cholesky-tensor element more accurately than simply discarding values whose magnitudes are below a threshold. The Cholesky tensors are diagonalized for DF but not for SPSF. Nevertheless, this truncation can be applied to both factorizations by taking $p$ and $n$ to index the nonzero elements of each Cholesky tensor in the corresponding basis.

Our truncation scheme explicitly considers the contribution to $\lambda_{\text{disc}}$ each time an element is removed. Unlike Eq.~\ref{eq:vBTruncation}, it does not double count the term $|L_p^\gamma L_n^\gamma|$ when $p=n$. We also recalculate the cost of removing elements from the $\gamma$th Cholesky tensor after every truncation because a subsequent removal cannot discard Hamiltonian terms that have already been eliminated. Specifically, we evaluate the cost
\begin{equation}
    |L^\gamma_m|\left(2\sum_{p=1}^{\Xi^\gamma}|L^\gamma_p|-|L^\gamma_m|\right)
    \label{eq:truncCost}
\end{equation}
of discarding the smallest element $L^\gamma_m$ in each Cholesky tensor and recalculate this cost for $L^\gamma$ whenever a value is discarded. At each step, we truncate the candidate $L^\gamma_m$ with the smallest cost according to Eq.~\ref{eq:truncCost}, stopping when $\lambda_\text{disc}$ reaches the prescribed threshold.

\begin{figure}[h]
    \centering
    \includegraphics[width=\linewidth]{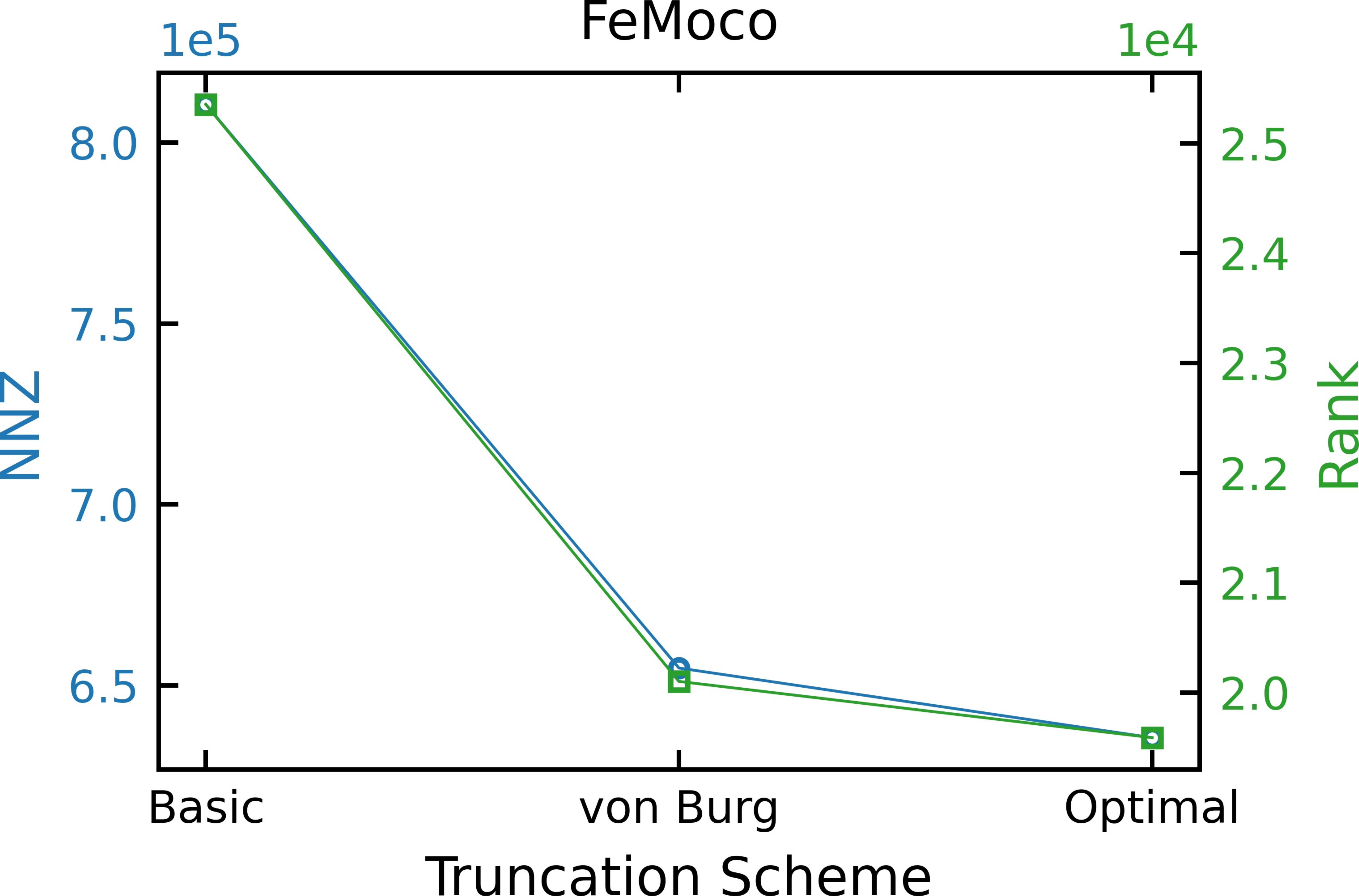}
    \caption{Comparison of coefficient data for SPSF (left axis) and DF (right axis) for different truncation schemes. $\lambda_\text{disc}$ is preserved between truncation methods applied to the same circuit.}
    \label{fig:truncation}
\end{figure}

We show in Fig.~\ref{fig:truncation} that this truncation scheme reduces the coefficient data further than the truncation scheme of von Burg \textit{et al.} from Eq.~\ref{eq:vBTruncation}, while $\lambda_\text{disc}$ remains constant between all truncation schemes for SPSF on the left axis and DF on the right axis. The truncation schemes considered are basic, where Cholesky tensor elements are truncated when their magnitudes are below some threshold; von Burg, where truncation follows Eq.~\ref{eq:vBTruncation}; and optimal, which follows Eq.~\ref{eq:truncCost}. $\lambda_\text{disc}$ is fixed for the SPSF circuit according to basic truncation with a threshold of $10^{-4}$, and it is fixed for the DF circuit according to von Burg truncation with a threshold of $1.25 \times 10^{-3}$. This truncation threshold for the DF circuit was determined to reach chemical accuracy for FeMoco~\cite{Lee2021THC}. However, the threshold for the SPSF circuit was chosen to be consistent with typical sparse thresholds. We note that it is not appropriate to compare $\lambda_\text{disc}$ between different decompositions of $H$ because the unitary strings associated with the discarded coefficients can vary significantly, so the same $\lambda_\text{disc}$ does not guarantee that the energy error due to truncation will be similar. Throughout the paper, we use the optimal truncation scheme unless otherwise stated, although we often present the truncation threshold in a more intuitive form by preserving $\lambda_\text{disc}$ between the optimal truncation and basic truncation for SPSF and von Burg truncation for DF.

\subsection{Quantum resource estimates}
\label{subsec:qre}

\begin{figure*}[t]
    \centering
    \includegraphics[width=\linewidth]{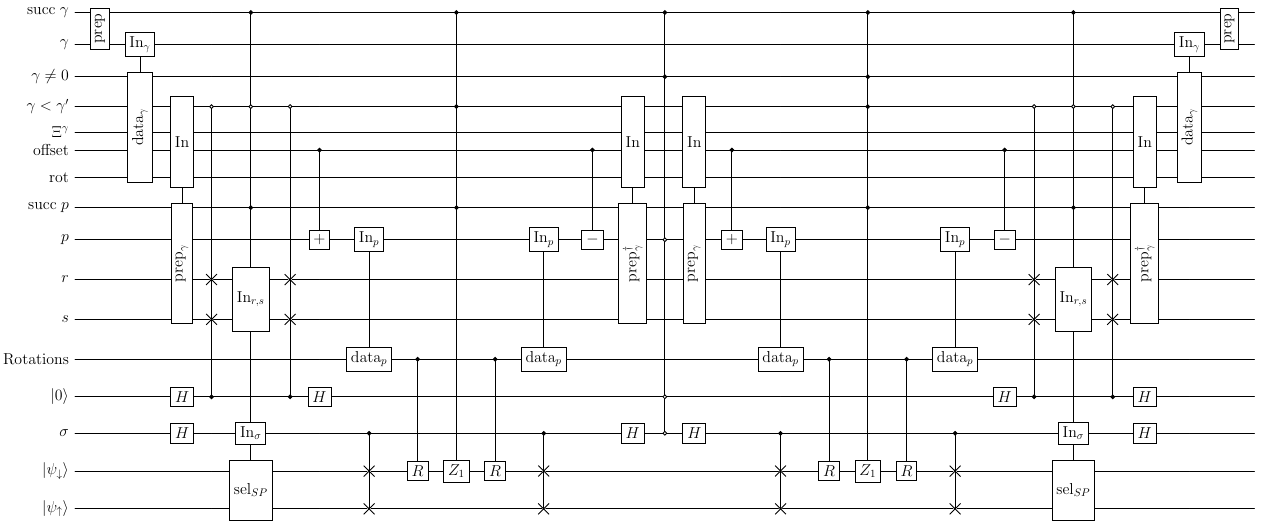}
    \caption{Schematic of the mixed SPSF/DF block encoding circuit. The register labeled $\gamma$ is the outer PREPARE register that indexes the Cholesky tensors. The register labeled $p$ is the inner PREPARE register that indexes nonzero elements or eigenvalues within each Cholesky tensor. The registers labeled succ $\gamma$ and succ $p$ flag successful preparation of the equal-superposition states on the $\gamma$ and $p$ registers, respectively. The $\gamma\neq 0$ register controls operations applied only to the two-body term. The $\gamma<\gamma'$ register is the SPSF/DF branch control. The $\Xi^\gamma$, offset, and rot registers store information used to prepare the $p$ register, controlled on the $\gamma$ register. The $r$ and $s$ registers store information needed for the SPSF SELECT subcircuit. The Rotations register stores information needed to perform the basis rotations for the DF SELECT subcircuit. The $|0\rangle$ and $\sigma$ registers generate the $r \leftrightarrow s$ and spin symmetries, respectively. The registers labeled $|\psi_\downarrow\rangle$ and $|\psi_\uparrow\rangle$ are the spin-down and spin-up components of the system.}
    \label{fig:hybridQC}
\end{figure*}

The freedom to place each Cholesky tensor on its optimal branch might be expected to reduce block-encoding costs. However, we find that implementing a mixed block-encoding circuit introduces important disadvantages. In this section, we outline the quantum resource estimates and explain the benefits and drawbacks of the mixed implementation.

The estimates of the Toffoli gates and qubits required to implement our mixed scheme are derived from the circuit shown in Fig.~\ref{fig:hybridQC}. The mixed and SPSF circuits are based largely on the architecture outlined in Appendices A, B, and C of Ref.~\cite{Lee2021THC}. Exact descriptions of circuit components and costs for the mixed and SPSF circuits are provided in the Supplemental Material. We do not reproduce the DF circuit because it is identical to the one detailed in Appendix C of Ref.~\cite{Lee2021THC}.

Our SPSF circuit is structured similarly to the DF circuit to make the two compatible within the mixed scheme. The $p$ register, which indexes the Cholesky-tensor eigenvalues in the DF circuit, instead indexes the nonzero elements of each Cholesky tensor in the SPSF circuit. The SELECT circuit and the inner PREPARE QROM differ between the two branches because the SELECT circuits require different inputs. The dominant cost of the SPSF circuit is the QROM that prepares the state on the $p$ register, whereas the dominant DF costs are the QROM that outputs the rotation data and the application of those rotations.

The circuit components in Fig.~\ref{fig:hybridQC} operate as follows. The prep box creates a superposition over the Cholesky index $\gamma$, and the $\text{In}_\gamma$ and $\text{data}_\gamma$ boxes output data for the next step. The $\text{In}$ and $\text{prep}_\gamma$ boxes then create a superposition over $p$, which indexes the retained eigenvalues or nonzero matrix elements in each Cholesky tensor. Controlled swaps complete PREPARE by exploiting the $r \leftrightarrow s$ symmetry of $L^\gamma_{r,s}$. The $\text{In}_{r,s}$, $\text{In}_\sigma$, and $\text{sel}_{SP}$ boxes implement SPSF SELECT, as in Fig.~13 of Ref.~\cite{Lee2021THC}. The controlled $+$ ($-$) box creates (uncomputes) a contiguous index for all eigenvalues on the DF branch. The $\text{In}_p$ and $\text{data}_p$ boxes output the rotation data, and the controlled $R$ box performs the DF basis rotations. The controlled $Z_1$ box applies the DF SELECT operation. Finally, the $\text{In}$ and $\text{prep}^\dagger_\gamma$ boxes uncompute the $p$-register preparation, and the central reflection is performed as part of OAA.

\begin{figure*}[t]
    \centering
    \includegraphics[width=\linewidth]{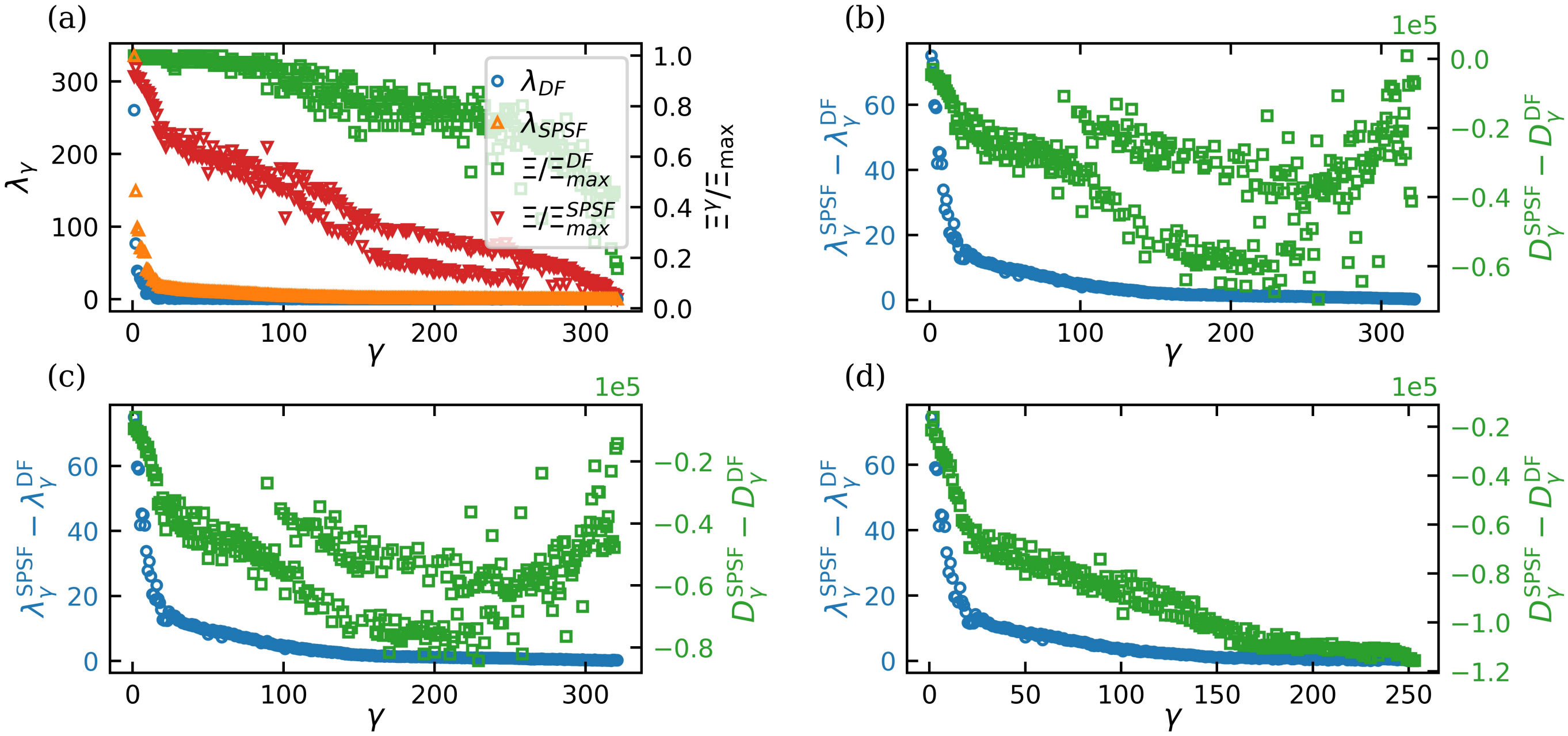}
    \caption{Comparison of the costs associated with each Cholesky tensor for FeMoco on the DF and SPSF branches. Panel (a) shows how $\lambda_\gamma$ and the normalized coefficient count $\Xi^\gamma/\Xi_{\text{max}}$ vary among individual Cholesky tensors on each branch. In panels (b)--(d), the left axes show how much larger $\lambda_\gamma$ is on the SPSF branch, and the right axes show how much less data $D^\gamma$ must be loaded on that branch. All panels use the optimal truncation scheme, with $\lambda_\text{disc}$ set by von Burg truncation at a threshold of $1.25 \times 10^{-3}$ for the DF data and by basic truncation at thresholds of $2 \times 10^{-4}$ [(a) and (c)], $10^{-4}$ (b), and $5 \times 10^{-4}$ (d) for the SPSF data.}
    \label{fig:CTmosaic}
\end{figure*}

The properties of individual Cholesky tensors in Fig.~\ref{fig:CTmosaic} illustrate why different tensors can benefit from different branches. Figure~\ref{fig:CTmosaic}(a) shows $\lambda_\gamma$ and the normalized coefficient count $\Xi^\gamma/\Xi_{\max}$. Here, $\Xi^\gamma$ denotes the retained rank on the DF branch and the number of retained matrix elements on the SPSF branch. The corresponding maximum values are $N/2$ and $N^2/8+N/4$, respectively. The tensors are truncated so that the pure DF and SPSF circuits have equal costs. The DF branch has substantially smaller $\lambda_\gamma$ for the Cholesky tensors with the largest 1-norm, whereas $\lambda_\gamma$ is comparable between the branches for tensors with smaller 1-norms. This occurs because diagonalization on the DF branch reduces the 1-norm, but the reduction in $\lambda_\gamma$ is modest when the original 1-norm is already small. The total QROM data $D=db$, where $d$ is the number of entries and $b$ is the output width, must also be considered to quantify the branch-dependent QROM costs.

Figures~\ref{fig:CTmosaic}(b)--\ref{fig:CTmosaic}(d) show the differences in $\lambda_\gamma$ and in the QROM data $D_\gamma=\Xi^\gamma b$ between the two branches. For the DF branch, we include data for both preparation of the $p$ register and the rotation angles. These QROMs have the same number of entries $\Xi^\gamma_{DF}$ but different output widths, $b_{DF}=n_{\Xi,DF}+\aleph_2+2$ and $b_{rot}=N\beth/2$. Thus, the total data loaded for each Cholesky tensor is $D^\gamma_{DF} = \Xi^\gamma_{DF}(b_{DF}+b_{rot})$. For the SPSF branch, the only QROM considered is the one that prepares the $p$ register. It loads $\Xi^\gamma_{sp}$ entries with output width $b_{sp}=4n_N+\aleph_2+2$, so $D^\gamma_{sp} = \Xi^\gamma_{sp}b_{sp}$. The DF truncation is the same in all panels, whereas the SPSF truncation is varied so that the pure circuits have equal costs in panel (c), DF is cheaper in panel (b), and SPSF is cheaper in panel (d). Cholesky tensors on the SPSF branch usually require less QROM data but always increase $\lambda$. Placing large-1-norm tensors on the DF branch and the remaining tensors on the SPSF branch can therefore decrease the QROM sizes without substantially increasing $\lambda$. These reductions, however, must overcome the mixed-circuit overheads discussed below.

\begin{figure*}[t]
    \centering
    \includegraphics[width=\linewidth]{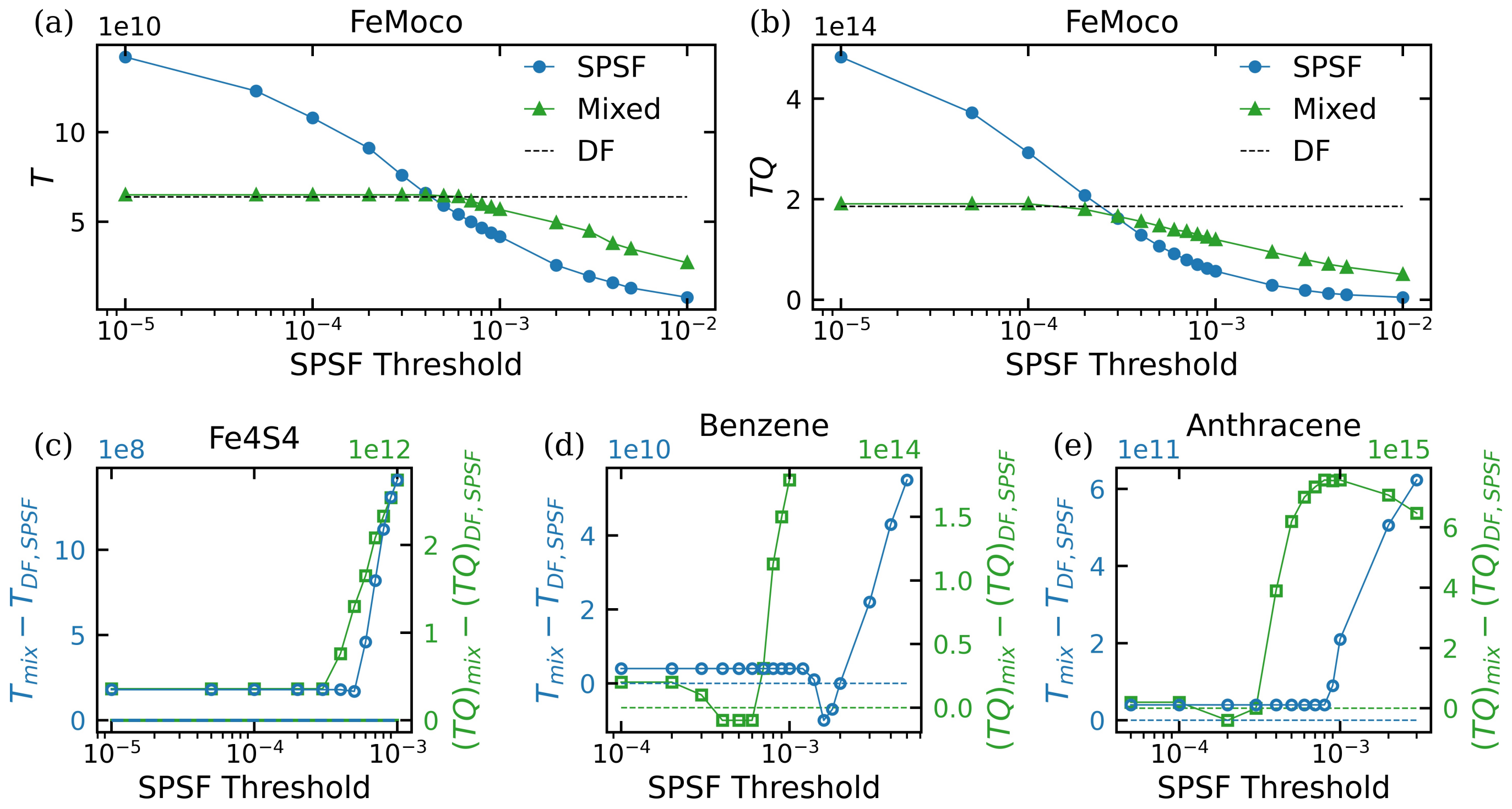}
    \caption{Cost comparison between the DF, SPSF, and mixed circuits for different molecules. Panels (a) and (b) show the Toffoli costs $T$ and the product of the Toffoli and qubit costs $TQ$, respectively, for FeMoco. Panels (c)--(e) show the difference between the cost of the mixed circuit and the lesser of the DF and SPSF circuits for $T$ on the left axes and $TQ$ on the right axes for \ce{Fe4S4}, benzene, and anthracene, respectively. All subplots use the optimal truncation with $\lambda_\text{disc}$ set according to the von Burg truncation with a threshold of $1.25 \times 10^{-3}$ for DF and $\lambda_\text{disc}$ set according to the basic truncation with the threshold listed on the x-axis for SPSF.}
    \label{fig:costMosaic}
\end{figure*}

Although the control costs needed to split the Cholesky tensors across the DF and SPSF branches are small, the mixed circuit requires significantly more Toffoli gates $T$ because most of the Hamiltonian data are loaded through two QROMs instead of one. These are the QROM that prepares the $p$ register for the SPSF branch and the QROM that outputs rotation angles for the DF branch. The QROM Toffoli cost is
\begin{equation}
    T_{\text{QROM}} = \bigg\lceil \frac{d}{k} \bigg\rceil + b(k-1),
\label{eq:qromTCost}
\end{equation}
where $k$ is a power of $2$ chosen to optimize QROM cost. Ignoring the ceiling function and allowing $k_{opt}=\sqrt{d/b}$ to be continuous gives the minimum QROM Toffoli cost as
\begin{equation}
    T_{\text{QROM}}^{\text{min}} \approx 2\sqrt{D}.
\end{equation}
The square root scaling causes the Toffoli cost for one QROM $T_{\text{QROM}}^{\text{one}}$ to increase by
\begin{equation}
    \frac{T_{\text{QROM}}^{\text{two}}}{T_{\text{QROM}}^{\text{one}}} = \sqrt{2}
    \label{eq:2QROMoverhead}
\end{equation}
when the same total data are split between two equally sized QROMs, where $T_{\text{QROM}}^{\text{one}}$ and $T_{\text{QROM}}^{\text{two}}$ denote the one- and two-QROM costs, respectively. This balanced regime is representative of a mixed circuit that meaningfully uses both branches. Because the QROMs dominate $T$, the mixed circuit is typically unable to outperform the pure DF or SPSF circuit in Toffoli count.

To potentially reduce the mixed-circuit Toffoli count, we consider combining the dominant QROMs for the two branches. This is challenging because the dominant SPSF QROM has output width $b_{\text{SPSF}}=4n_N+\aleph_2+2$, where $n_N = \lceil \log(N/2) \rceil$, whereas the dominant DF QROM has output width $b_{\text{rot}}=\beth N/2$. Typical values are $\aleph_2=10$ and $\beth=20$, so $b_{\text{rot}} \gg b_{\text{SPSF}}$. To combine these QROMs without padding the SPSF entries with large numbers of zeros, we consider packing $c=\lfloor b_{\text{rot}}/b_{\text{SPSF}}\rfloor$ SPSF entries into a single entry with output width $\approx b_{\text{rot}}$. This reduces the number of SPSF entries by a factor of $c$ and substantially decreases the combined QROM cost. However, extracting $c$ ordinary SPSF entries of width $b_{\text{SPSF}}$ from a packed entry by controlled swaps has Toffoli cost
\begin{equation}
    C_{\text{unpack}} \approx b_{\text{SPSF}}(c-1) \approx 10N
\end{equation}
for each application of the combined QROM. We take the repacking cost to be zero when measurement-based uncomputation is used. The SPSF entries must be unpacked twice in the circuit, giving a total Toffoli unpacking cost of $\approx 20N$. In our benchmarks, this overhead prevents the mixed circuit from significantly outperforming the pure DF or SPSF circuit in Toffoli count.

One advantage of not combining QROMs is that they can share temporary ancilla qubits with each other. QROMs generally use $b$ persistent qubits and $b(k-1) + \lceil\log(d/k)\rceil$ ancilla qubits. Using $k_{opt}$, the qubits needed for the largest QROM are
\begin{equation}
    Q_{\text{QROM}} \approx \sqrt{D},
\end{equation}
while other QROMs require only $b$ qubits because they can reuse ancilla qubits from the larger QROM. When the dominant QROMs have equal data and share ancilla qubits, we halve $D$ for our largest QROM, so the dominant qubit source $Q_{\text{QROM}}$ is reduced by a factor of $1/\sqrt{2}$.

The logical-qubit reduction provides an opportunity for the mixed circuit to outperform the pure DF or SPSF circuits. Although Toffoli count is generally the main driver of circuit time, freeing additional qubits may accelerate the calculation by allowing more qubits to be used for magic-state factories. We consider a qubitization cost function of the form
\begin{equation}
    C(\alpha)=TQ^\alpha,
\end{equation}
where $\alpha$ specifies how important $Q$ is. We then compare this cost function for DF and mixed circuits with uncombined dominant QROMs. A simple but illustrative model is
\begin{align}
    \label{eq:dfCostModel}
    C_{\text{DF}}(\alpha) &= \big(4\sqrt{D}+A_{\text{DF}}\big)\big(\sqrt{D}+B_{\text{DF}}\big)^\alpha, \text{ and} \\
    C_{\text{mix}}(\alpha) &= \Bigg(8\sqrt{\frac{D}{2}}+A_{\text{mix}}\Bigg)\Bigg(\sqrt{\frac{D}{2}}+B_{\text{mix}}\Bigg)^\alpha.
    \label{eq:mixedCostModel}
\end{align}
Here, $D$ is the data loaded by the largest QROMs. $A_{\text{DF}} \lesssim A_{\text{mix}}$ are the remaining Toffoli costs and $B_{\text{DF}} \lesssim B_{\text{mix}}$ are the remaining qubits, which are only slightly larger for the mixed circuit because of control logic for the branches and the SPSF select subcircuit. We assume the mixed circuit has equal $D$ for the QROMs on each branch. Plugging in values for the Li FeMoco Hamiltonian~\cite{li_etal_2019a} where the DF circuit has $D = 3.0 \times 10^7$, $A_{\text{DF}} = 1.6 \times 10^4$, and $B_{\text{DF}} = 3 \times 10^2$, we expect the mixed circuit to outperform the DF circuit for $\alpha > 0.66$, even without considering that the mixed circuit can lower the total QROM data $D$.

Figure~\ref{fig:costMosaic} compares the costs of the pure DF and SPSF circuits with that of the mixed circuit. For FeMoco, Figs.~\ref{fig:costMosaic}(a) and~\ref{fig:costMosaic}(b) show the Toffoli count $T$ and the product of the Toffoli count and logical qubit count $TQ$, respectively, as functions of the SPSF thresholds that match $\lambda_{\text{disc}}$ for the optimal truncation. When the SPSF threshold is small, every Cholesky tensor is best placed on the DF branch, so the mixed-circuit costs are slightly larger than the DF costs. Once the SPSF threshold is sufficiently large, the mixed circuit can lower its costs by placing some tensors on the SPSF branch. For $T$, however, the DF branch dominates the mixed circuit even after the SPSF circuit becomes the cheapest pure circuit. This behavior results from the two-QROM overhead in Eq.~\ref{eq:2QROMoverhead}. The mixed-circuit value of $T$ also cannot approach the SPSF value because we always place the one-body term on the DF branch, whose SELECT operation remains costly even when that branch contains little data. We observe a slight decrease in $TQ$, as predicted by Eqs.~\ref{eq:dfCostModel} and~\ref{eq:mixedCostModel}. The decrease is modest, however, and occurs only in the narrow regime where the DF and SPSF circuits have similar costs. In general, truncation thresholds are chosen to be as large as possible while retaining chemical accuracy, so the two pure circuits will not reliably have similar costs for arbitrary molecules.

Figures~\ref{fig:costMosaic}(c)--\ref{fig:costMosaic}(e) show results for \ce{Fe4S4}, benzene, and anthracene. The left axes show the mixed-circuit Toffoli count relative to the lower of the two pure-circuit counts, and the right axes show the corresponding difference in $TQ$. When a solid curve falls below the dashed zero line, the mixed circuit outperforms both pure circuits. For our smallest system, \ce{Fe4S4}, the mixed circuit never outperforms the pure circuits. For benzene, the mixed circuit can reduce both $T$ and $TQ$, but the improvement is modest and occurs only within a narrow range of truncation thresholds. For anthracene, we observe no reduction in $T$ and only a modest reduction in $TQ$ over a narrow threshold range. Thus, the Toffoli reduction is not a general feature of acenes.

While there are several barriers that impede the ability of mixed block encoding circuits to outperform the individual circuits that comprise them, this work shows that the mixed circuit can outperform DF and SPSF circuits when the pure circuits have comparable costs and when optimizing for lower qubit count in addition to lower Toffoli count. Furthermore, it suggests a route toward resource savings from mixed circuits with more compatible compression schemes, where the large QROMs on each branch can be efficiently combined.

\section{Conclusion}
\label{sec:conclusion}

We considered a mixed DF--SPSF block-encoding scheme and found that mixed Hamiltonian decompositions typically do not reduce the Toffoli count relative to the pure decompositions. Two primary barriers prevent meaningful reductions. First, loading most of the data through two comparably sized QROMs rather than one dominant QROM increases the Toffoli count. Although sharing ancilla qubits between the QROMs can reduce the circuit's logical-qubit requirement, reducing the Toffoli count is typically the higher priority for early FTQCs. Second, it is difficult to obtain comparable pure-circuit costs while maintaining chemical accuracy under aggressive truncation. When the pure decompositions do not have comparable costs, placing Cholesky tensors on the more expensive branch offers little benefit. The resulting reductions in total QROM data or in the LCU coefficient 1-norm $\lambda$ are then too modest for the mixed decomposition to outperform the cheaper pure decomposition. Furthermore, although we took measures to prevent a Cholesky tensor from being placed on a branch that would truncate it heavily, we did not explicitly calculate the QPE energy error due to truncation. The mixed decomposition, which assigns tensors to minimize circuit cost, may therefore have a larger truncation error than the pure decompositions.

Although combining DF and SPSF generally does not reduce the Toffoli count, novel block-encoding circuits could require fewer quantum resources if their dominant QROMs can be combined efficiently. Doing so would eliminate one of the primary barriers to reducing the cost of a mixed circuit. This approach is not feasible for the mixed DF--SPSF circuit studied here because the largest QROMs in the two pure circuits have dramatically different output widths. Moreover, engineering compatible dominant QROMs may itself complicate the design of low-cost block-encoding circuits. Our study thus reveals circuit-level challenges that limit the ability of mixed schemes to improve the feasibility of QPE on early FTQCs.

\begin{acknowledgments}
The research presented in this article was supported by the Laboratory Directed Research and Development (LDRD) program of Los Alamos National Laboratory (LANL). 
This research used computational resources provided by the Institutional Computing (IC) Program and the Darwin testbed at Los Alamos National Laboratory (LANL), funded by the Computational Systems and Software Environments subprogram of LANL's Advanced Simulation and Computing program. 
LANL is operated by Triad National Security, LLC, for the National Nuclear Security Administration of the U.S. Department of Energy (contract no. 89233218CNA000001). 
\end{acknowledgments}

\bibliography{quantcomput_ref}

\clearpage
\begin{widetext}
\appendix

 \begin{center}
 \large\bf Supplemental Material for ``\mytitle"
 \end{center}

\def\CGBSupplementBodyOnly{1}
\ifdefined\CGBSupplementBodyOnly
\newcommand{\CGBStandaloneEnd}{}
\else

\documentclass[10pt,
               aps,
               prx]
              {revtex4-2}
\bibliographystyle{apsrev4-2}

\usepackage{macros-design}
\usepackage{macros-phys}

\begin{document}

\title{
Supplemental Material for ``Quantum resource trade-offs of mixed sparse single and double factorization for qubitization-based phase estimation"
}
\author{Troy Losey}
\author{Jacob Z. Williams}
\author{Avanish Mishra}
\author{Jian-Xin Zhu}
\author{Yu Zhang}
\email{zhy@lanl.gov}
\affiliation{%
Theoretical Division, Los Alamos National Laboratory, Los Alamos, NM, 87545
}%

\maketitle
\newcommand{\CGBStandaloneEnd}{\end{document}}
\fi

\renewcommand{\thefigure}{S\arabic{figure}}
\setcounter{figure}{0}

\section{Branch optimization}
\label{sec:SI:branchOpt}

The strength of a mixed block-encoding scheme is the freedom to place each Cholesky tensor on the branch that minimizes the total circuit cost. The reduction can arise from lowering either the coefficient 1-norm $\lambda$ or the amount of coefficient data loaded by QROM. In the main text, we characterize a tensor $L^\gamma$ by its contribution $\lambda_\gamma$ to the total two-body normalization, $\lambda^V=\sum_{\gamma=1}^{N_\gamma}\lambda_\gamma$, and by the coefficient data retained after truncation. The optimal branch cannot, however, be determined for each $L^\gamma$ independently because it depends on the assignments of all other Cholesky tensors. Here we describe the procedure used to assign every tensor to either the DF or the SPSF branch.

The first issue to address is that the error due to truncation can increase significantly if we perform the DF and SPSF truncations independently on all of the Cholesky tensors and then place each $L^\gamma$ on the branch that most lowers the total cost of the circuit. This is because one truncation scheme may heavily truncate certain Cholesky tensors while the other truncation scheme heavily truncates different Cholesky tensors. Generally, all truncation schemes will agree that $L^\gamma$ with large 1-norms should not be truncated much and $L^\gamma$ with small 1-norms should be truncated aggressively, but there are some intermediate Cholesky tensors where the truncation schemes can disagree. This can result in the final decomposition of the Hamiltonian having too many heavily truncated Cholesky tensors for the error due to truncation to remain below the allowed threshold for chemical accuracy.

To make the different truncation schemes more consistent with which $L^\gamma$ are heavily truncated, we remove any Cholesky tensors from the Hamiltonian that are left empty by either one of the truncation schemes. In practice, this means that we perform the DF truncation on all Cholesky tensors and discard all $L^\gamma$ that had all eigenvalues truncated. Then we perform SPSF truncation on the set of $L^\gamma$ that weren't discarded by DF truncation and discard all $L^\gamma$ that had all elements truncated by SPSF truncation to get the final set of Cholesky tensors that will be used in the block encoding. We also perform DF truncation again on this final set of $L^\gamma$, but this time it is guaranteed that no Cholesky tensors will have all eigenvalues truncated, so no more $L^\gamma$ are removed and there is no need to perform SPSF truncation again. This forces both truncation schemes to end up with the same set of $L^\gamma$, and that set of $L^\gamma$ will not include any of the Cholesky tensors that were most heavily truncated by either scheme. Restricting both truncation schemes to the same set of nonempty Cholesky tensors reduces their individual truncation efficiency, but it is necessary to control truncation error when the schemes are combined in the mixed circuit.

The other challenge for choosing the optimal branch for each Cholesky tensor is that the optimal branch for each Cholesky tensor depends on the branch of every other Cholesky tensor. Because the QPE cost depends on the product of $\lambda$ and the single-walk cost $C_\text{walk}$, the preferred assignment of one tensor depends on both the coefficient data and the contributions to $\lambda$ from all other tensors, as well as the one-body term. Moreover, for $N_\gamma$ Cholesky tensors and $2$ branches, there are $2^{N_\gamma}$ different branch assignments, so it is typically not feasible to check the cost of every assignment.

We represent an assignment by a bit string such as $|0110\cdots10\rangle$, ordered by the Cholesky index $\gamma$, with $0$ denoting DF and $1$ denoting SPSF. Motivated by two-site sweeping algorithms, we sweep through pairs of adjacent tensors and evaluate all four branch assignments for each pair. We adopt a new assignment whenever it lowers the circuit cost and repeat the sweeps until a complete sweep produces no change. The resulting assignment is a local minimum. Because it is not guaranteed to be the global minimum, we repeat the procedure from several initial assignments and compare the converged costs.

The initial states include the branch assignment estimated by the procedure in the main text, alternating branches like $|0101...01\rangle$, all Cholesky tensors on the same branch like $|00...00\rangle$ and $|11...11\rangle$, and the first third of the sites on the DF branch and the rest on the SPSF branch like $|00...1111\rangle$. This last initial state is inspired by the fact that the SPSF branch typically has less coefficient data, but the DF branch has a dramatically smaller $\lambda_\gamma$ for the first Cholesky tensors. For these $5$ initial states, the branch distribution that results in the lowest cost for the circuit is typically found by most of the initial states, which gives us confidence that we are finding the true optimal branch distribution of Cholesky tensors that minimizes the cost of the mixed circuit. 

\section{Circuit Resources}

\subsection{SPSF circuit}

The detailed Toffoli and qubit costs of the SPSF circuit are derived from architecture outlined in the appendices of Ref.~\cite{Lee2021THC} and detailed below. Since these steps also appear in the mixed circuit, we refer to the mixed circuit shown Fig.~\ref{fig:hybridQC} the main text to show these steps.

\begin{enumerate}
    \item We begin by preparing a superposition state over the $\gamma$ register, where $\gamma$ indexes the Cholesky tensors from the first factorization. This is the box labeled "prep" in the mixed circuit figure in the main text and is performed with the following steps.
    \begin{enumerate}
        \item Prepare an equal superposition over $N_\gamma+1$ basis states with Toffoli cost $3n_{N_\gamma}-3\eta+2b_r-9$. This is for $N_\gamma$ Cholesky tensors and the 1-body term which is indexed by $\gamma=0$. $n_{N_\gamma} = \lceil \log(N_\gamma+1) \rceil$ is the number of states in the $\gamma$ register, $\eta$ is the largest integer where $N_\gamma$ is divisible by $2^\eta$, and $b_r$ are bits of precision for a rotation on an ancilla qubit. We use $b_r = 8$ as a typical value.
        \item Use a small QROM with alt value size $n_{N_\gamma}$ and keep value size $\aleph_1$ to prepare the state on the $\gamma$ register. This gives output size $b_{p1} = n_{N_\gamma}+\aleph_1$, where we use $\aleph_1 = 10$ as a typical value. The QROM cost is then
        \begin{equation}
            \bigg\lceil \frac{N_\gamma+1}{k_{p1}} \bigg\rceil + b_{p1}(k_{p1}-1).
            \label{eq:outPrepQROM}
        \end{equation}
        \item Perform an inequality test with cost $\aleph_1$.
        \item Perform a controlled swap with cost $n_{N_\gamma}$.
    \end{enumerate}
    \item Output data into additional registers used to prepare the $p$ register. This step is indicated by the boxes labeled "$\text{In}_\gamma$" and "$\text{data}_\gamma$" in the mixed-circuit figure. The small QROM takes the $\gamma$ register as input and outputs a one-bit flag for $\gamma=0$, the number $\Xi^\gamma$ of nonzero entries in $n_\Xi$ bits, an offset in $n_{N_\gamma,\Xi}$ bits, and $b_r$ bits for a rotation on an ancilla. The total output width is $b_o = n_\Xi+n_{N_\gamma,\Xi}+b_r+1$, where $n_\Xi = \lceil \log(\max_\gamma[\Xi^\gamma]) \rceil$ and $n_{N_\gamma,\Xi} = \lceil \log(N_\gamma\Xi+\Xi^{(0)}) \rceil$. Here $N_\gamma\Xi=\sum_{\gamma=1}^{N_\gamma}\Xi^\gamma$ is the total number of nonzero Cholesky-tensor entries and $\Xi^{(0)}=N^2/8+N/4$ is the number of symmetry-unique entries in the one-body term. The QROM cost is
    \begin{equation}
        \bigg\lceil \frac{N_\gamma+1}{k_o} \bigg\rceil + b_o(k_o-1).
        \label{eq:dataQROM}
    \end{equation}
    \item Now we prepare the state on the $p$ register by using the data output in the previous step. This operation is shown by the boxes labeled "In" and "$\text{prep}_\gamma$ in the mixed circuit figure. This is by far the most expensive step in the circuit because the QROM used is large. The steps for the inner preparation are as follows.
    \begin{enumerate}
        \item Create an equal superposition state on the $p$ register. This time the cost is $7n_\Xi+2b_r-6$ because the operation is controlled on the $\gamma$ register.
        \item Add the offset to the $p$ register to create a contiguous register to go into the QROM. These additions cost $n_{N_\gamma,\Xi}-1$.
        \item Use QROM with output size $b_{\text{sp}} = 4n_N + \aleph_2 + 2$, where we use the typical value $\aleph_2=10$. The $+2$ is for the sign bit and its alt value. This QROM has $N_\gamma\Xi$ entries for all nonzeroes in the ERI tensor and $\Xi^{(0)}$ entries for the 1-body term. The QROM cost is then
        \begin{equation}
            \bigg\lceil \frac{N_\gamma\Xi + \Xi^{(0)}}{k_{sp}} \bigg\rceil + b_{sp}(k_{sp}-1).
            \label{eq:inPrepQROM}
        \end{equation}
        \item Perform an inequality test and a controlled swap with costs $\aleph_2$ and $n_\Xi$, respectively.
    \end{enumerate}
    \item Next we perform the first iteration of the select subcircuit as part of oblivious amplitude amplification (OAA). This is shown as swaps controlled on the $|0\rangle$ register, then boxes labeled "$\text{In}_{r,s}$", "$\text{In}_\sigma$", and "$\text{sel}_\text{SP}$", and then more swaps controlled on the $|0\rangle$ register. The costs are as follows.
    \begin{enumerate}
        \item Apply controlled swaps on the $r$ and $s$ registers to exploit the symmetry of the Cholesky tensors to finish preparing the state with cost $n_N$.
        \item Perform sparse select. It does not need to be controlled, so the cost is $2N-4$.
        \item Invert the controlled swaps on the $r$ and $s$ registers with cost $n_N$.
    \end{enumerate}
    \item Now we invert the preparation of the $p$ register. This appears in the mixed circuit figure as boxes labeled "In" and "$\text{prep}^\dagger_\gamma$". The costs are the same as step 3 except we can invert the QROM much cheaper by using measurement-based uncomputation, so inverting the QROM costs
    \begin{equation}
        \bigg\lceil \frac{N_\gamma\Xi+\Xi^{(0)}}{k'_{sp}} \bigg\rceil + k'_{sp},
        \label{eq:invInPrepQROM}
    \end{equation}
    instead of the cost in step 3(c).
    \item Reflect on the following qubits that were prepared in step 3.
    \begin{enumerate}
        \item The $n_\Xi$ qubits in the $p$ register.
        \item The $\aleph_2$ qubits used for coherent alias sampling for the equal superposition state for the $p$ register.
        \item The $|0\rangle$ qubit used to control the swaps that generate the $p,q$ symmetry.
        \item The $\sigma$ qubit used to control the swaps that generate the spin symmetry.
        \item One qubit rotated for OAA.
    \end{enumerate}
    These reflections on $n_\Xi+\aleph_2+3$ qubits must be controlled on the succ $\gamma$ and $\gamma\neq0$ qubits, so the total cost for these reflections is $n_\Xi+\aleph_2+3$. This appears in the mixed circuit figure as a controlled $Z_1$ gate in the center of the circuit.
    \item Repeat steps 3-5 without the 1-body term. Neglecting the one body term reduces the number of entries for the QROM and inverse QROM in Eq.~\eqref{eq:inPrepQROM} and Eq.~\eqref{eq:invInPrepQROM}, respectively, by $\Xi^{(0)}$. It also means that the SPSF select in step 4 needs to be controlled on the $\gamma\neq0$ register, so its cost increases by $2$.
    \item Invert steps 1 and 2. In the mixed circuit figure, these are the boxes labeled "$\text{In}_\gamma$" and "$\text{data}_\gamma$" to invert step 2 and the box labeled "prep" to invert step 1. The costs are the same except it is cheaper to invert the QROM, so Eq.~\eqref{eq:outPrepQROM} changes to
    \begin{equation}
        \bigg\lceil \frac{N_\gamma+1}{k'_{p1}} \bigg\rceil + k'_{p1}
    \end{equation}
    and Eq.~\eqref{eq:dataQROM} changes to
    \begin{equation}
        \bigg\lceil \frac{N_\gamma+1}{k'_o} \bigg\rceil + k'_o.
    \end{equation}
    \item The following qubits must be reflected to finish the quantum walk.
    \begin{enumerate}
        \item The $n_{N_\gamma}$ qubits in the $\gamma$ register.
        \item The $n_\Xi$ qubits in the $p$ register.
        \item The $\aleph_1$ qubits used for coherent alias sampling for the equal superposition state for the $\gamma$ register.
        \item The $\aleph_2$ qubits used for coherent alias sampling for the equal superposition state for the $p$ register.
        \item The $|0\rangle$ qubit used to control the swaps that generate the $p,q$ symmetry.
        \item The $\sigma$ qubit used to control the swaps that generate the spin symmetry.
        \item The two qubits rotated for OAA.
    \end{enumerate}
    These reflections are on $n_{N_\gamma}+n_\Xi+\aleph_1+\aleph_2+4$ qubits, so the Toffoli cost is $n_{N_\gamma}+n_\Xi+\aleph_1+\aleph_2+2$.
    \item The quantum walk steps must be controlled with unary iteration on an ancilla qubit. The cost of unary iteration and controlling the reflection are $2$.
\end{enumerate}

The total Toffoli cost for a step of the quantum walk for the SPSF circuit is then
\begin{equation}
\begin{split}
    &4N+4n_N+9n_{N_\gamma}+34n_\Xi+4n_{N_\gamma,\Xi}+3\aleph_1+6\aleph_2+12b_r-6\eta-45 \\
    &+\bigg\lceil\frac{N_\gamma+1}{k_{p1}}\bigg\rceil+b_{p1}(k_{p1}-1)+\bigg\lceil\frac{N_\gamma+1}{k'_{p1}}\bigg\rceil+k'_{p1}+\bigg\lceil\frac{N_\gamma+1}{k_{o}}\bigg\rceil+b_o(k_o-1)+\bigg\lceil\frac{N_\gamma+1}{k'_{o}}\bigg\rceil+k'_o \\
    &+\bigg\lceil\frac{N_\gamma\Xi+\Xi^{(0)}}{k_{sp}}\bigg\rceil+\bigg\lceil\frac{N_\gamma\Xi}{k_{sp}}\bigg\rceil+2b_{sp}(k_{sp}-1)+\bigg\lceil\frac{N_\gamma\Xi+\Xi^{(0)}}{k'_{sp}}\bigg\rceil+\bigg\lceil\frac{N_\gamma\Xi}{k'_{sp}}\bigg\rceil+2k'_{sp}
\end{split}
\end{equation}

The qubits required for the SPSF circuit are as follows.

\begin{enumerate}
    \item There are $\lceil\log(\mathcal{I}+1)\rceil$ qubits for the control register for the phase estimation and $\lceil\log(\mathcal{I}+1)\rceil-1$ qubits for the unary iteration.
    \item The system uses $N$ qubits.
    \item The $\gamma$ register uses $n_{N_\gamma}$ qubits, one qubit is needed to record the success of preparation on the $\gamma$ register, and one qubit is rotated during this state preparation.
    \item The QROM used to prepare the $\gamma$ register has an output size of $b_{p1}=n_{N_\gamma}+\aleph_1$, uses $\aleph_1$ for the inequality test, and needs one qubit to record the output of the inequality test.
    \item The QROM that outputs data needed to prepare the $p$ register has an output size of $b_o = n_\Xi+n_{N_\gamma,\Xi}+b_r+1$.
    \item The $p$ register has $n_\Xi$ qubits and two more are needed to flag the success of preparation on the $p$ register and be rotated during state preparation.
    \item The QROM used to prepare the $p$ register has an output size of $b_{sp}=4n_N+\aleph_2+2$, uses $\aleph_2$ for the inequality test, and needs one qubit to record the output of the inequality test. There are also
    \begin{equation}
        b_{sp}(k_{sp}-1)+\bigg\lceil\log\bigg(\frac{N_\gamma\Xi+\Xi^{(0)}}{k_{sp}}\bigg)\bigg\rceil
    \end{equation}
    temporary ancilla qubits needed for this QROM. Temporary ancillas from many other parts of the circuit can be neglected because this QROM requires substantially more of them.
    \item The phase-gradient state uses $b_r$ qubits.
    \item Two qubits are required for the $|0\rangle$ and $\sigma$ registers.
    \item There is one qubit needed to perform controlled Hadamards with Toffoli gates.
\end{enumerate}

The total number of qubits needed for the SPSF circuit is then
\begin{equation}
\begin{split}
    &2\lceil\log(\mathcal{I}+1)\rceil + N + 2n_{N_\gamma}+2\aleph_1+\aleph_2 + 2n_\Xi+n_{N_\gamma,\Xi} \\
    &+ b_{sp}k_{sp}+\bigg\lceil\log\bigg(\frac{N_\gamma\Xi+\Xi^{(0)}}{k_{sp}}\bigg)\bigg\rceil + 2b_r + 9.
\end{split}
\end{equation}

\subsection{Mixed circuit}

The costs for the mixed circuit are derived from architecture outlined in the appendices of Ref.~\cite{Lee2021THC} and similar to those of the SPSF circuit. The differences are that the mixed circuit has added architecture from the DF circuit and there are a few additional controls for the branches. The mixed circuit is shown in Fig.~\ref{fig:hybridQC}, and the differences in Toffoli costs between the mixed and SPSF circuits are the following.

\begin{enumerate}
    \item Same as step 1 for the SPSF circuit.
    \item Same as step 2 for the SPSF circuit except the QROM output size increases by $1$ to additionally output the branch control $\gamma<\gamma'$, so $b_o^\text{mix} = n_\Xi+n_{N_\gamma,\Xi}+b_r+2$. Note that the definitions of $n_\Xi$ and $n_{N_\gamma,\Xi}$ haven't change even though $\Xi^\gamma$ is the number of nonzero eigenvalues of Cholesky tensors when $1\leq \gamma <\gamma'$ and the number of nonzero elements of non-diagonal Cholesky tensors when $\gamma\geq \gamma'$. However, we always place the 1-body term on the DF branch, so $\Xi^{(0)}=M$ for the mixed circuit.
    \item There are several differences in the costs of the preparation of the $p$ register between the SPSF and mixed circuits.
    \begin{enumerate}
        \item The superposition state controlled on $\gamma$ is created the same way as before. The cost is $7n_\Xi+2b_r-6$.
        \item The contiguous register is made the same way, so the cost is $n_{N_\gamma,\Xi}-1$.
        \item Here we must output data with different QROMs for each branch because the inner prep QROM for the DF circuit has a smaller output size than the inner prep QROM for the SPSF circuit. The costs are
        \begin{equation}
            \Bigg\lceil\frac{(N_\gamma\Xi)_{sp}}{k_{sp}}\Bigg\rceil + b_{sp}(k_{sp}-1) + 1
            \label{eq:mixedInPrepSPSF}
        \end{equation}
        for the SPSF branch and
        \begin{equation}
            \Bigg\lceil\frac{(N_\gamma\Xi)_{DF}+\Xi^{(0)}}{k_{DF}}\Bigg\rceil + b_{DF}(k_{DF}-1) + 1
            \label{eq:mixedInPrepDF}
        \end{equation}
        for the DF branch, where $(N_\gamma\Xi)_{sp}=\sum_{\gamma=\gamma'}^{N_\gamma}\Xi^\gamma$ and $(N_\gamma\Xi)_{DF}=\sum_{\gamma=1}^{\gamma'-1}\Xi^\gamma$. $b_{sp}=4n_N+\aleph_2+2$ like before, but $b_{DF}=n_{\Xi,DF}+\aleph_2+2$, where $n_{\Xi,DF}=\lceil\log(\max_{\gamma\in DF}[\Xi^\gamma])\rceil$ and $\gamma\in DF$ means all $\gamma$ in the DF branch except $\gamma=0$. The added $+1$ for each QROM cost comes from controlling the QROMs on each branch.
        \item Next we perform an inequality test across both branches with cost $\aleph_2$ like before. However, we perform controlled swaps separately on each branch with cost $n_{\Xi,DF}+1$ and $n_{\Xi,sp}+1$. The $+1$s come from additionally controlling the swaps on each branch and $n_{\Xi,sp}=\lceil\log(\max_{\gamma\in sp}[\Xi^\gamma])\rceil$, where $\gamma\in sp$ means all $\gamma$ in the SPSF branch.
    \end{enumerate}
    \item For SPSF SELECT, the controlled swaps cost $2n_N+2$ because they are additionally controlled on the SPSF branch, and the branch-controlled $\text{sel}_{SP}$ box costs $2N-2$. The DF SELECT costs are as follows.
    \begin{enumerate}
        \item Create a contiguous register for the DF branch with cost $n_{\Xi,DF}-1$. This is shown as a controlled box labeled "$+$" in the mixed circuit figure.
        \item Use QROM to output rotation angles for the DF select. This is by far the largest QROM in the DF circuit and can be larger than the SPSF inner prep QROM in the mixed circuit depending on the branch distribution of the Cholesky tensors. It is indicated by the boxes labeled "$\text{In}_p$" and "$\text{data}_p$" in the mixed circuit figure. The cost is then
        \begin{equation}
            \Bigg\lceil\frac{(N_\gamma\Xi)_{DF}+\Xi^{(0)}}{k_{rot}}\Bigg\rceil + b_{rot}(k_{rot}-1).
            \label{eq:DFrotQROM}
        \end{equation}
        The output size for the QROM $b_{rot}=N\beth/2$, where we use $\beth=20$ as a typical value.
        \item Perform swaps controlled on $\sigma$ to generate the spin symmetry with cost $N/2$.
        \item Apply controlled rotations to rotate to the basis where the Cholesky tensors on the DF branch are diagonal. This has cost $N(\beth-2)$ and is indicated by the controlled box labeled "$R$" in the mixed circuit figure.
        \item Perform the $Z_1$ operation, controlled on the DF branch and the success of the preparation of the $\gamma$ and $p$ registers. The controls have cost $2$.
        \item Reverse the controlled rotations and controlled swaps with costs $N(\beth-2)$ and $N/2$, respectively.
        \item Invert the QROM with cost
        \begin{equation}
            \Bigg\lceil\frac{(N_\gamma\Xi)_{DF}+\Xi^{(0)}}{k'_{rot}}\Bigg\rceil + k'_{rot}.
        \end{equation}
        \item Invert the contiguous register with cost $n_{\Xi,DF}-1$.
    \end{enumerate}
    \item Invert the preparation of the $p$ register. The cost is the same as step 3 for this circuit except the QROMs are cheaper to invert, so those costs become
    \begin{equation}
        \Bigg\lceil\frac{(N_\gamma\Xi)_{sp}}{k'_{sp}}\Bigg\rceil + k'_{sp} + 1
    \end{equation}
    for the SPSF branch QROM and
    \begin{equation}
        \Bigg\lceil\frac{(N_\gamma\Xi)_{DF}+\Xi^{(0)}}{k'_{DF}}\Bigg\rceil + k'_{DF} + 1
        \label{eq:mixedInPrepInvDF}
    \end{equation}
    for the DF branch QROM.
    \item Reflect on the same qubits as the SPSF circuit in step 6. The cost is $n_\Xi+\aleph_2+3$.
    \item Repeat steps 3-5 without the 1-body term. Neglecting the one body term reduces the number of entries for the QROM and inverse QROM in Eq.~\eqref{eq:mixedInPrepDF} and Eq.~\eqref{eq:mixedInPrepInvDF}, respectively, by $\Xi^{(0)}$.
    \item Invert steps 1 and 2. In the mixed circuit figure, these are the boxes labeled "$\text{In}_\gamma$" and "$\text{data}_\gamma$" to invert step 2 and the box labeled "prep" to invert step 1. This time, the costs are exactly the same as step 8 for the SPSF circuit.
    \item Reflect on the same qubits as the SPSF circuit in step 9 to finish constructing the quantum walk. The total cost is $n_{N_\gamma}+n_\Xi+\aleph_1+\aleph_2+2$.
    \item Like the SPSF circuit, the quantum walk steps must be controlled with unary iteration on an ancilla qubit. The cost of unary iteration and controlling the reflection are $2$.
\end{enumerate}

The total Toffoli cost for a step of the quantum walk for the mixed circuit is then
\begin{equation}
\begin{split}
    &2N(2\beth-1) + 4n_N+9n_{N_\gamma}+30n_\Xi + 8n_{\Xi,DF}+4n_{\Xi,sp} + 4n_{N_\gamma,\Xi} + 3\aleph_1+6\aleph_2+12b_r-6\eta-23 \\
    &+\bigg\lceil\frac{N_\gamma+1}{k_{p1}}\bigg\rceil+b_{p1}(k_{p1}-1)+\bigg\lceil\frac{N_\gamma+1}{k'_{p1}}\bigg\rceil+k'_{p1}+\bigg\lceil\frac{N_\gamma+1}{k_{o}}\bigg\rceil+b_o^\text{mix}(k_o-1)+\bigg\lceil\frac{N_\gamma+1}{k'_{o}}\bigg\rceil+k'_o \\
    &+2\bigg\lceil\frac{(N_\gamma\Xi)_{sp}}{k_{sp}}\bigg\rceil+2b_{sp}(k_{sp}-1) + \bigg\lceil\frac{(N_\gamma\Xi)_{DF}+\Xi^{(0)}}{k_{DF}}\bigg\rceil+\bigg\lceil\frac{(N_\gamma\Xi)_{DF}}{k_{DF}}\bigg\rceil+2b_{DF}(k_{DF}-1) \\
    &+ \bigg\lceil\frac{(N_\gamma\Xi)_{DF}+\Xi^{(0)}}{k_{rot}}\bigg\rceil+\bigg\lceil\frac{(N_\gamma\Xi)_{DF}}{k_{rot}}\bigg\rceil+2b_{rot}(k_{rot}-1) + \bigg\lceil\frac{(N_\gamma\Xi)_{DF}+\Xi^{(0)}}{k'_{rot}}\bigg\rceil+\bigg\lceil\frac{(N_\gamma\Xi)_{DF}}{k'_{rot}}\bigg\rceil \\
    &+ 2k'_{rot} + 2\bigg\lceil\frac{(N_\gamma\Xi)_{sp}}{k'_{sp}}\bigg\rceil+2k'_{sp} + \bigg\lceil\frac{(N_\gamma\Xi)_{DF}+\Xi^{(0)}}{k'_{DF}}\bigg\rceil+\bigg\lceil\frac{(N_\gamma\Xi)_{DF}}{k'_{DF}}\bigg\rceil+2k'_{DF}
\end{split}
\end{equation}

Like with Toffolis, the qubits required are largely the same between the mixed and SPSF circuits. The differences in qubits between the mixed and SPSF circuits are detailed below.

\begin{enumerate}
    \item The qubits used in steps 1-4, 6, 9, and 10 for the SPSF circuit are also used in the mixed circuit.
    \item The QROM that outputs data needed to prepare the $p$ register has an output size of $b_o^\text{mix} = n_\Xi+n_{N_\gamma,\Xi}+b_r+2$.
    \item The preparation of the $p$ register during the inner preparation requires $\aleph_2$ qubits for the inequality test and $1$ qubit to store the result.
    \item The phase-gradient state uses $\beth$ qubits.
    \item We reuse ancilla from the largest QROM, so the qubit cost of a QROM changes depending on whether it is the largest QROM. In many circuits, one QROM is almost always the largest. However, in the mixed circuit, the largest QROM changes between the SPSF inner prep QROM and the DF rotations QROM depending on how many Cholesky tensors are placed on each branch. Therefore, we account for the different qubit costs according to which of these QROMs requires more qubits.
    \begin{enumerate}
        \item The temporary ancillas used by the SPSF inner prep QROM are
        \begin{equation}
            b_{sp}(k_{sp}-1)+\bigg\lceil\log\bigg(\frac{(N_\gamma\Xi)_{sp}}{k_{sp}}\bigg)\bigg\rceil.
            \label{eq:spsfInPrepMixedAncillas}
        \end{equation}
        When this is larger than Eq.~\eqref{eq:dfRotMixedAncillas}, the circuit uses these temporary ancillas.
        \item The temporary ancillas used by the DF rotations QROM are
        \begin{equation}
            b_{rot}(k_{rot}-1),
            \label{eq:dfRotMixedAncillas}
        \end{equation}
        where we neglect the small $\log$ term. When this is larger than Eq.~\eqref{eq:spsfInPrepMixedAncillas}, the circuit uses these temporary ancillas.
    \end{enumerate}
In addition to the temporary ancilla qubits from either Eq.~\eqref{eq:spsfInPrepMixedAncillas} or Eq.~\eqref{eq:dfRotMixedAncillas}, the circuit needs $b_{sp} = 4n_N+\aleph_2+2$ persistent qubits for the inner preparation QROMs and $b_{rot} = N\beth/2$ persistent for the DF rotations QROM. Note that the DF inner prep QROM has a smaller output size than the SPSF inner prep QROM, so the qubit cost of the DF inner prep QROM is $0$ because it outputs into the same registers as the SPSF inner prep QROM. The outputs are just stored on their own respective branches.
\end{enumerate}

The total number of qubits required for the mixed circuit is then
\begin{equation}
\begin{split}
    &2\lceil\log(\mathcal{I}+1)\rceil + N(\beth/2+1) + 2n_{N_\gamma} + 4n_N + 2\aleph_1+2\aleph_2 + \beth + 2n_\Xi+n_{N_\gamma,\Xi} \\
    &+ \max\bigg[ b_{sp}(k_{sp}-1)+\bigg\lceil\log\bigg(\frac{(N_\gamma\Xi)_{sp}}{k_{sp}}\bigg)\bigg\rceil,\:\: b_{rot}(k_{rot}-1) \bigg] + b_r + 12.
\end{split}
\end{equation}

\section{Ranks and sparsity of the Cholesky tensors for all systems}

The following figures show the relative ranks and the average sparsities of the Cholesky decomposed electron repulsion integrals for all the systems studied. For the iron--sulfur clusters~\cite{li_chan_2017, li_2026a}, FeMoco~\cite{li_etal_2019a, li_2026}, the P-cluster of nitrogenase~\cite{li_etal_2019b, li_2024}, and the heme center of cytochrome P450~\cite{goings_etal_2022, goings_etal_2022a}, the Cholesky tensors are represented in the localized basis specifying their respective active spaces. For the polyacenes (benzene--hexacene), the tensors are represented in the symmetrically orthogonalized (L\"owdin)~\cite{lowdin_1950} double-zeta Dunning basis set~\cite{dunning_1989}.

In each figure, (a) plots the ranks of each Cholesky tensor $\Op{L}_\gamma$ in descending order, divided by the dimension of $\Op{L}_\gamma$ so that a value of $1$ indicates full rank. (b) plots the averaged matrix elements $\bar{L}_{ij}^\gamma = \frac{1}{N_\gamma} \sum_\gamma^{N_\gamma} L_{ij}^\gamma$ on a diverging-color logarithmic scale to reveal the sparsity pattern: red pixels are above the sparsity threshold of $10^{-5}$ (except for the P cluster of nitrogenase, whose threshold is $10^{-4}$), and blue pixels are below it. (c) is as (b), but only the $\Op{L}^\gamma$ with a relative rank of at least $1/2$ are averaged.

\begin{figure*}[ht]
    \centering
    \subfloat[]{\includegraphics[width=0.29\linewidth]{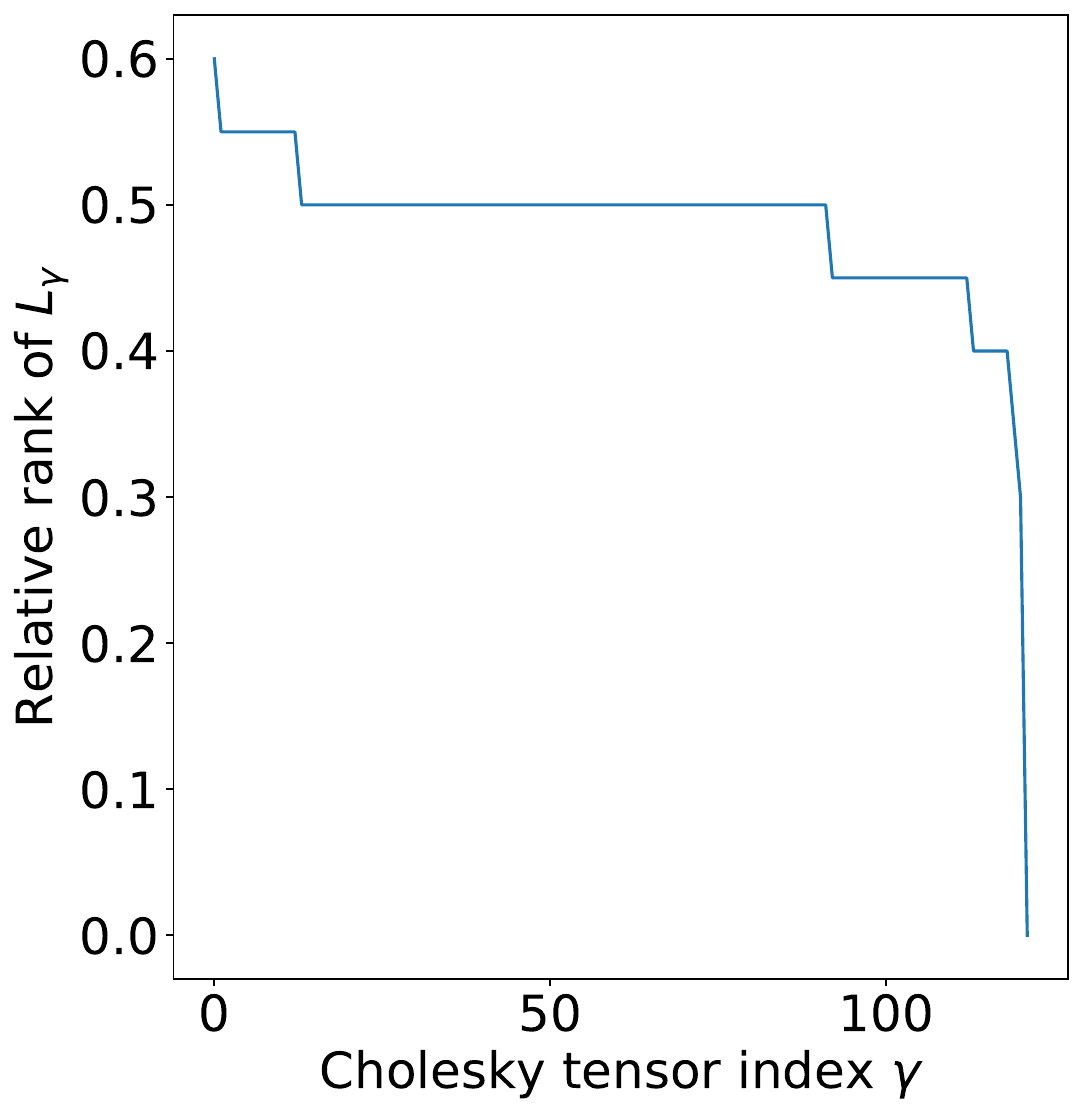}}
    \hfill
    \subfloat[]{\includegraphics[width=0.35\linewidth]{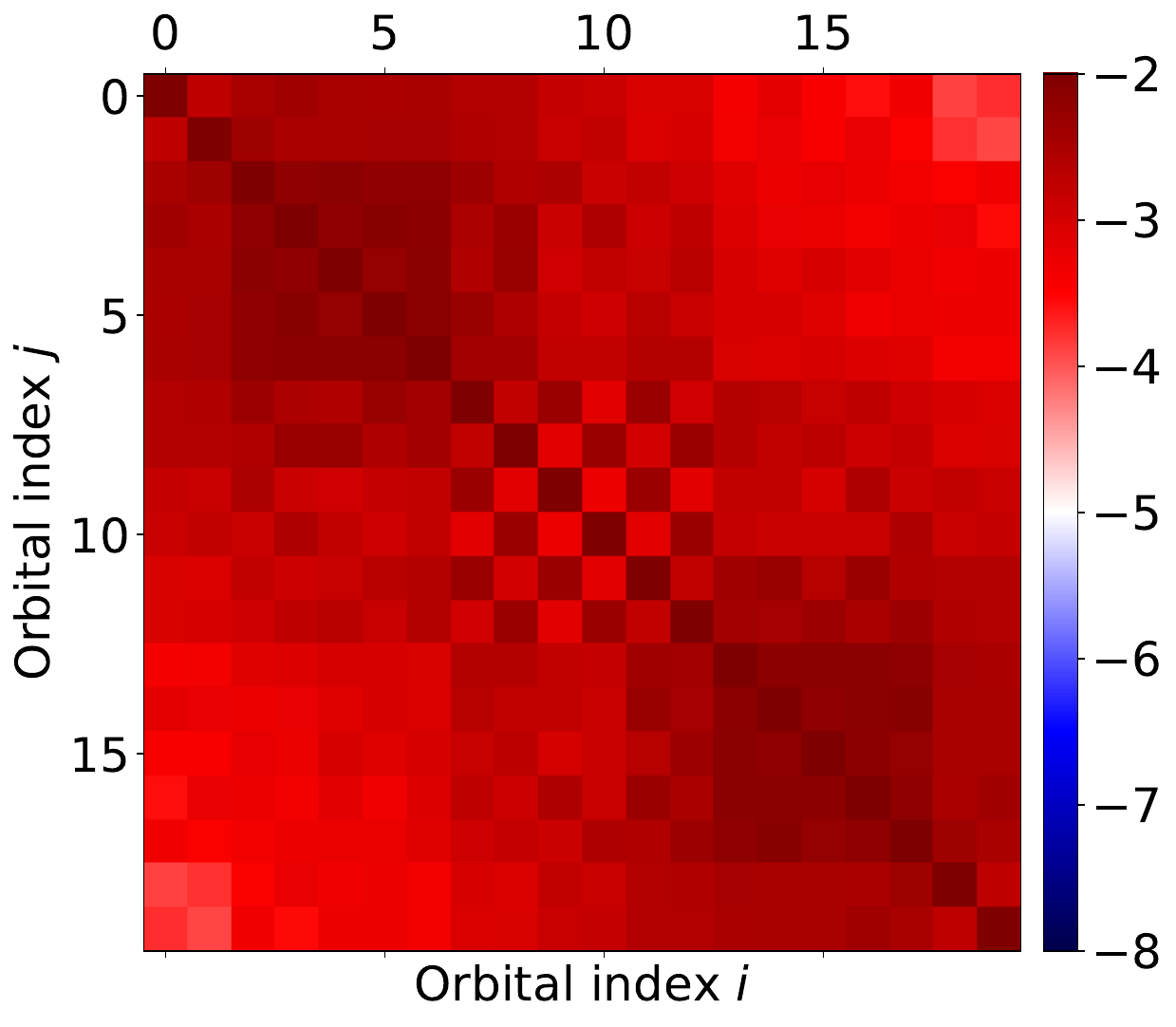}}
    \hfill
    \subfloat[]{\includegraphics[width=0.35\linewidth]{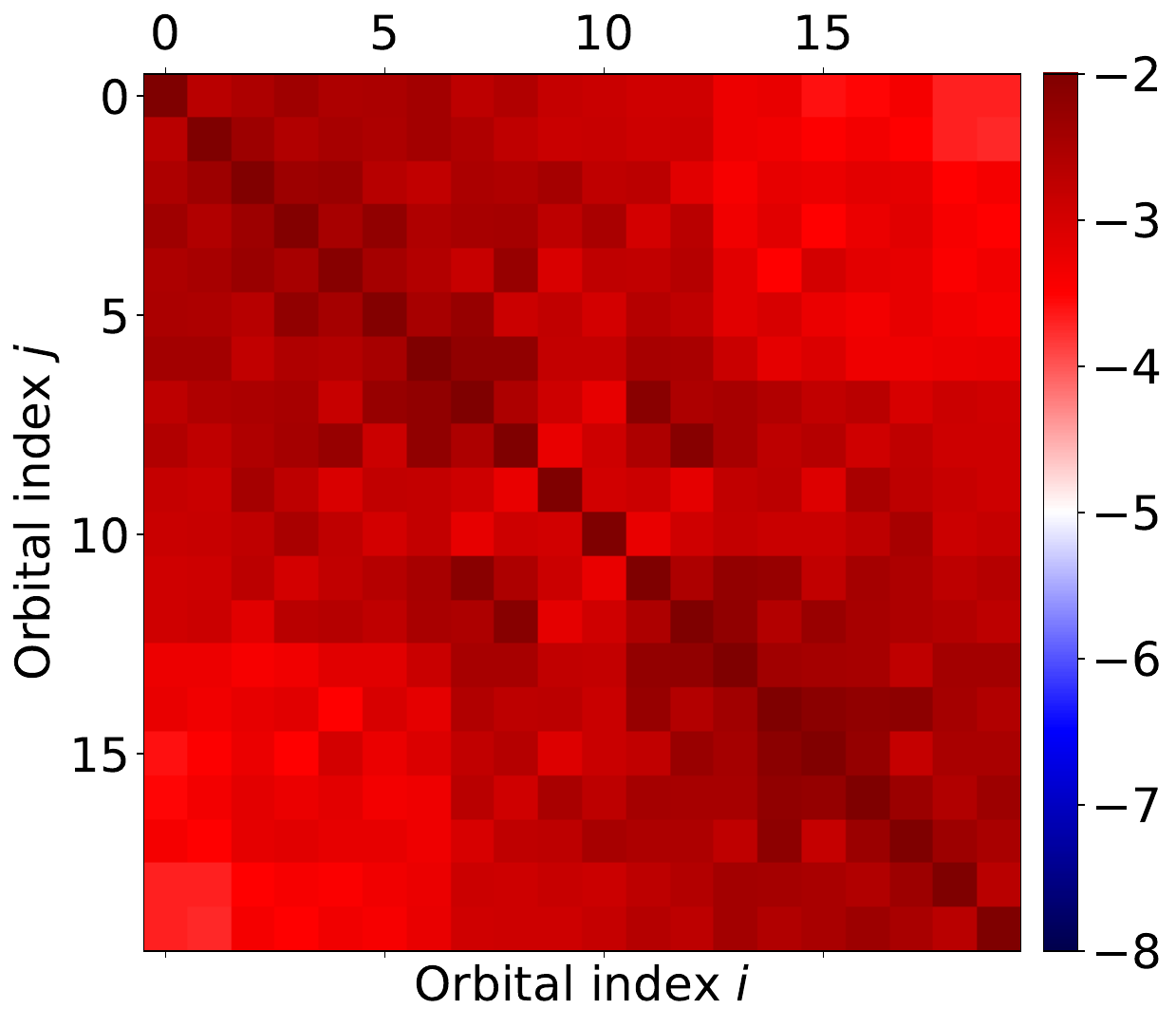}}
    \caption{(a) Relative ranks,
             (b--c) average sparsity of the Cholesky tensors for the \ce{[Fe2S2]} cluster.
             (c) averages only over Cholesky tensors with relative rank $\geq 1/2$.}
    \label{fig:supp-df-sp-fe2s2}
\end{figure*}

\begin{figure*}[ht]
    \centering
    \subfloat[]{\includegraphics[width=0.29\linewidth]{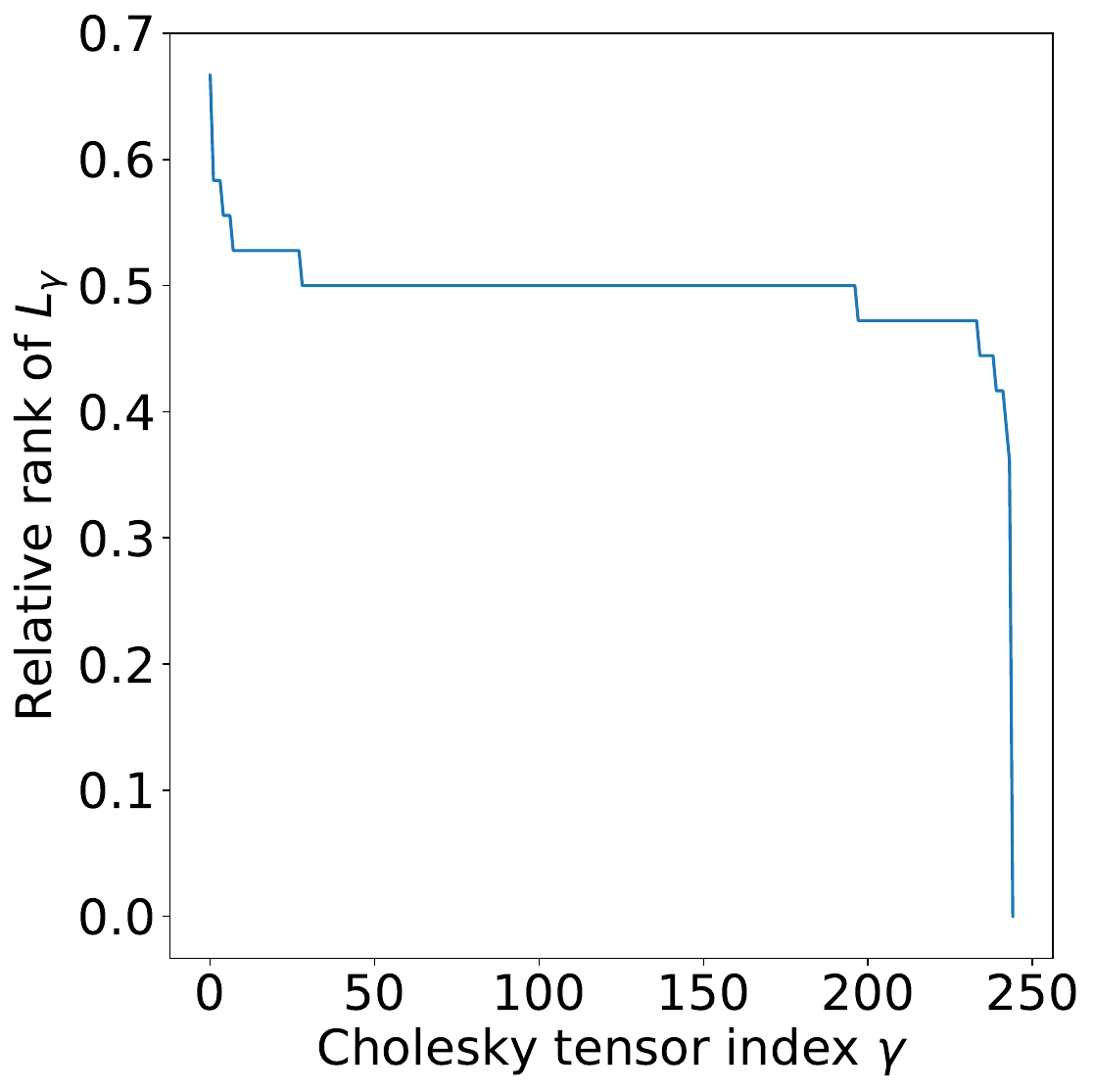}}
    \hfill
    \subfloat[]{\includegraphics[width=0.35\linewidth]{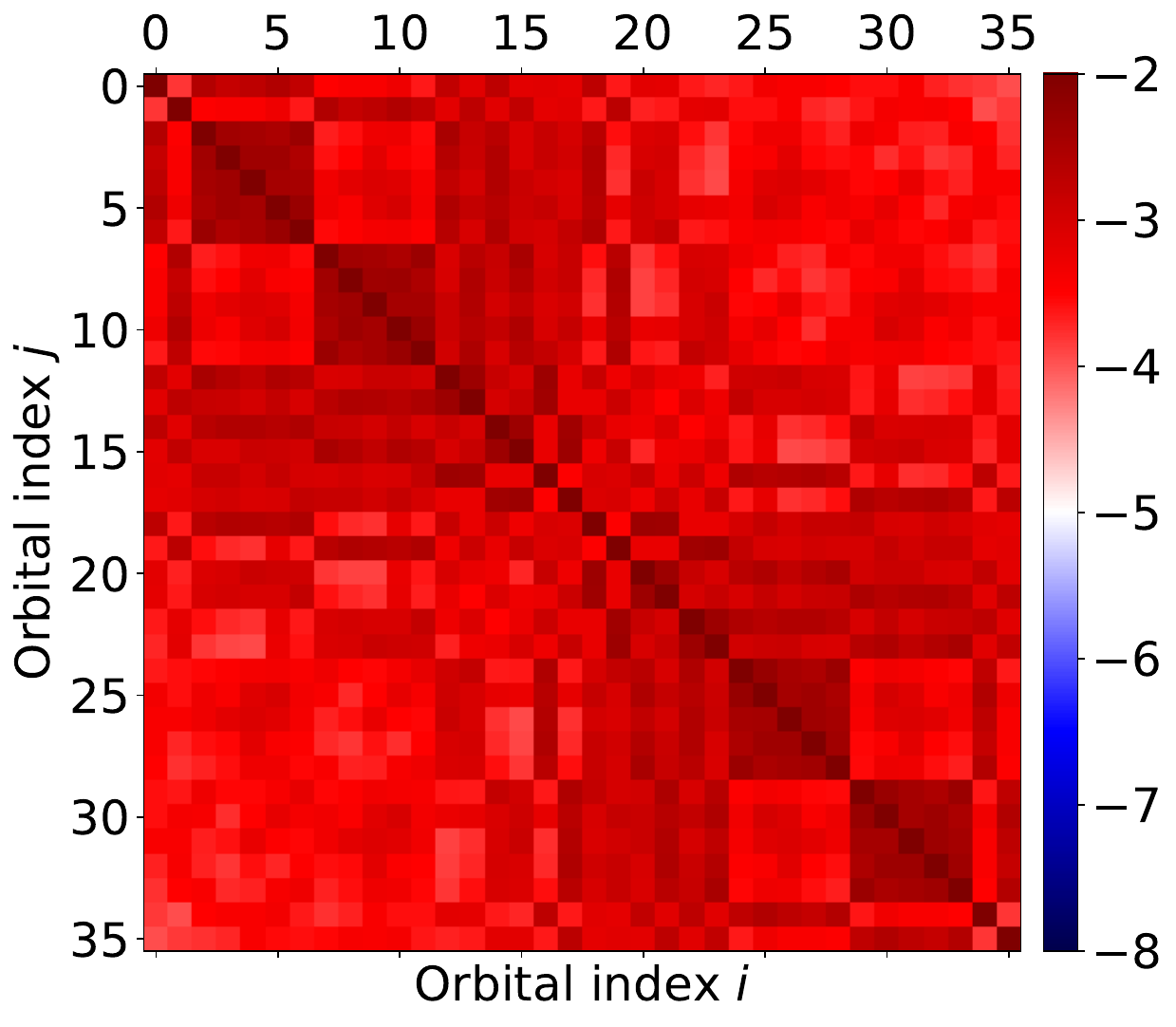}}
    \hfill
    \subfloat[]{\includegraphics[width=0.35\linewidth]{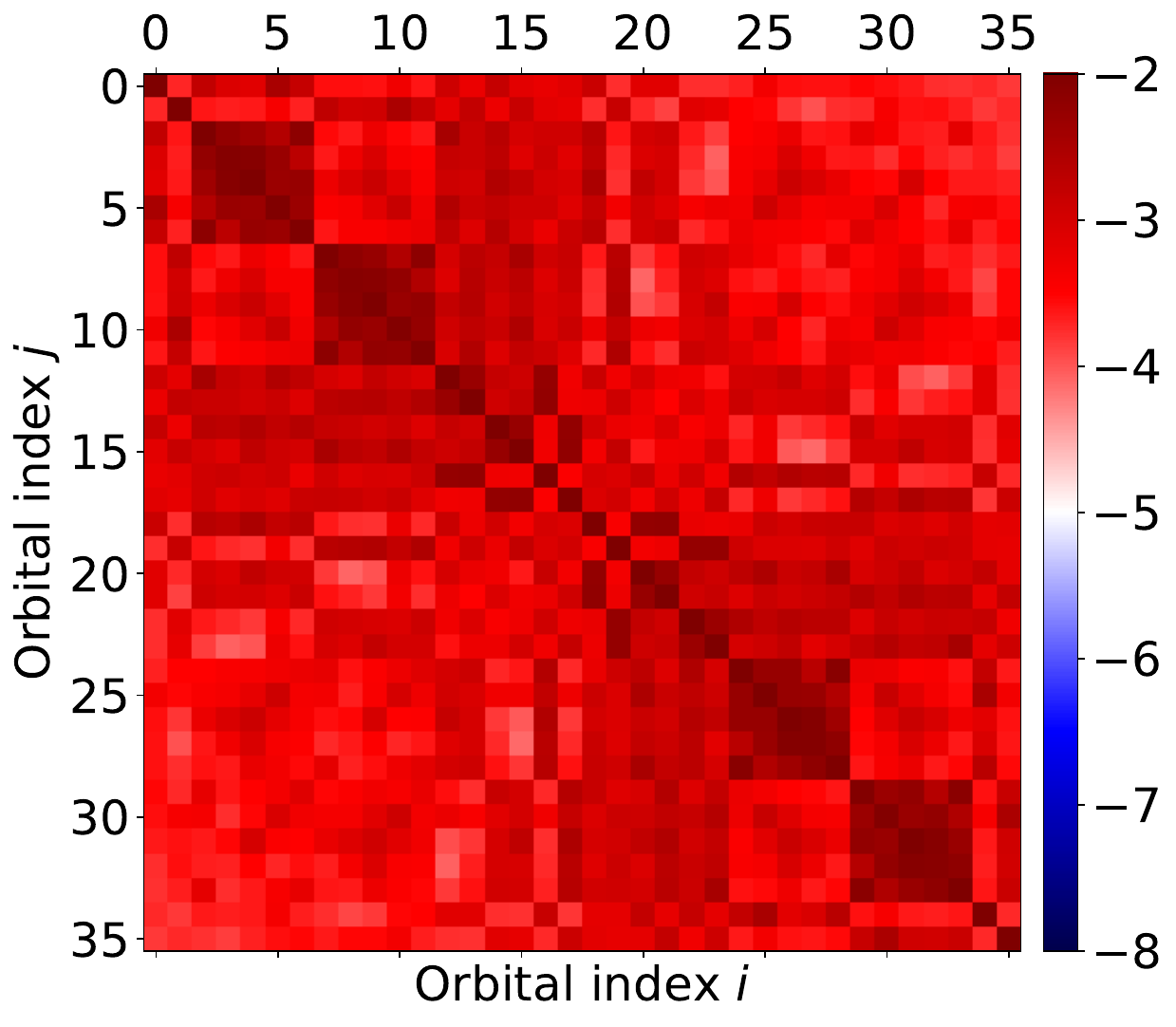}}
    \caption{(a) Relative ranks,
             (b--c) average sparsity of the Cholesky tensors for the \ce{[Fe4S4]} cluster.
             (c) averages only over Cholesky tensors with relative rank $\geq 1/2$.}
    \label{fig:supp-df-sp-fe4s4}
\end{figure*}

\begin{figure*}[ht]
    \centering
    \subfloat[]{\includegraphics[width=0.29\linewidth]{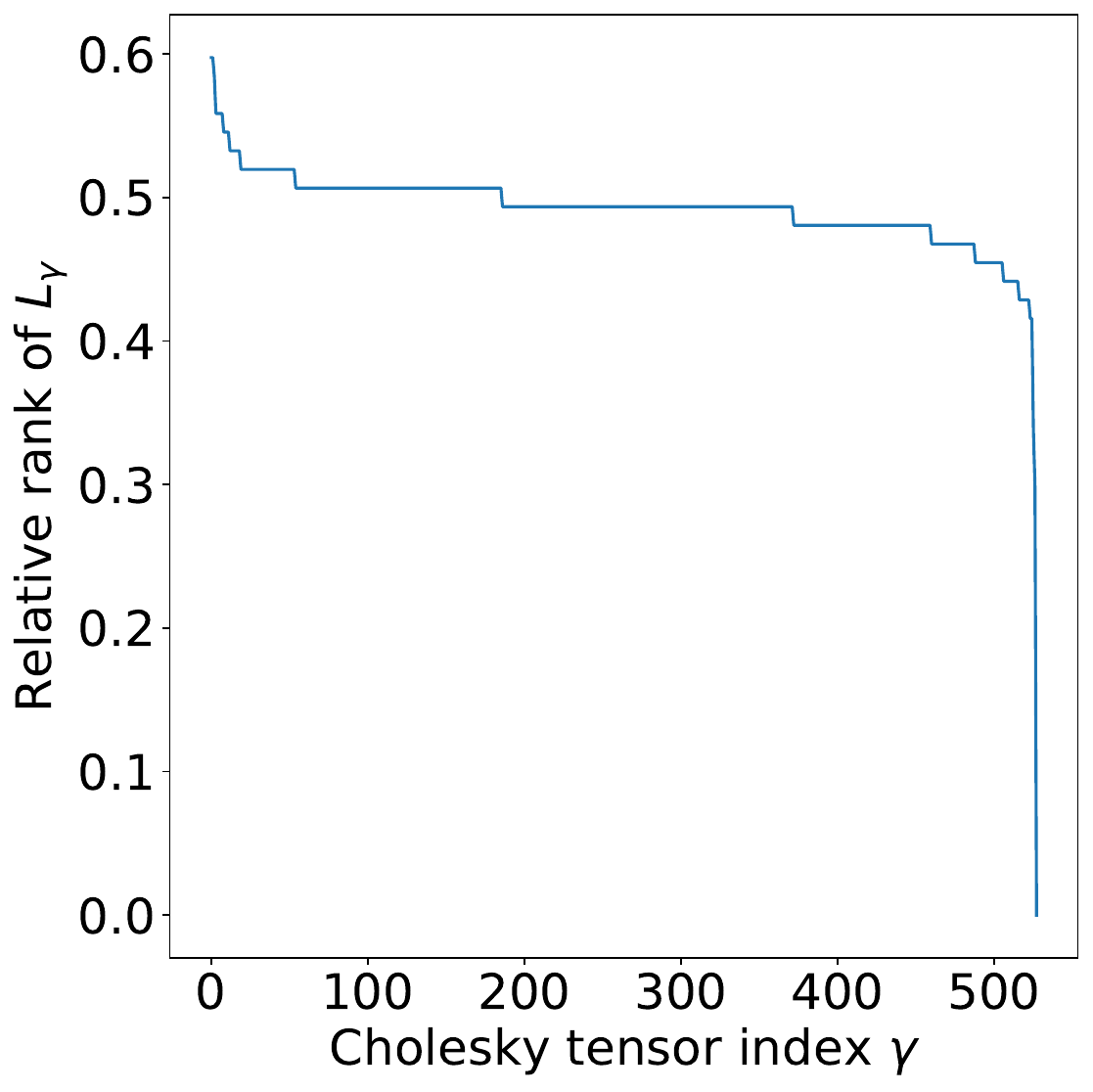}}
    \hfill
    \subfloat[]{\includegraphics[width=0.35\linewidth]{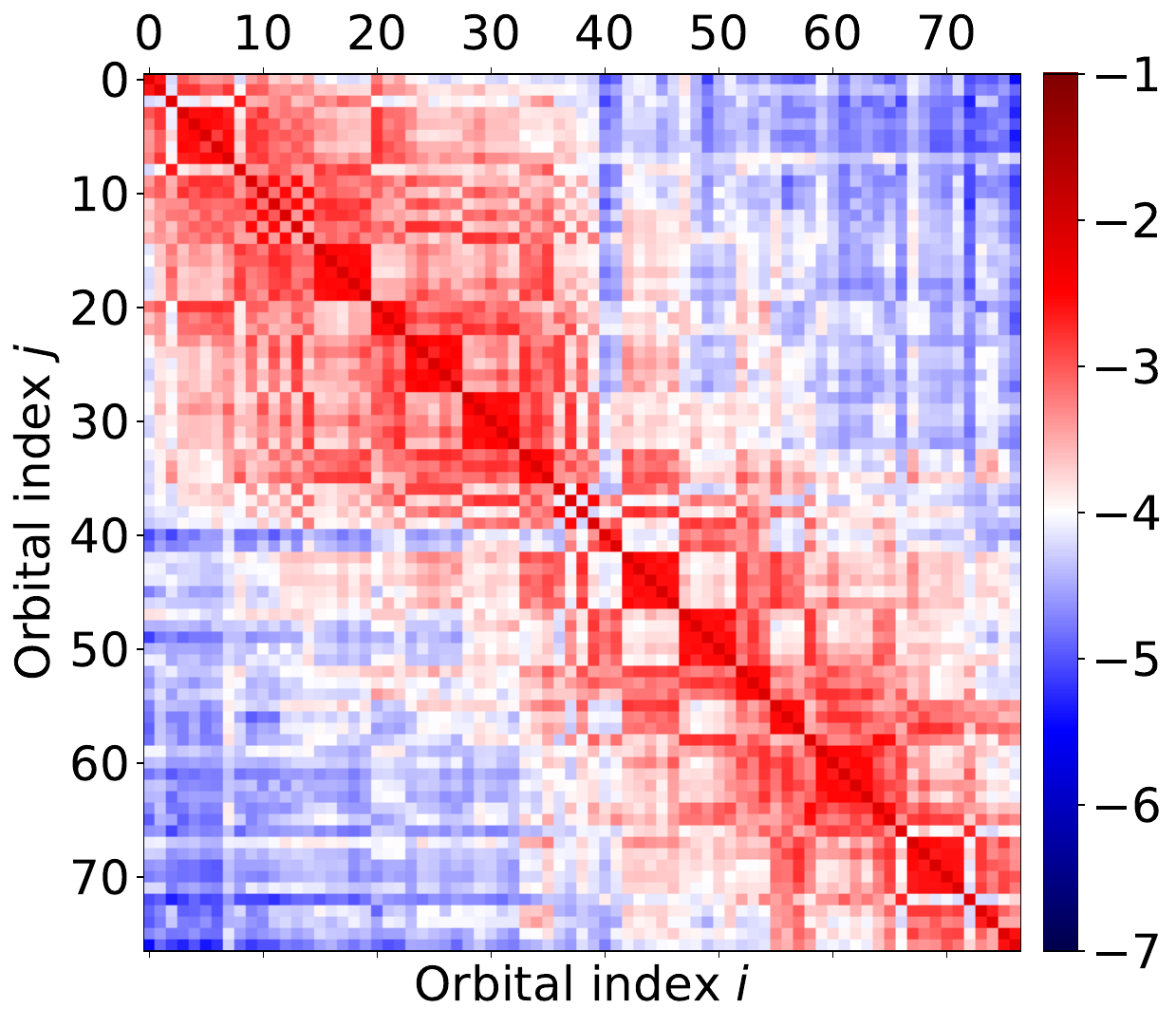}}
    \hfill
    \subfloat[]{\includegraphics[width=0.35\linewidth]{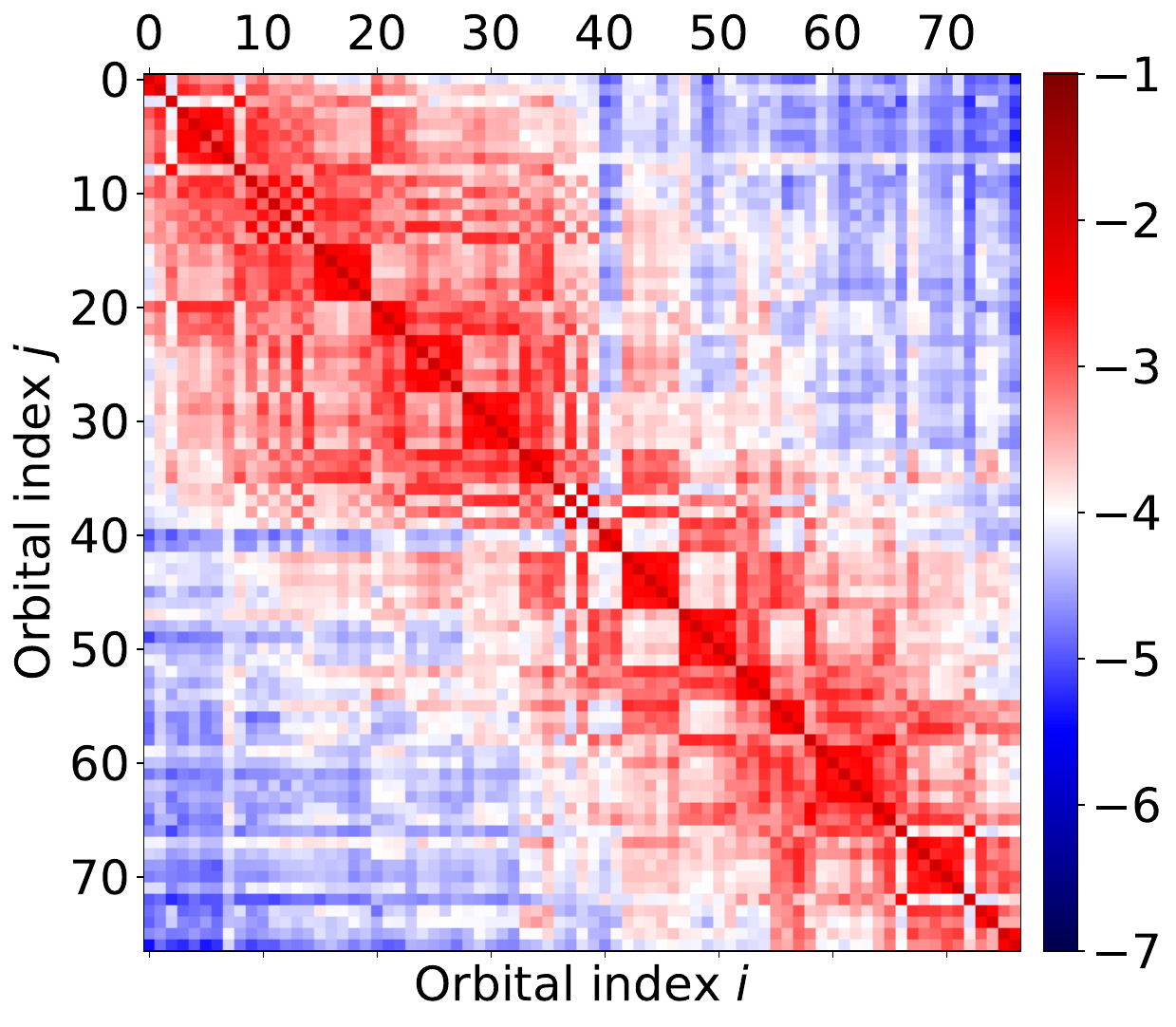}}
    \caption{(a) Relative ranks,
             (b--c) average sparsity of the Cholesky tensors for the P-cluster of nitrogenase.
             (c) averages only over Cholesky tensors with relative rank $\geq 1/2$.}
    \label{fig:supp-df-sp-p-cluster}
\end{figure*}

\begin{figure*}[ht]
    \centering
    \subfloat[]{\includegraphics[width=0.29\linewidth]{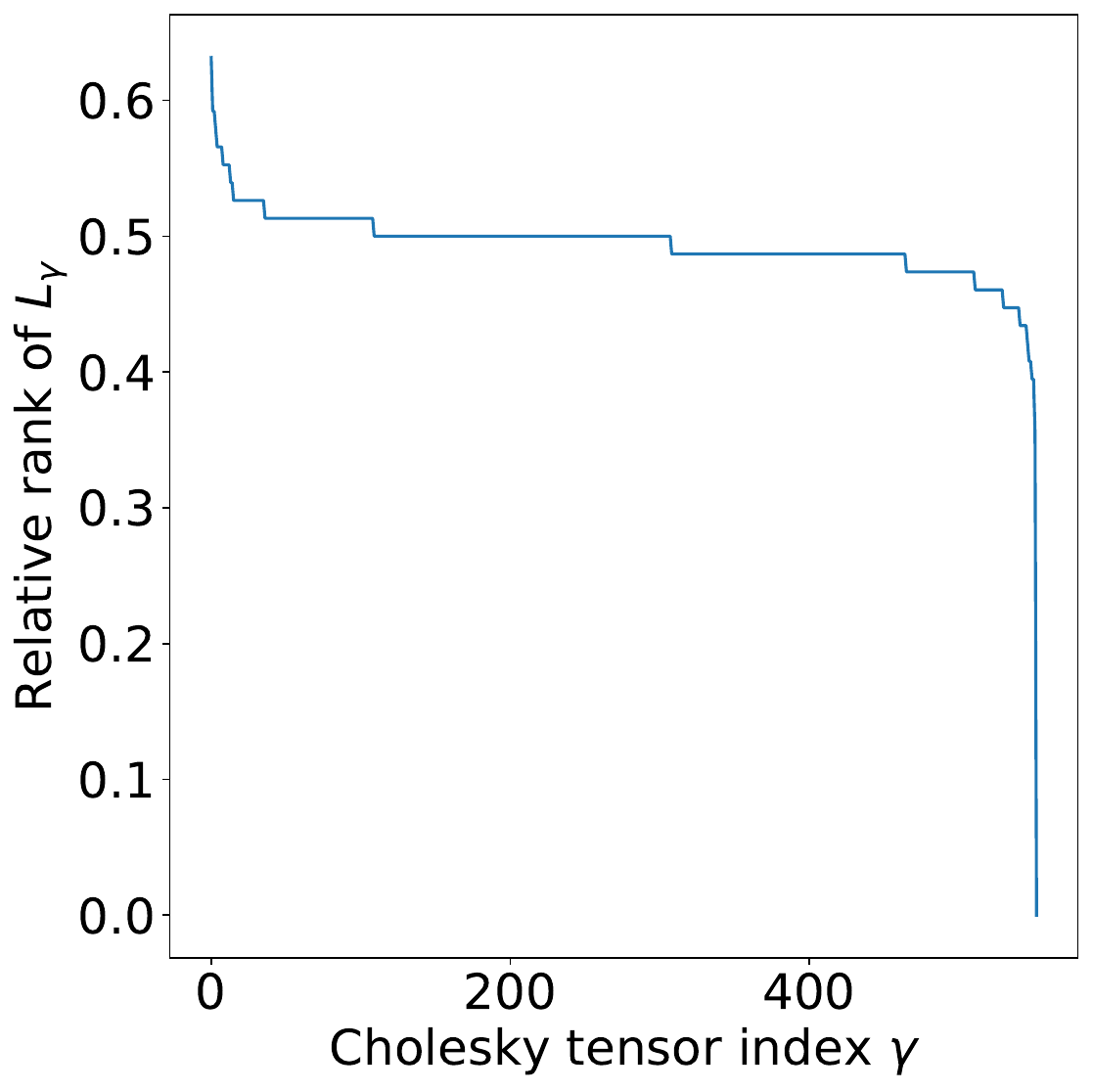}}
    \hfill
    \subfloat[]{\includegraphics[width=0.35\linewidth]{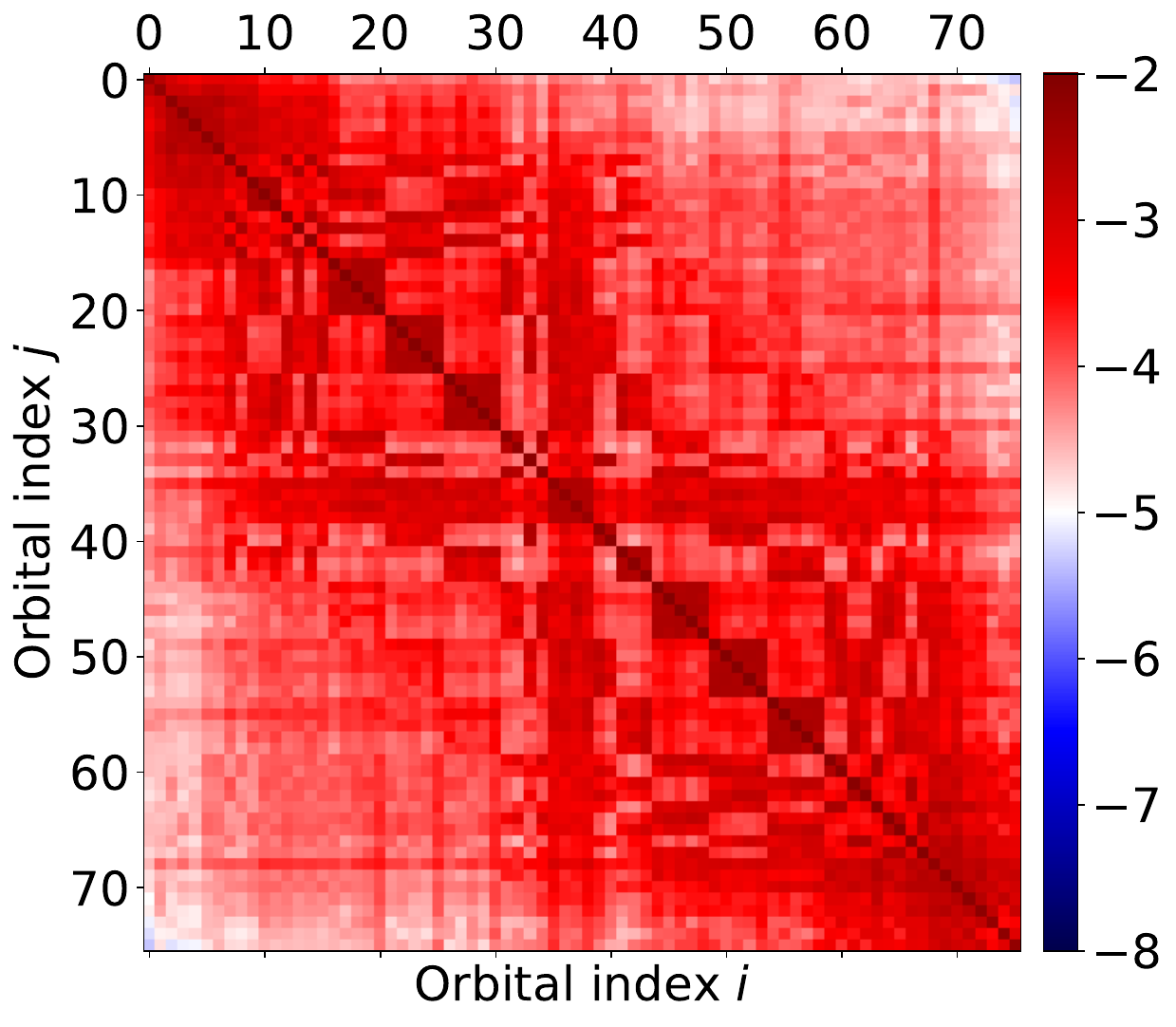}}
    \hfill
    \subfloat[]{\includegraphics[width=0.35\linewidth]{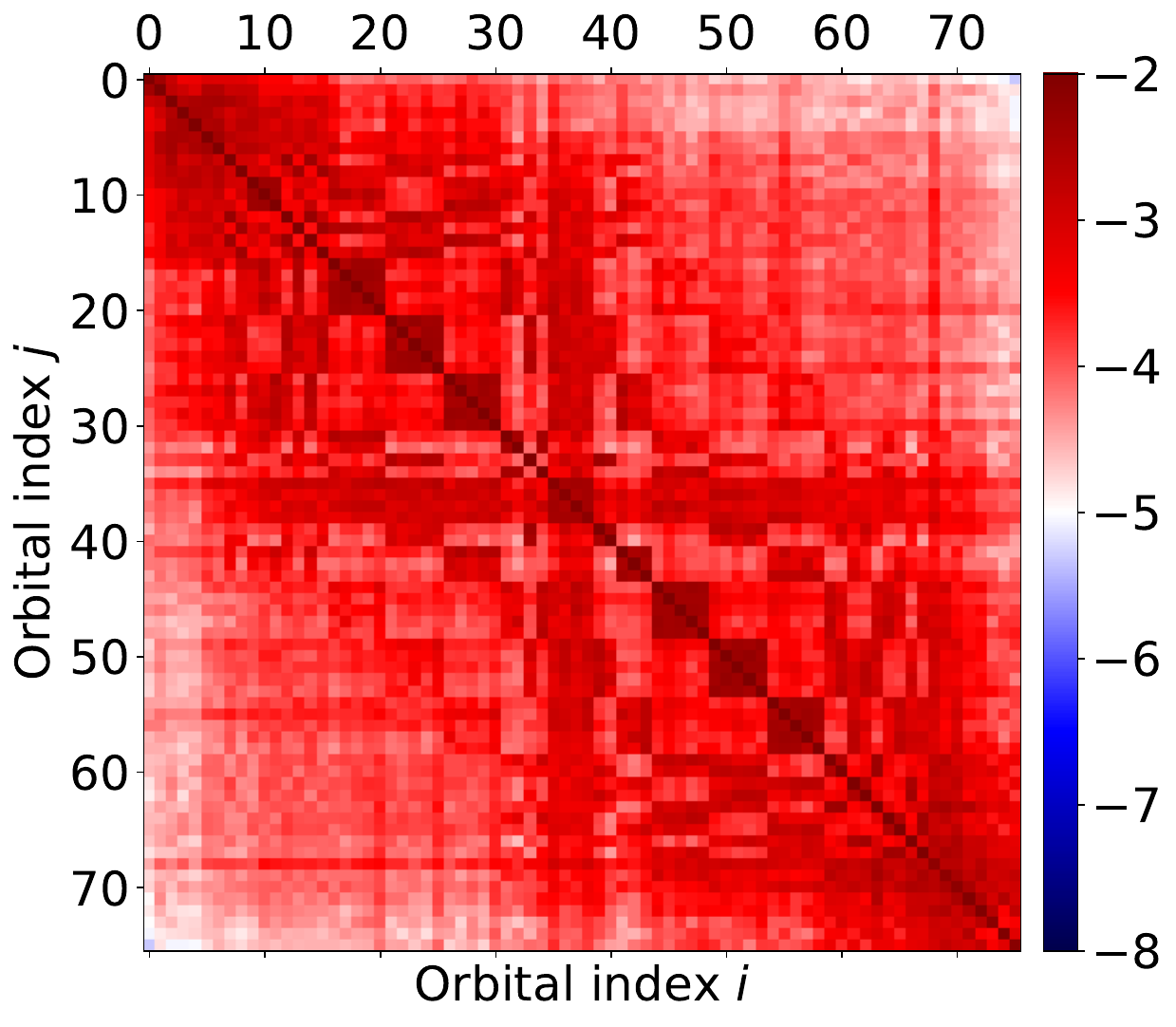}}
    \caption{(a) Relative ranks,
             (b--c) average sparsity of the Cholesky tensors for the Fe--Mo cofactor of nitrogenase.
             (c) averages only over Cholesky tensors with relative rank $\geq 1/2$.}
    \label{fig:supp-df-sp-femoco}
\end{figure*}

\begin{figure*}[ht]
    \centering
    \subfloat[]{\includegraphics[width=0.29\linewidth]{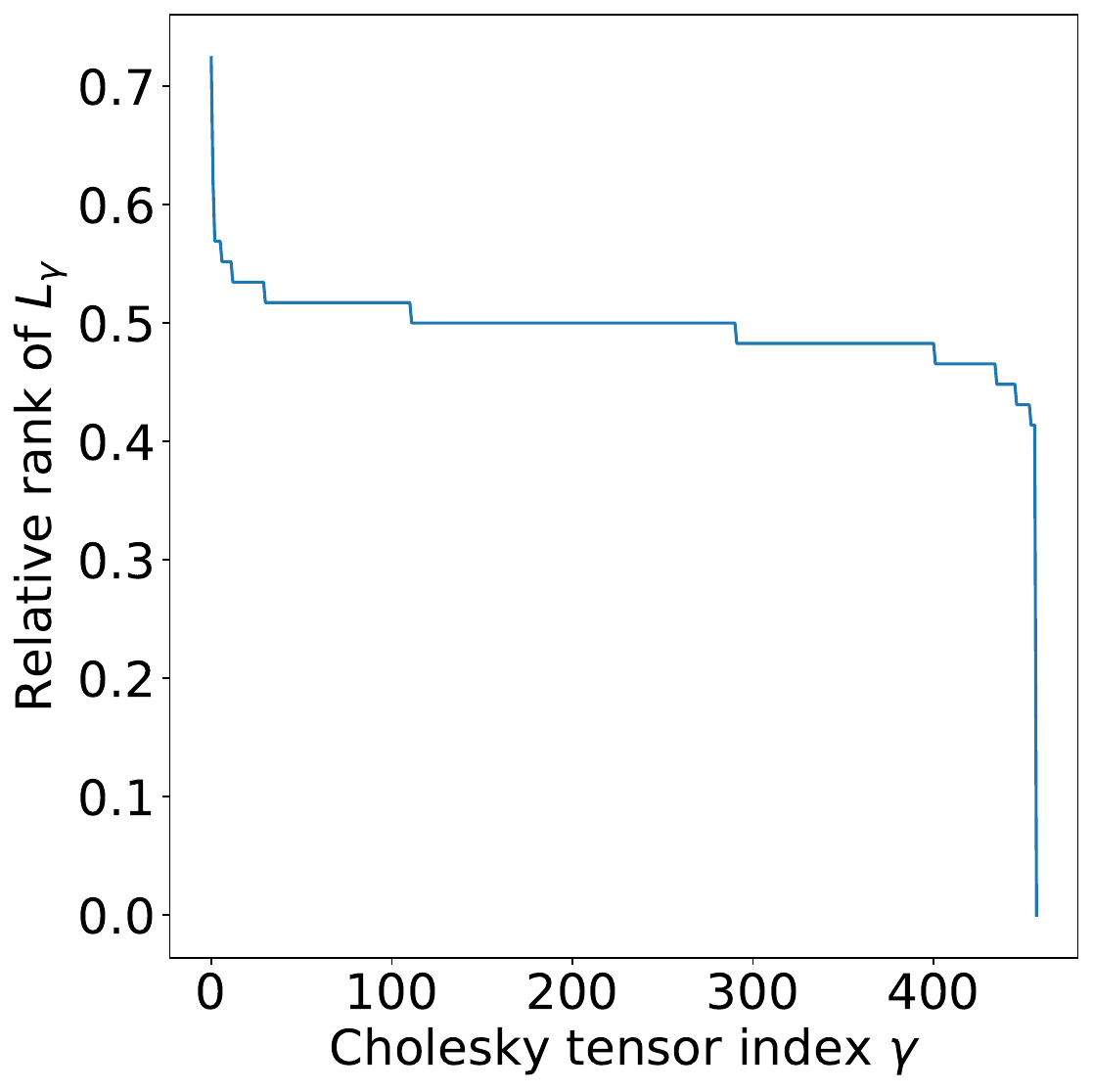}}
    \hfill
    \subfloat[]{\includegraphics[width=0.35\linewidth]{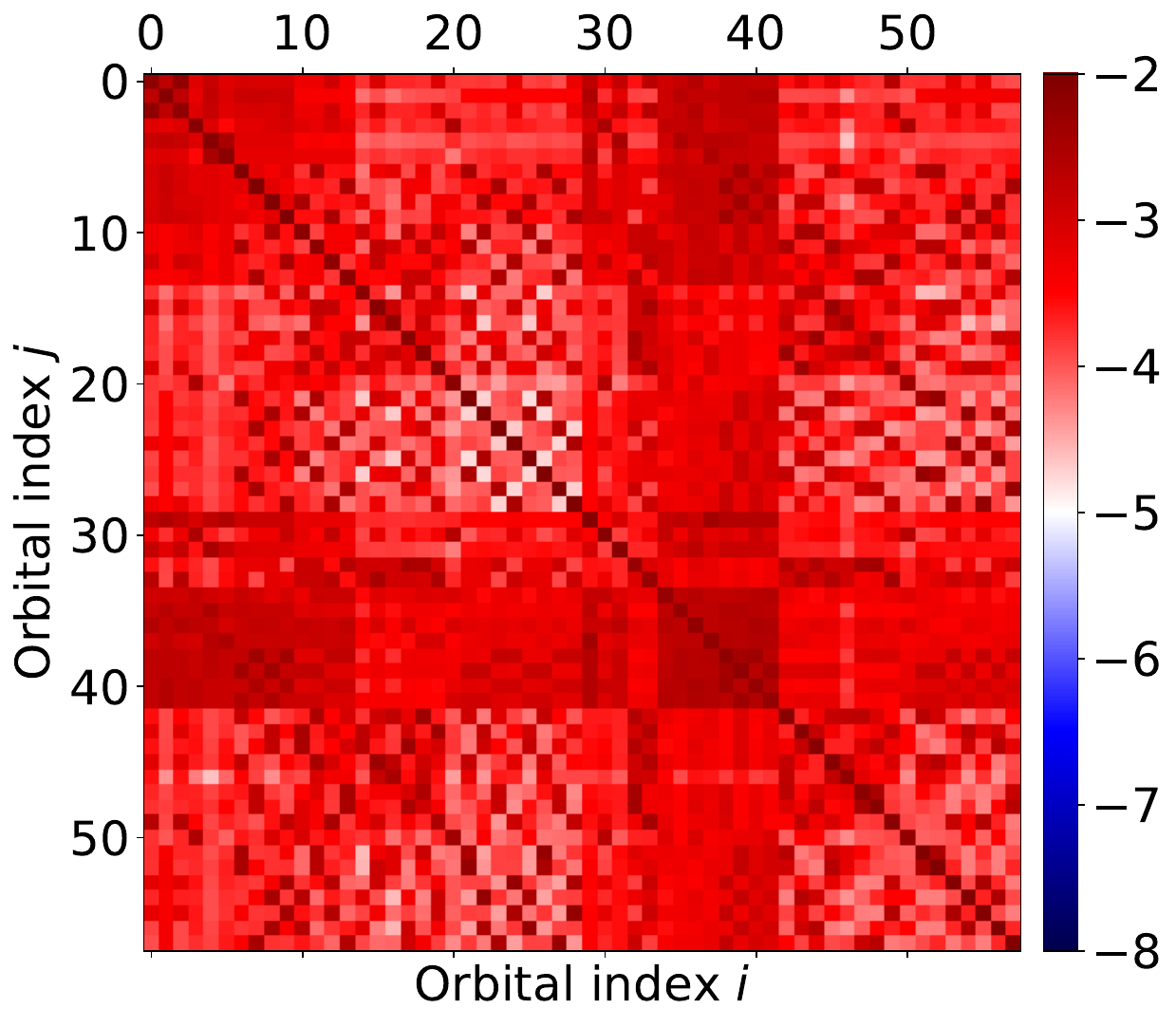}}
    \hfill
    \subfloat[]{\includegraphics[width=0.35\linewidth]{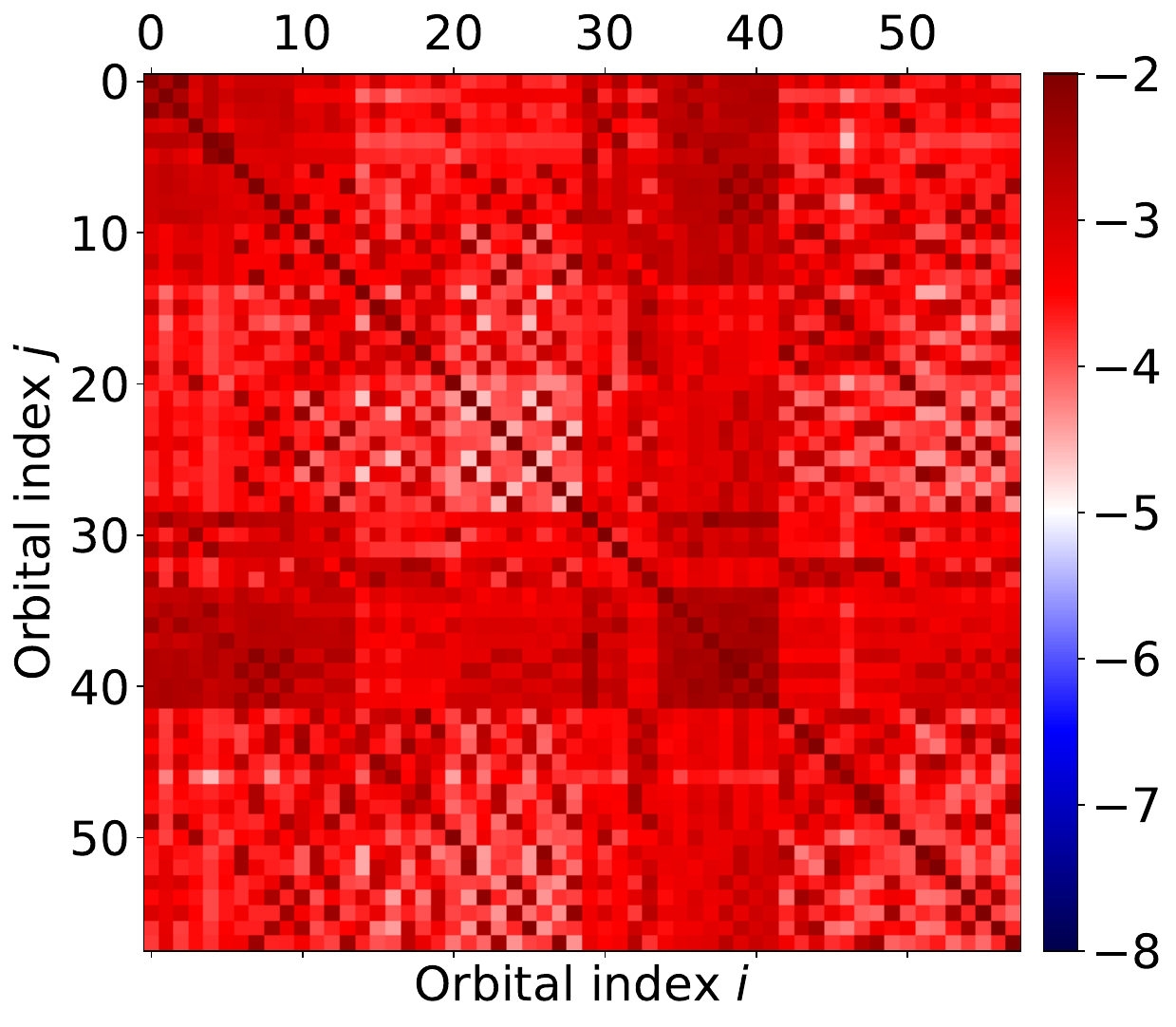}}
    \caption{(a) Relative ranks,
             (b--c) average sparsity of the Cholesky tensors for the heme unit of cytochrome P450.
             (c) averages only over Cholesky tensors with relative rank $\geq 1/2$.}
    \label{fig:supp-df-sp-p450}
\end{figure*}

\begin{figure*}[ht]
    \centering
    \subfloat[]{\includegraphics[width=0.29\linewidth]{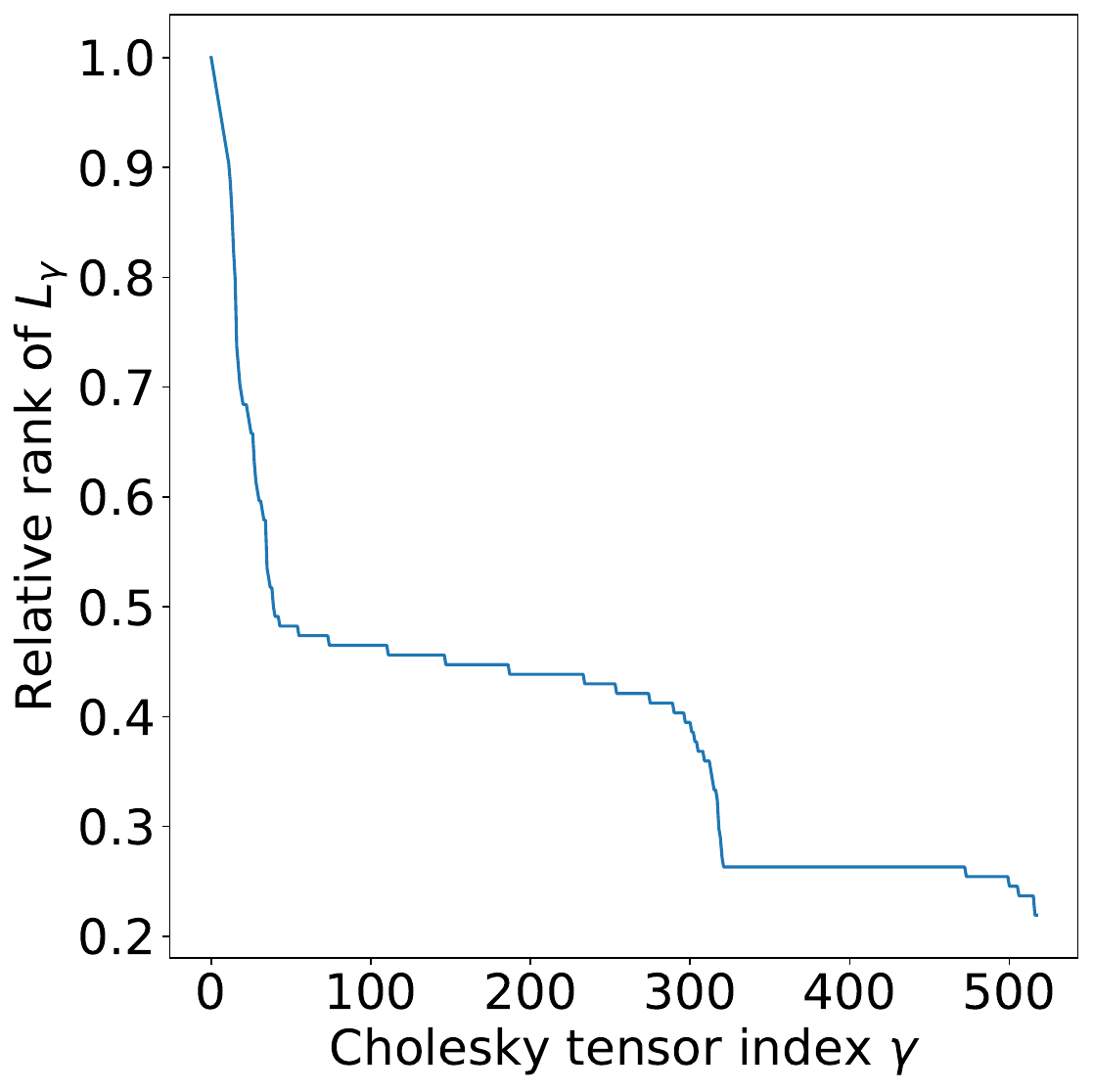}}
    \hfill
    \subfloat[]{\includegraphics[width=0.35\linewidth]{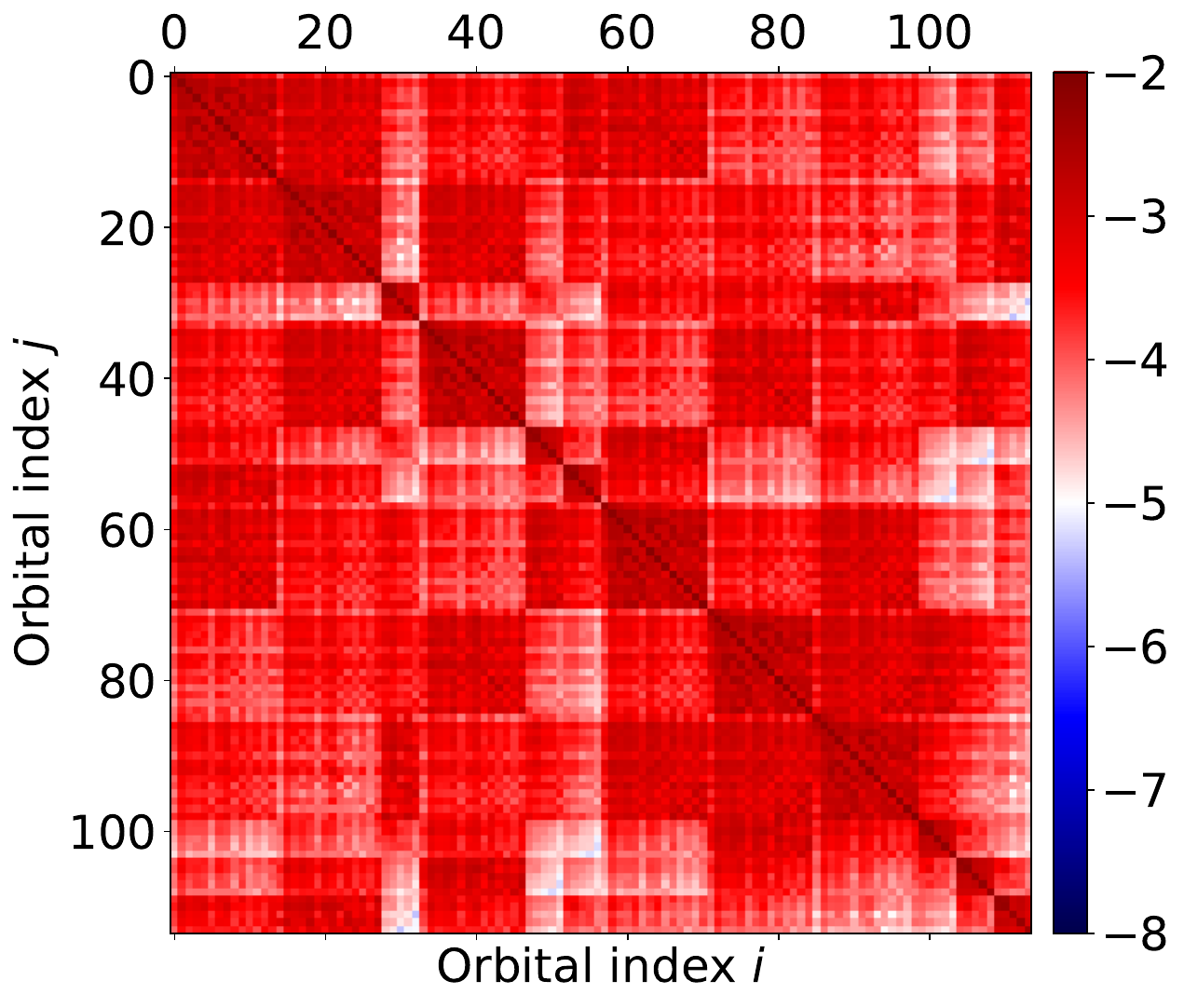}}
    \hfill
    \subfloat[]{\includegraphics[width=0.35\linewidth]{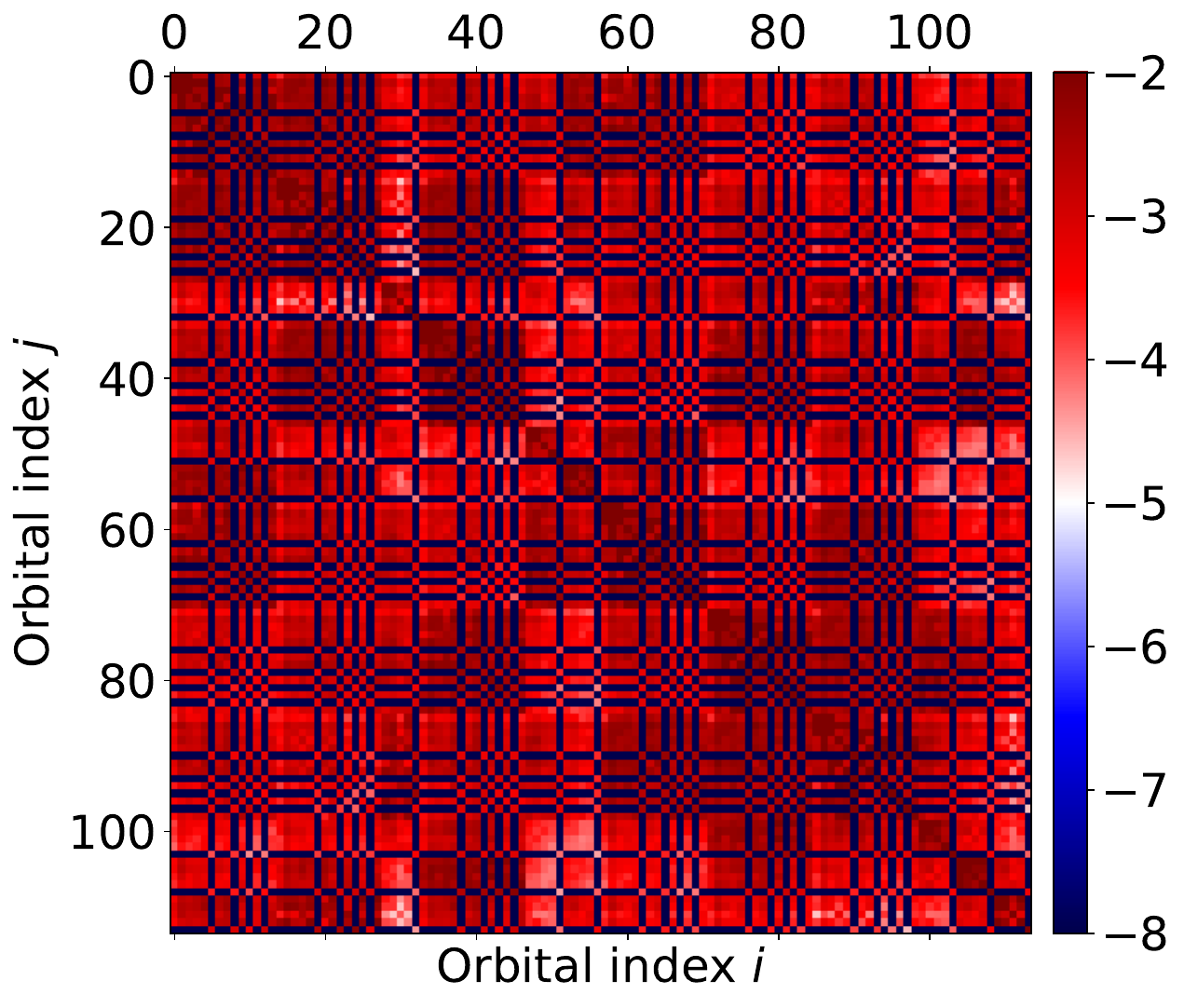}}
    \caption{(a) Relative ranks,
             (b--c) average sparsity of the Cholesky tensors for 1-acene (benzene).
             (c) averages only over Cholesky tensors with relative rank $\geq 1/2$.}
    \label{fig:supp-df-sp-1acene}
\end{figure*}

\begin{figure*}[ht]
    \centering
    \subfloat[]{\includegraphics[width=0.29\linewidth]{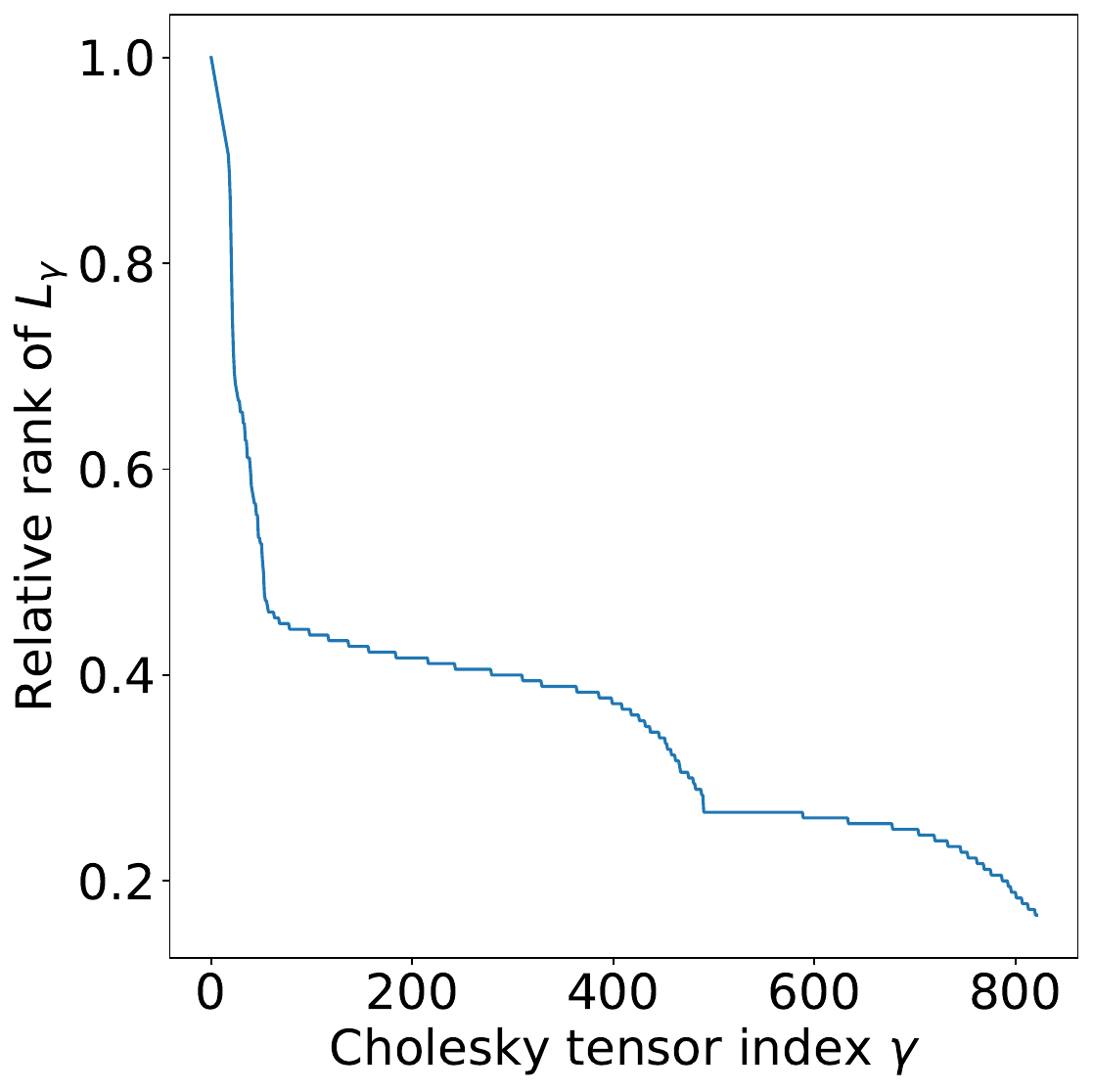}}
    \hfill
    \subfloat[]{\includegraphics[width=0.35\linewidth]{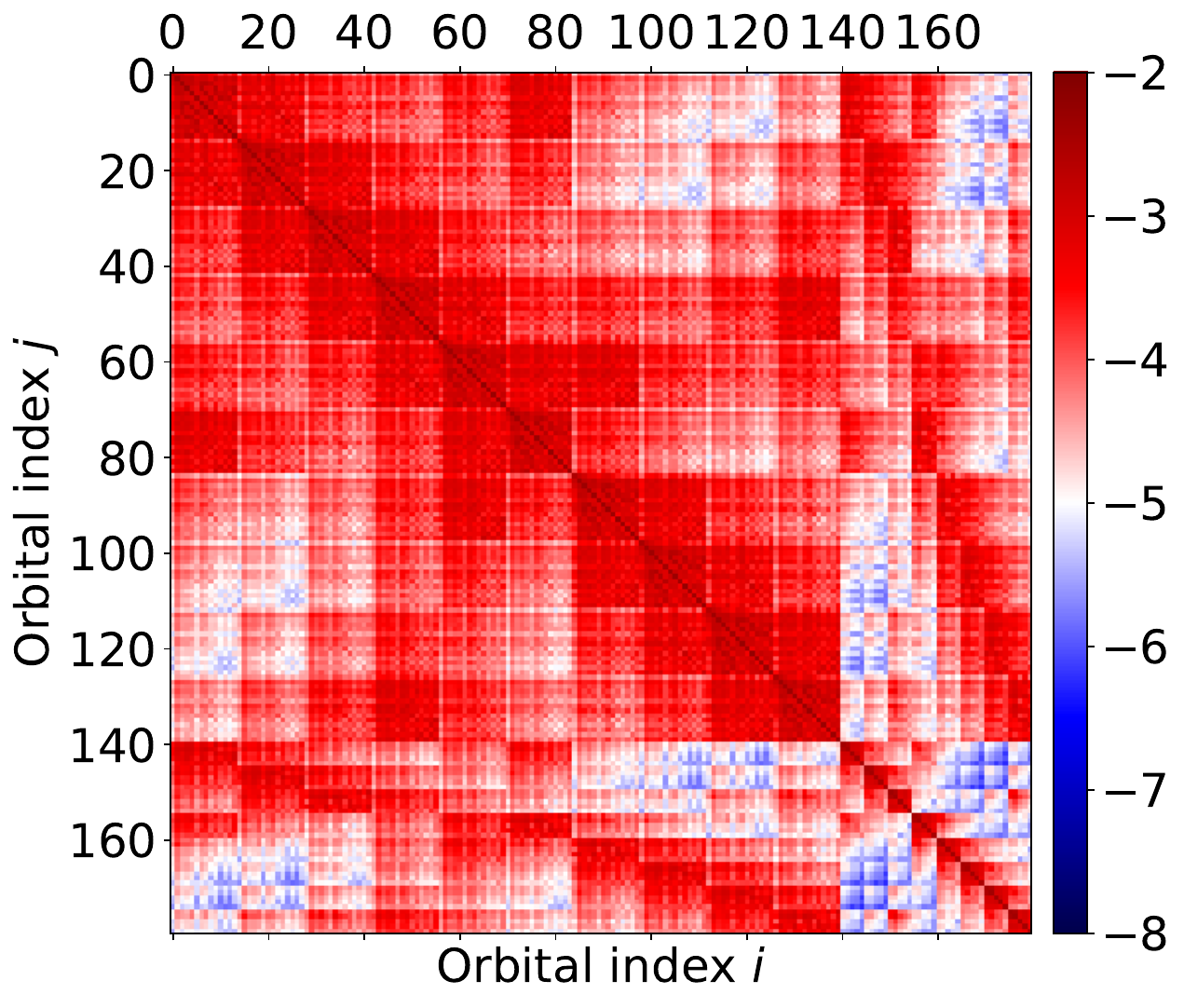}}
    \hfill
    \subfloat[]{\includegraphics[width=0.35\linewidth]{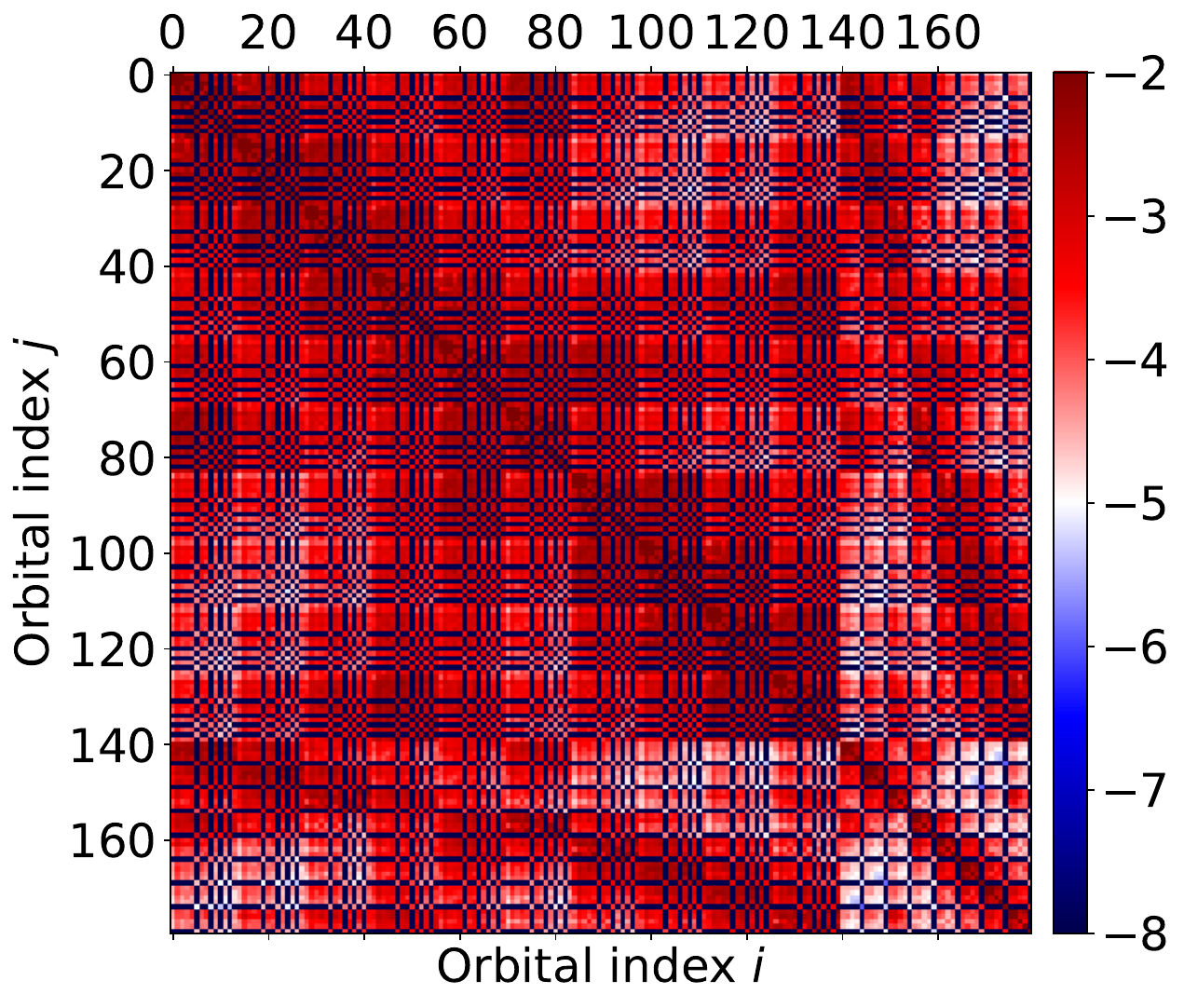}}
    \caption{(a) Relative ranks,
             (b--c) average sparsity of the Cholesky tensors for 2-acene (naphthalene).
             (c) averages only over Cholesky tensors with relative rank $\geq 1/2$.}
    \label{fig:supp-df-sp-2acene}
\end{figure*}

\begin{figure*}[ht]
    \centering
    \subfloat[]{\includegraphics[width=0.29\linewidth]{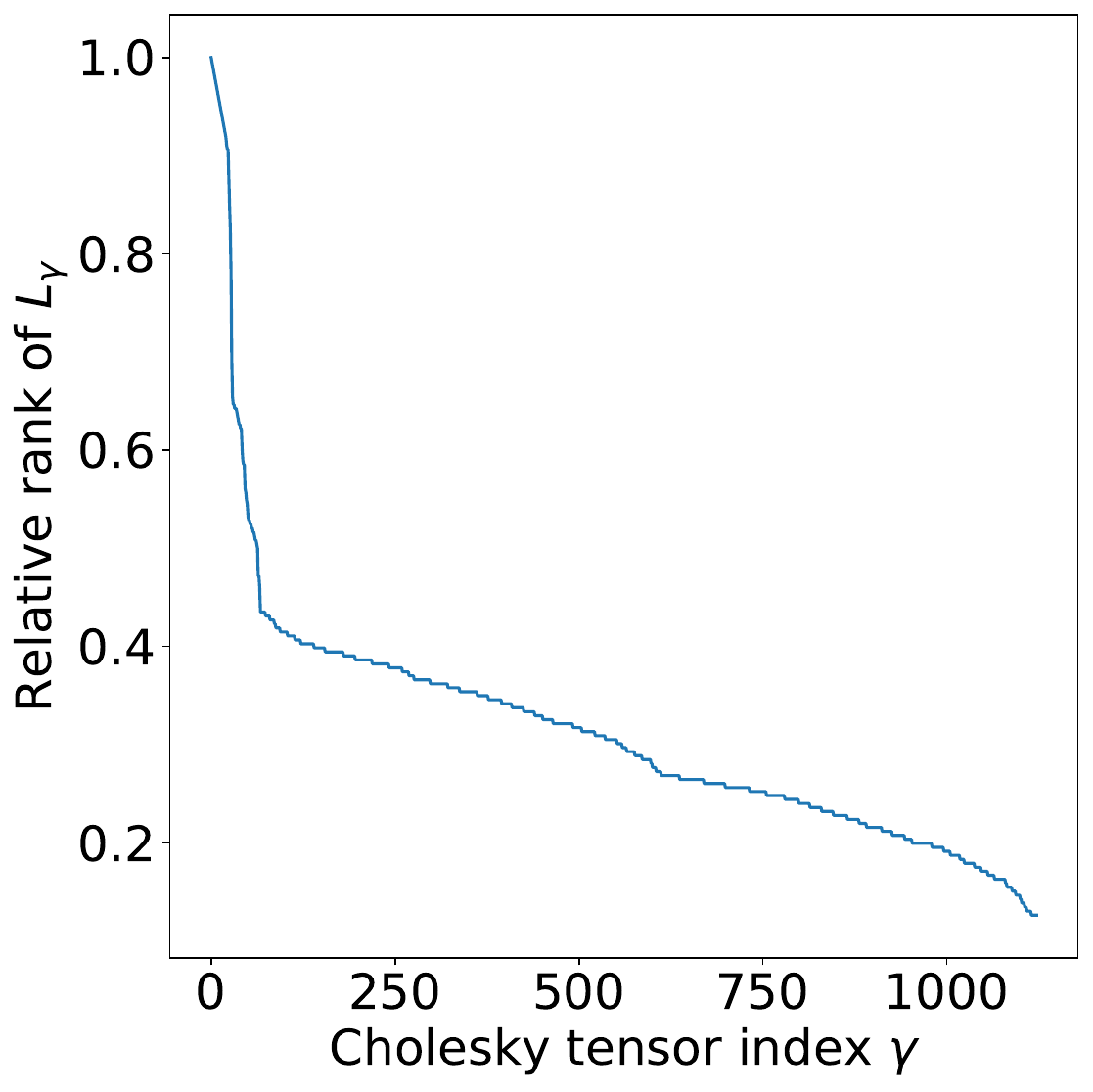}}
    \hfill
    \subfloat[]{\includegraphics[width=0.35\linewidth]{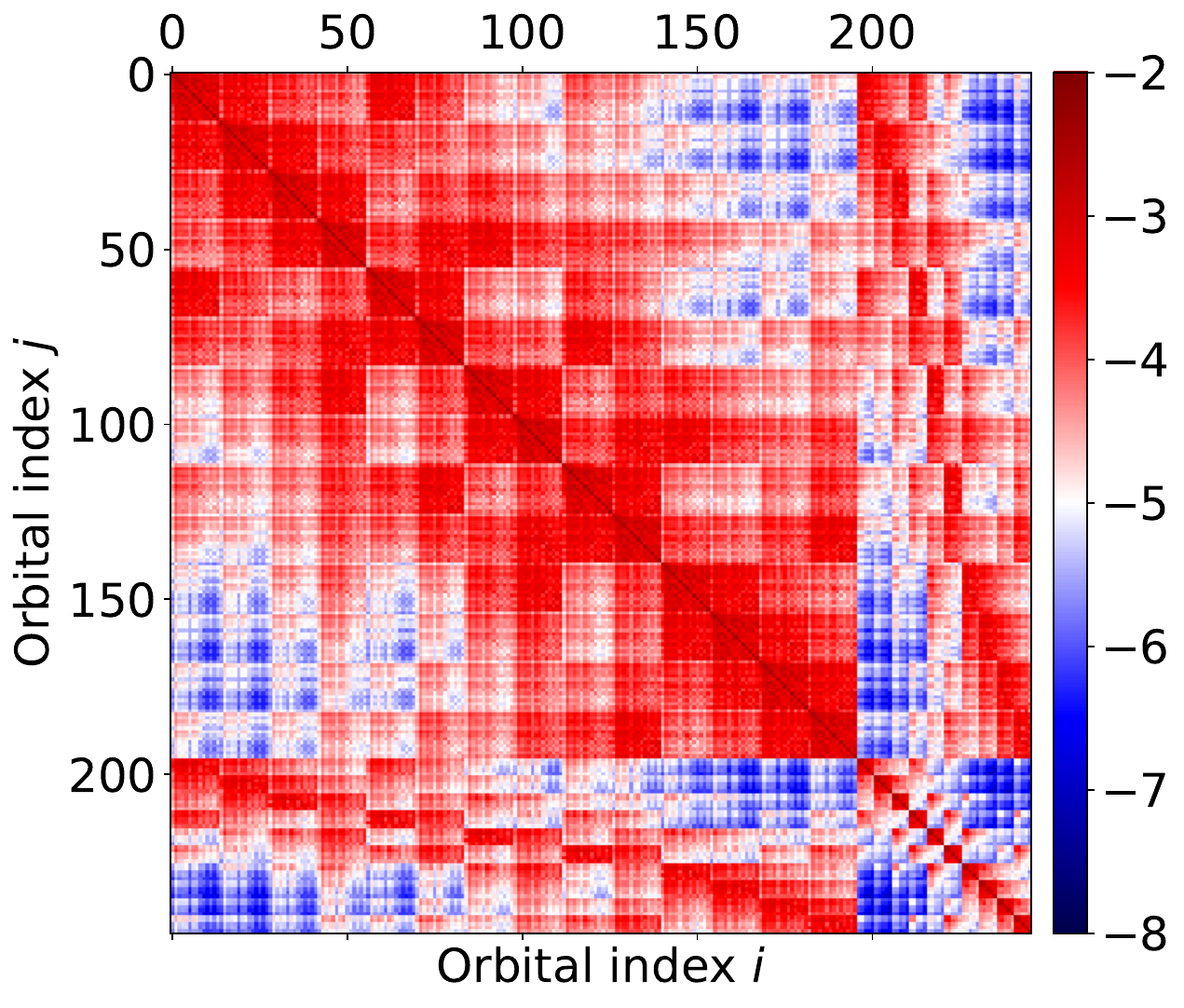}}
    \hfill
    \subfloat[]{\includegraphics[width=0.35\linewidth]{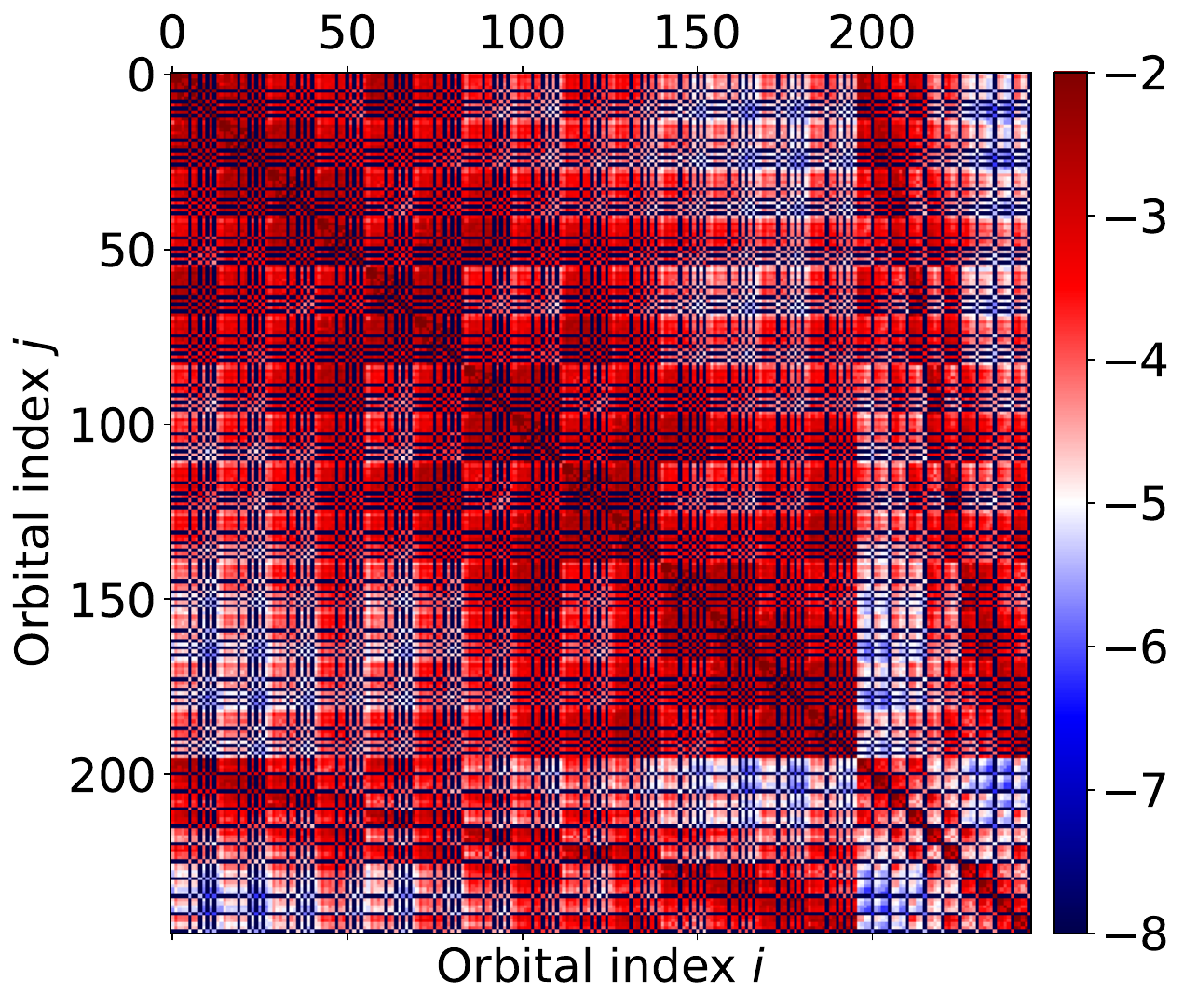}}
    \caption{(a) Relative ranks,
             (b--c) average sparsity of the Cholesky tensors for 3-acene (anthracene).
             (c) averages only over Cholesky tensors with relative rank $\geq 1/2$.}
    \label{fig:supp-df-sp-3acene}
\end{figure*}

\begin{figure*}[ht]
    \centering
    \subfloat[]{\includegraphics[width=0.29\linewidth]{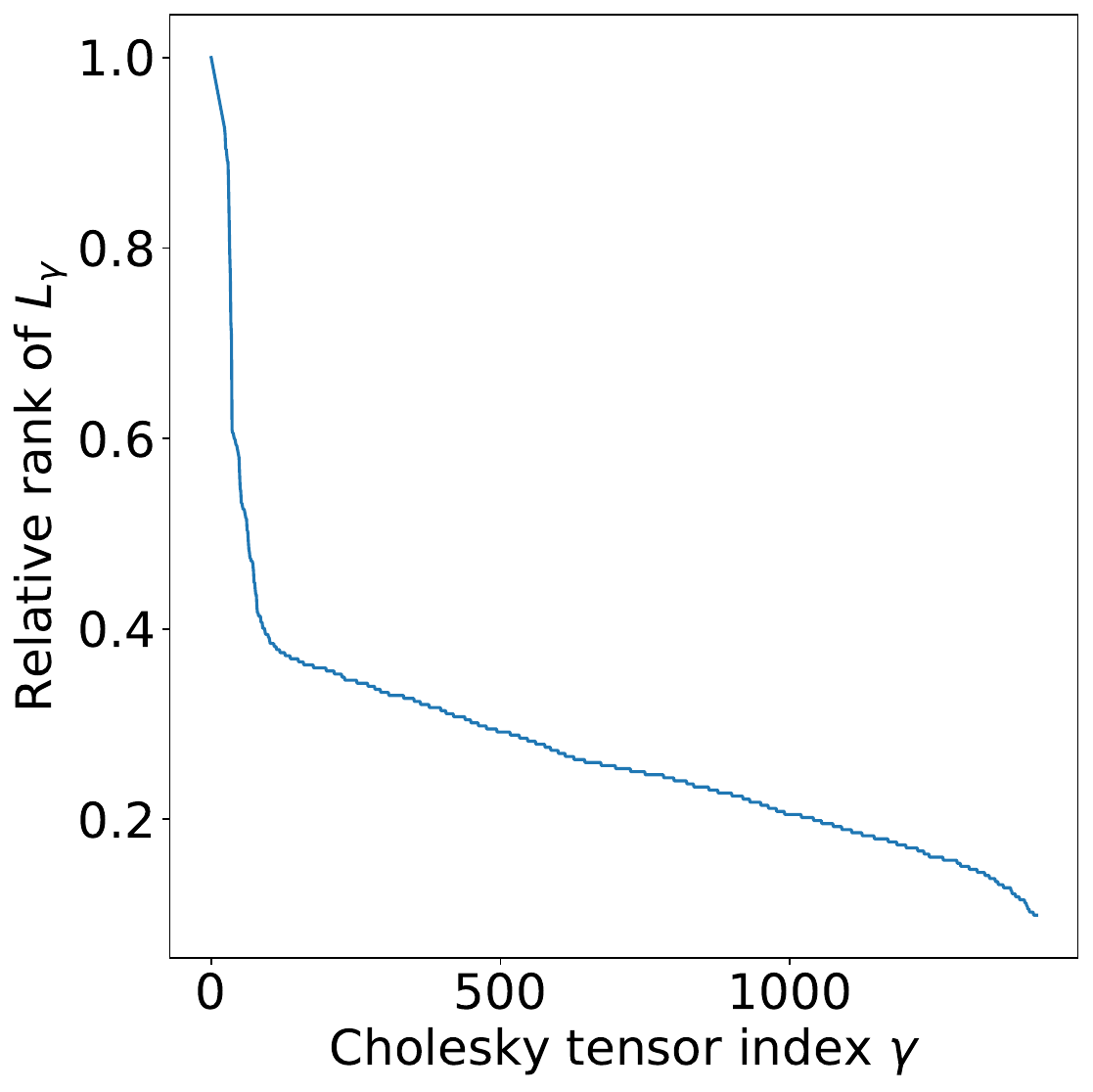}}
    \hfill
    \subfloat[]{\includegraphics[width=0.35\linewidth]{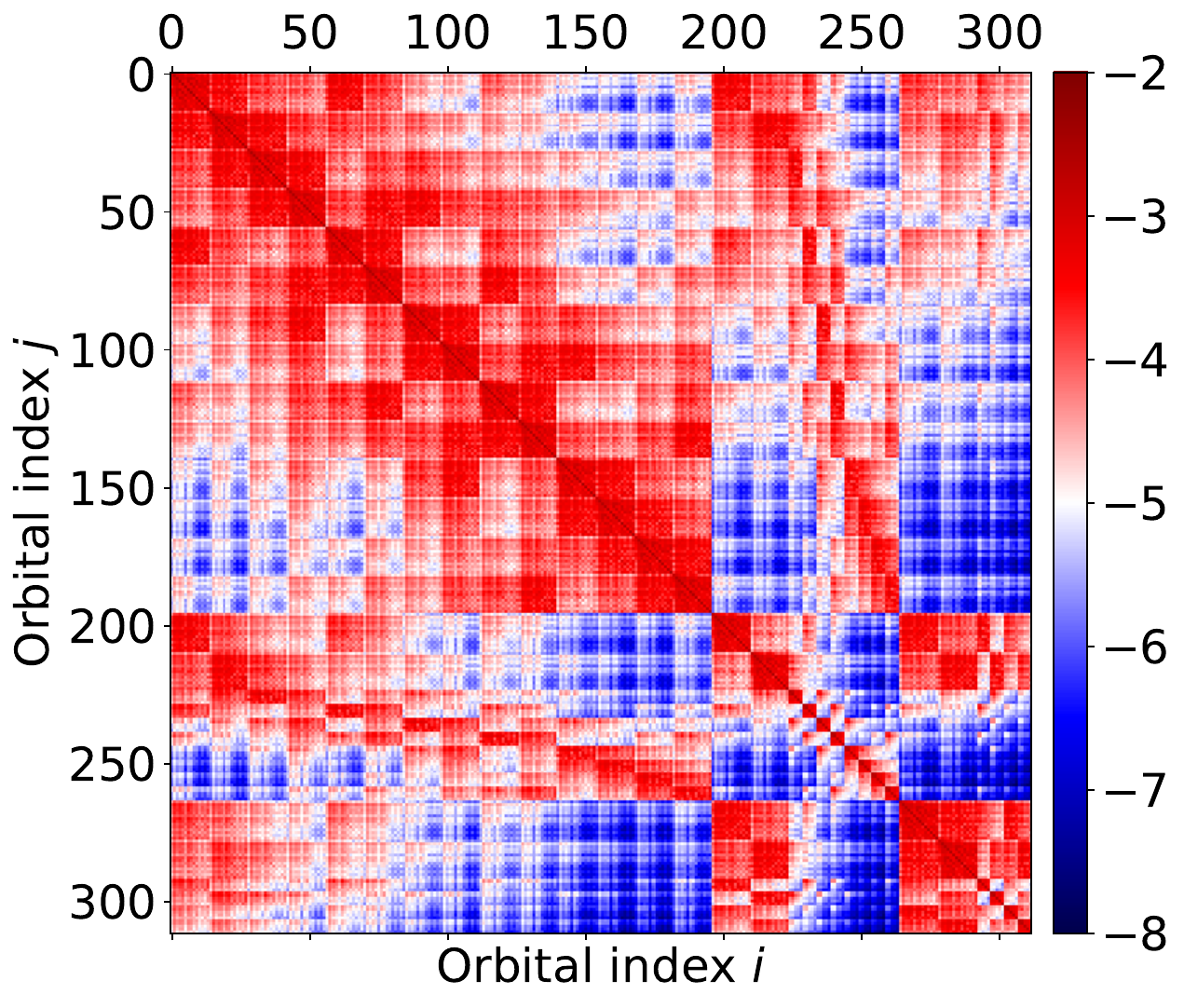}}
    \hfill
    \subfloat[]{\includegraphics[width=0.35\linewidth]{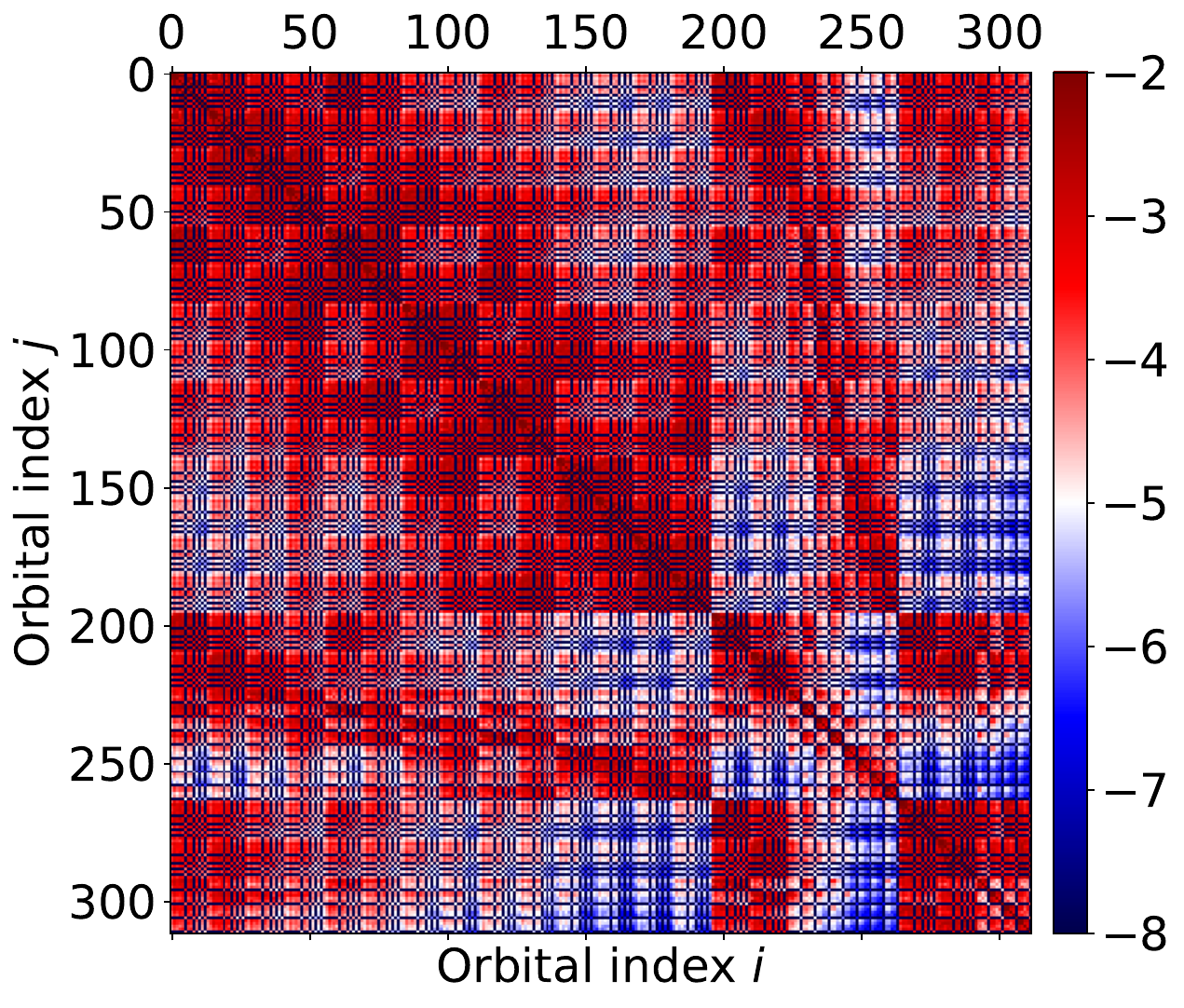}}
    \caption{(a) Relative ranks,
             (b--c) average sparsity of the Cholesky tensors for 4-acene (tetracene).
             (c) averages only over Cholesky tensors with relative rank $\geq 1/2$.}
    \label{fig:supp-df-sp-4acene}
\end{figure*}

\begin{figure*}[ht]
    \centering
    \subfloat[]{\includegraphics[width=0.29\linewidth]{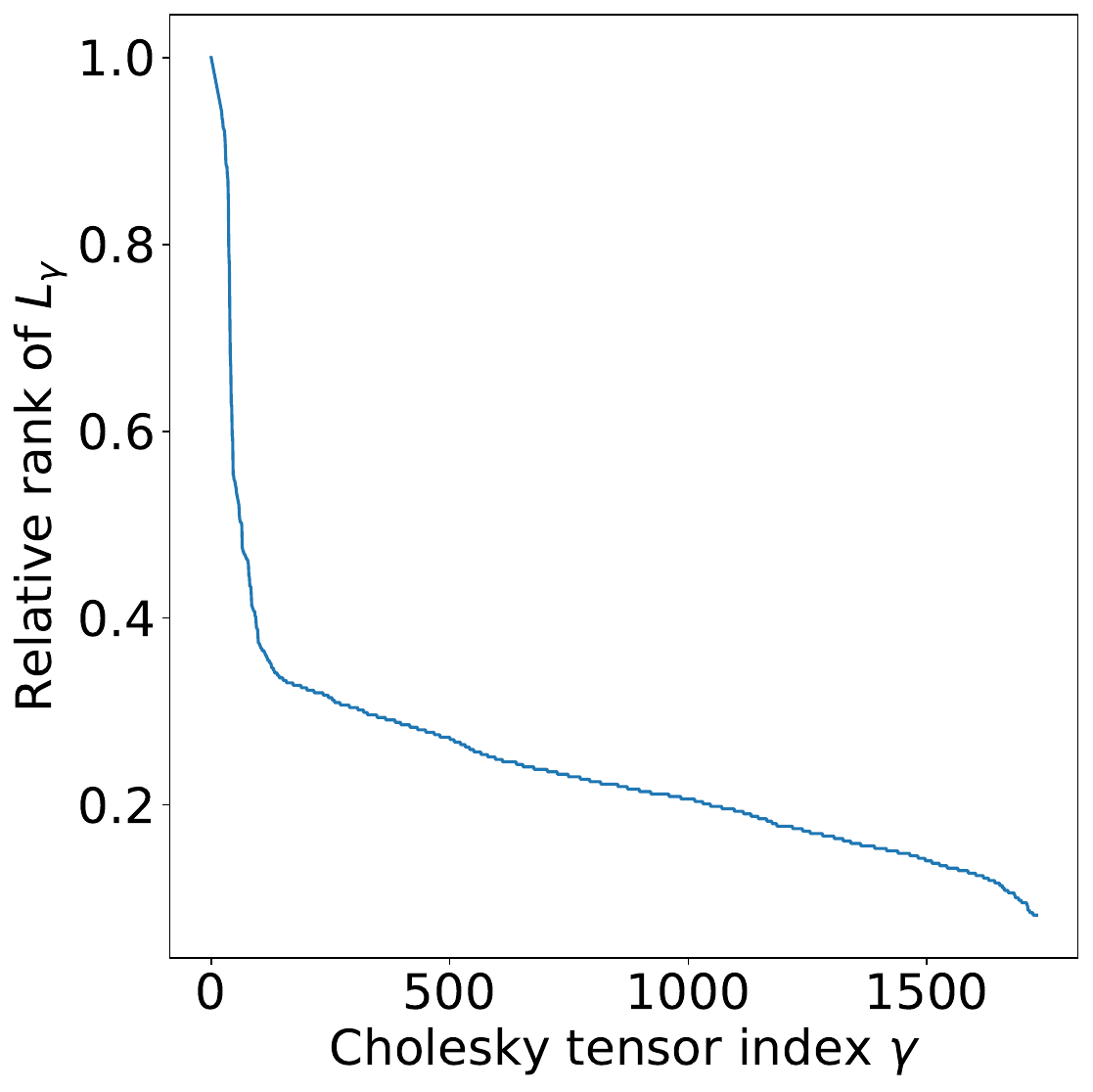}}
    \hfill
    \subfloat[]{\includegraphics[width=0.35\linewidth]{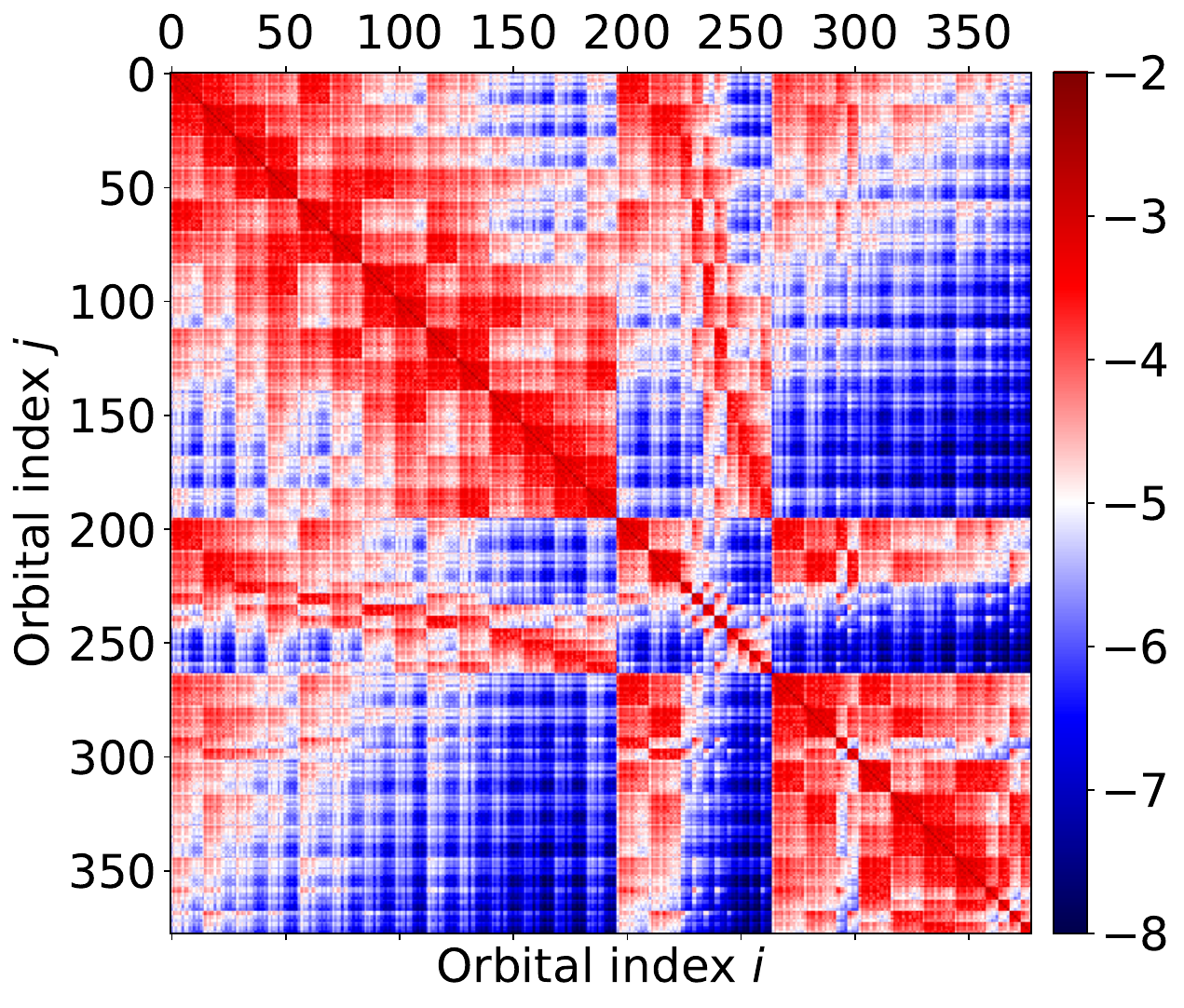}}
    \hfill
    \subfloat[]{\includegraphics[width=0.35\linewidth]{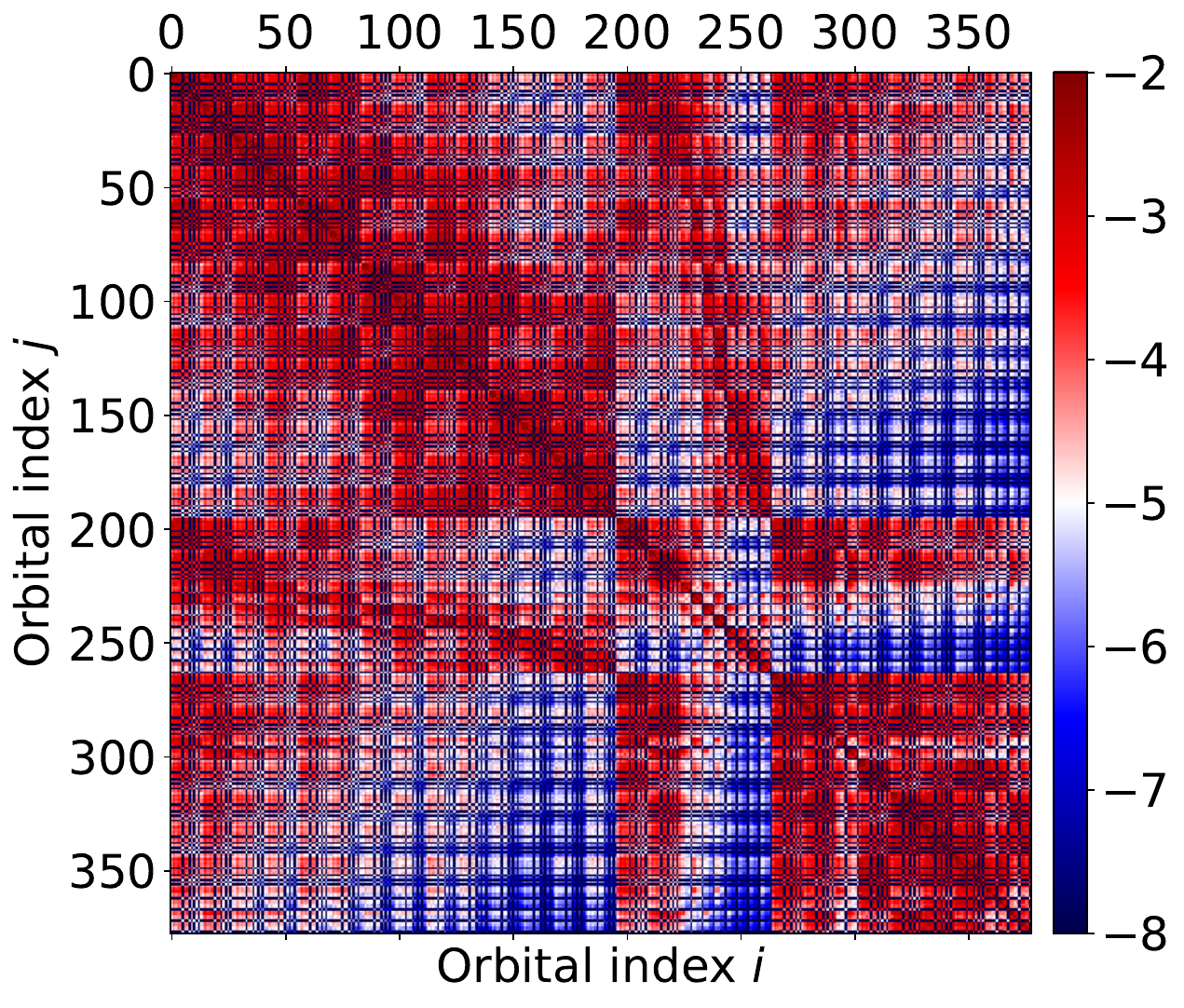}}
    \caption{(a) Relative ranks,
             (b--c) average sparsity of the Cholesky tensors for 5-acene (pentacene).
             (c) averages only over Cholesky tensors with relative rank $\geq 1/2$.}
    \label{fig:supp-df-sp-5acene}
\end{figure*}

\begin{figure*}[ht]
    \centering
    \subfloat[]{\includegraphics[width=0.29\linewidth]{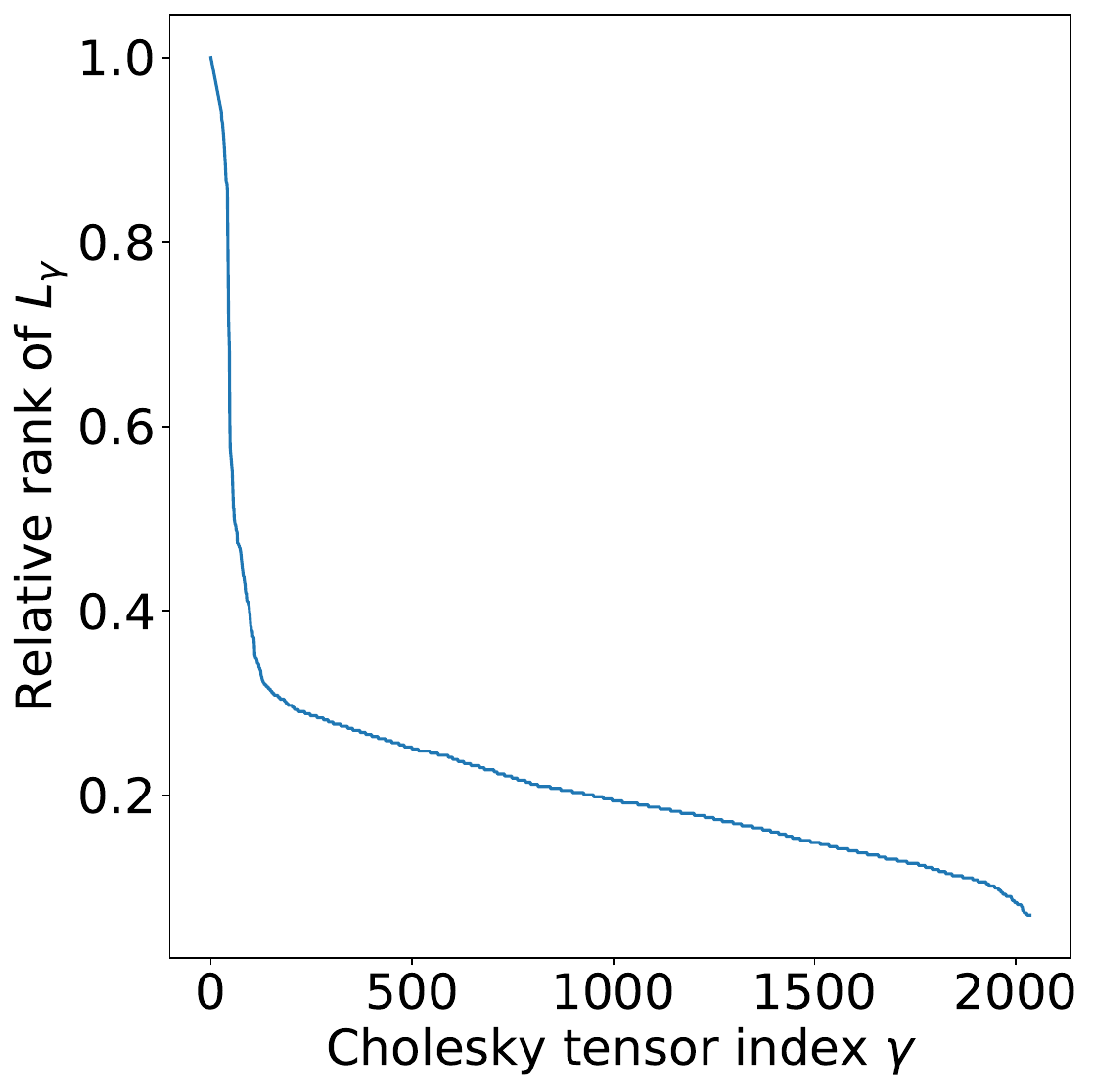}}
    \hfill
    \subfloat[]{\includegraphics[width=0.35\linewidth]{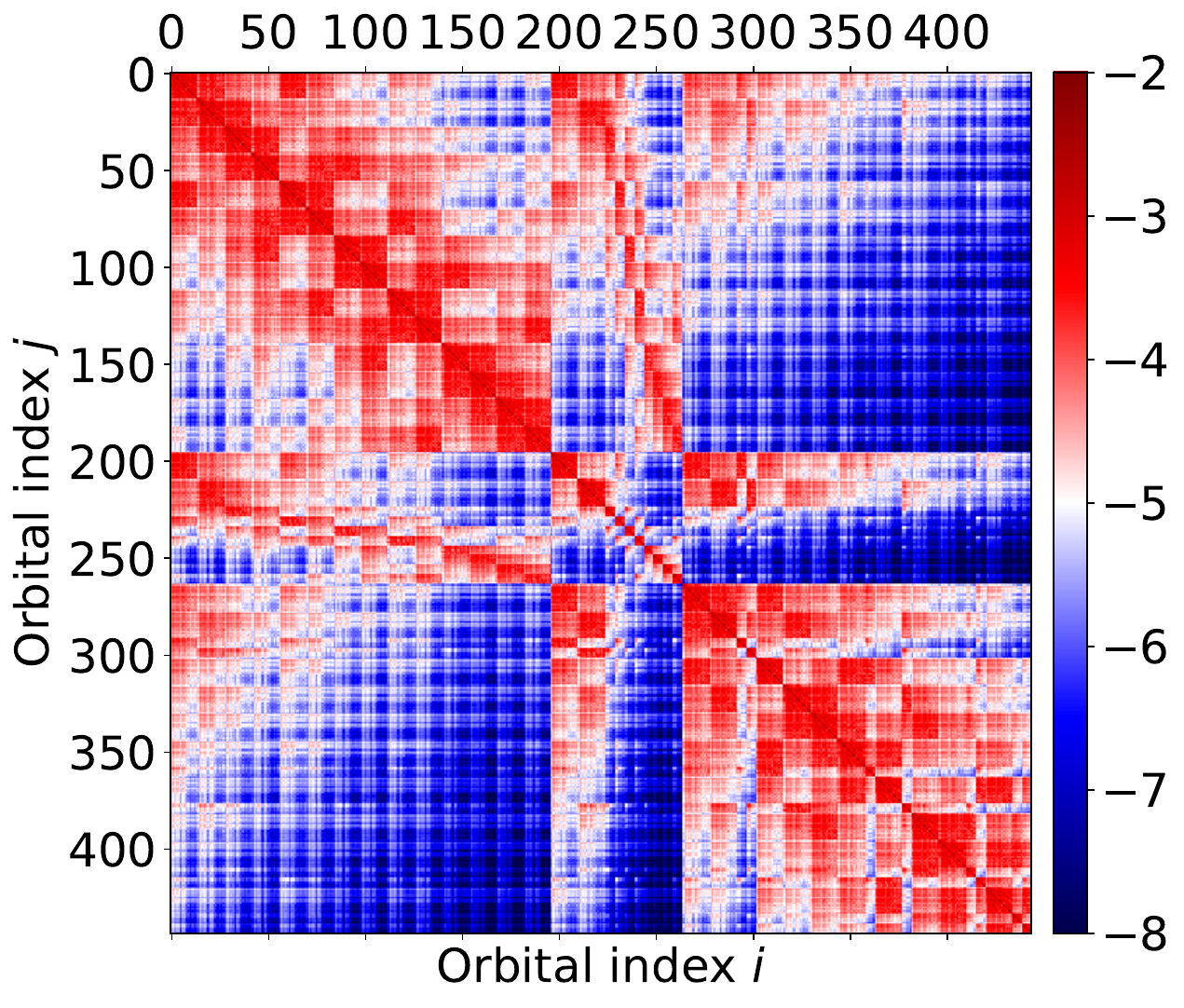}}
    \hfill
    \subfloat[]{\includegraphics[width=0.35\linewidth]{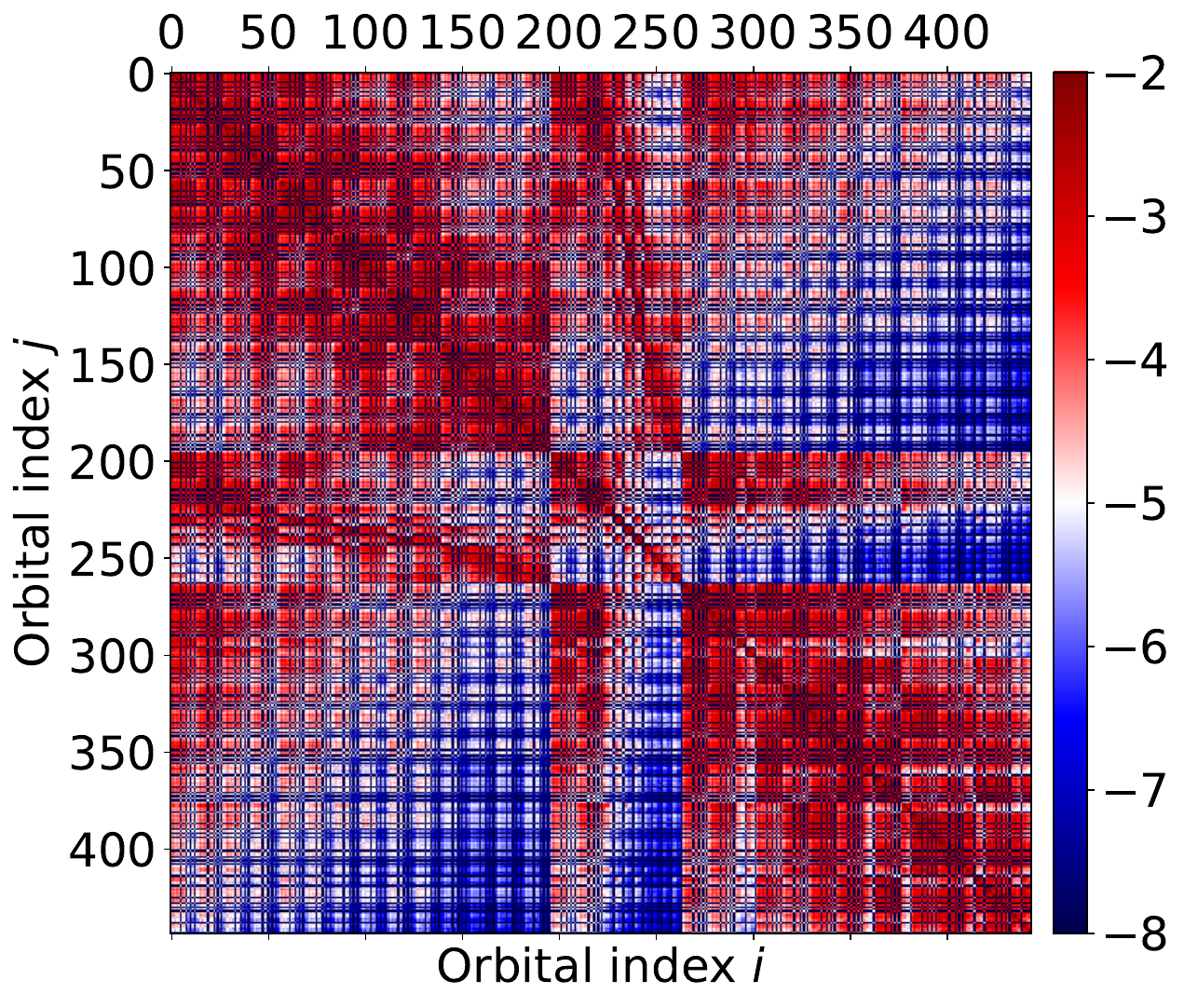}}
    \caption{(a) Relative ranks,
             (b--c) average sparsity of the Cholesky tensors for 6-acene (hexacene).
             (c) averages only over Cholesky tensors with relative rank $\geq 1/2$.}
    \label{fig:supp-df-sp-6acene}
\end{figure*}


\ifdefined\CGBSupplementBodyOnly
\else
\bibliography{quantcomput_ref}
\fi

\CGBStandaloneEnd
\let\CGBSupplementBodyOnly\undefined
\let\CGBStandaloneEnd\undefined
\let\CGBFigureFour\undefined
\let\CGBFigureFive\undefined

\end{widetext}

\end{document}